\documentclass[submission,copyright,creativecommons]{eptcs}
\makeatletter
\@ifundefined{lhs2tex.lhs2tex.sty.read}%
  {\@namedef{lhs2tex.lhs2tex.sty.read}{}%
   \newcommand\SkipToFmtEnd{}%
   \newcommand\EndFmtInput{}%
   \long\def\SkipToFmtEnd#1\EndFmtInput{}%
  }\SkipToFmtEnd

\newcommand\ReadOnlyOnce[1]{\@ifundefined{#1}{\@namedef{#1}{}}\SkipToFmtEnd}
\usepackage{amstext}
\usepackage{amssymb}
\usepackage{stmaryrd}
\DeclareFontFamily{OT1}{cmtex}{}
\DeclareFontShape{OT1}{cmtex}{m}{n}
  {<5><6><7><8>cmtex8
   <9>cmtex9
   <10><10.95><12><14.4><17.28><20.74><24.88>cmtex10}{}
\DeclareFontShape{OT1}{cmtex}{m}{it}
  {<-> ssub * cmtt/m/it}{}

\DeclareFontShape{OT1}{cmtt}{bx}{n}
  {<5><6><7><8>cmtt8
   <9>cmbtt9
   <10><10.95><12><14.4><17.28><20.74><24.88>cmbtt10}{}
\DeclareFontShape{OT1}{cmtex}{bx}{n}
  {<-> ssub * cmtt/bx/n}{}

\newcommand{\Conid}[1]{\mathit{#1}}
\newcommand{\Varid}[1]{\mathit{#1}}
\newcommand{\anonymous}{\kern0.06em \vbox{\hrule\@width.5em}}

\newcommand{\bind}{\mathbin{>\!\!\!>\mkern-6.7mu=}}

\renewcommand{\leq}{\leqslant}
\renewcommand{\geq}{\geqslant}
\usepackage{polytable}

\@ifundefined{mathindent}%
  {\newdimen\mathindent\mathindent\leftmargini}%
  {}%

\def\resethooks{%
  \global\let\SaveRestoreHook\empty
  \global\let\ColumnHook\empty}
\newcommand*{\savecolumns}[1][default]%
  {\g@addto@macro\SaveRestoreHook{\savecolumns[#1]}}
\newcommand*{\restorecolumns}[1][default]%
  {\g@addto@macro\SaveRestoreHook{\restorecolumns[#1]}}
\newcommand*{\aligncolumn}[2]%
  {\g@addto@macro\ColumnHook{\column{#1}{#2}}}

\resethooks

\newcommand{\onelinecommentchars}{\quad-{}- }
\newcommand{\commentbeginchars}{\enskip\{-}
\newcommand{\commentendchars}{-\}\enskip}

\newcommand{\visiblecomments}{%
  \let\onelinecomment=\onelinecommentchars
  \let\commentbegin=\commentbeginchars
  \let\commentend=\commentendchars}

\newcommand{\invisiblecomments}{%
  \let\onelinecomment=\empty
  \let\commentbegin=\empty
  \let\commentend=\empty}

\visiblecomments

\newlength{\blanklineskip}
\newcommand{\hsindent}[1]{\quad}
\let\hspre\empty
\let\hspost\empty

\EndFmtInput
\makeatother
\renewcommand{\commentbeginchars}{\quad(\ensuremath{\ast\;}}
\renewcommand{\commentendchars}{\ensuremath{\;\ast})\quad}
\renewcommand{\commentbegin}{\commentbeginchars}
\renewcommand{\commentend}{\commentendchars}
\newcommand{\Tyvarid}[1]{\Varid{\!\!\mbox{}^\shortmid\!\!#1}}
\newcommand{\Keyword}[1]{\mathbf{#1}}
\renewcommand{\Conid}[1]{\mathrm{#1}}
\ReadOnlyOnce{polycode.fmt}%
\makeatletter

\newcommand{\hsnewpar}[1]%
  {{\parskip=0pt\parindent=0pt\par\vskip #1\noindent}}

\newcommand{\hscodestyle}{}

\newcommand{\sethscode}[1]%
  {\expandafter\let\expandafter\hscode\csname #1\endcsname
   \expandafter\let\expandafter\endhscode\csname end#1\endcsname}

  {\par\noindent
   \advance\leftskip\mathindent
   \hscodestyle
   \let\\=\@normalcr
   \let\hspre\(\let\hspost\)%
   \pboxed}%
  {\endpboxed\)%
   \par\noindent
   \ignorespacesafterend}

  {\hsnewpar\abovedisplayskip
   \advance\leftskip\mathindent
   \hscodestyle
   \let\hspre\(\let\hspost\)%
   \pboxed}%
  {\endpboxed%
   \hsnewpar\belowdisplayskip
   \ignorespacesafterend}

  {\hsnewpar\abovedisplayskip
   \advance\leftskip\mathindent
   \hscodestyle
   \let\\=\@normalcr
   \(\pboxed}%
  {\endpboxed\)%
   \hsnewpar\belowdisplayskip
   \ignorespacesafterend}

\newcommand{\plainhs}{\sethscode{plainhscode}}

\plainhs

  {\hsnewpar\abovedisplayskip
   \advance\leftskip\mathindent
   \hscodestyle
   \let\\=\@normalcr
   \(\parray}%
  {\endparray\)%
   \hsnewpar\belowdisplayskip
   \ignorespacesafterend}

  {\parray}{\endparray}

  {\(\parray}{\endparray\)}

\def\codeframewidth{\arrayrulewidth}
\RequirePackage{calc}

  {\parskip=\abovedisplayskip\par\noindent
   \hscodestyle
   \arrayrulewidth=\codeframewidth
   \tabular{@{}|p{\linewidth-2\arraycolsep-2\arrayrulewidth-2pt}|@{}}%
   \hline\framedhslinecorrect\\{-1.5ex}%
   \let\endoflinesave=\\
   \let\\=\@normalcr
   \(\pboxed}%
  {\endpboxed\)%
   \framedhslinecorrect\endoflinesave{.5ex}\hline
   \endtabular
   \parskip=\belowdisplayskip\par\noindent
   \ignorespacesafterend}

\newcommand{\framedhslinecorrect}[2]%
  {#1[#2]}

  {\(\def\column##1##2{}%
   \let\>\undefined\let\<\undefined\let\\\undefined
   \newcommand\>[1][]{}\newcommand\<[1][]{}\newcommand\\[1][]{}%
   \def\fromto##1##2##3{##3}%
   }{\) }%

  {\let\orighscode=\hscode
   \let\origendhscode=\endhscode
   \def\endhscode{\def\hscode{\endgroup\def\@currenvir{hscode}\\}\begingroup}
   \orighscode\def\hscode{\endgroup\def\@currenvir{hscode}}}%
  {\origendhscode
   \global\let\hscode=\orighscode
   \global\let\endhscode=\origendhscode}%

\makeatother
\EndFmtInput
\providecommand{\event}{OCaml 2016} 
\usepackage{breakurl}             

\usepackage{ucs}
\usepackage[utf8x]{inputenc}

\usepackage[T1]{fontenc} 
\usepackage{fixltx2e}
\usepackage{MnSymbol}
\usepackage{enumitem}

\usepackage{listings}
\usepackage{color}
\newlength{\commentwidth}
\newcommand{\lstemph}[1]{\textrm{\textit{#1}}\hspace{0.1em}}
\newlength{\lstskip}
\newcommand{\lstmainstyle}{\sffamily}
\newcommand{\underscore}{\underbar{\ }}

\lstdefinestyle{inline}
  { basicstyle=\lstmainstyle}

\lstdefinestyle{knuth}
  { basicstyle=\rmfamily
  , identifierstyle=\itshape
  , columns=fullflexible
  , keepspaces=true 
  , morecomment=[n][\upshape]{(*}{*)}
  }

\newcommand{\ocaml}{\textsc{OCaml}}
\usepackage[textsize=ssmall]{todonotes}
\presetkeys{todonotes}{color=blue!5}{}

\title{Generic Programming in \ocaml\footnotetext{This work was partially supported by the Secure-OCaml FUI project.}
}

\author{Florent Balestrieri
\institute{ENSTA-ParisTech, Université Paris-Saclay}
\email{Florent.Balestrieri@ensta-paristech.fr}
\and
Michel Mauny
\institute{Inria Paris}
\email{Michel.Mauny@inria.fr}
}
\def\titlerunning{Generic Programming in OCaml}
\def\authorrunning{F.\ Balestrieri \& M.\ Mauny}

\begin{document}

\maketitle
\begin{abstract}
We present a library for generic programming in \ocaml{}, adapting some techniques borrowed from other functional languages.
The library makes use of three recent additions to \ocaml{}: generalised abstract datatypes are essential to reflect types, extensible variants allow this reflection to be open for new additions, and extension points provide syntactic sugar and generate boiler plate code that simplify the use of the library.
The building blocks of the library can be used to support many approaches to generic programming through the concept of view.
Generic traversals are implemented on top of the library and provide powerful combinators to write concise definitions of recursive functions over complex tree types.
Our case study is a type-safe deserialisation function that respects type abstraction.
\end{abstract}

\section{Introduction}
Typed functional programming languages come with rich type systems guaranteeing strong safety properties for the programs.
However, the restrictions imposed by types, necessary to banish wrong programs, may prevent us from generalizing  over some particular programming patterns, thus leading to boilerplate code and duplicated logic.
Generic programming allows us to recover the loss of flexibility by adding an extra expressive layer to the language.

The purpose of this article is to describe the user interface and explain the implementation of a generic programming library\footnote{The library is available at \url{https://github.com/balez/generic}} for the language \ocaml{}. We illustrate its usefulness with an implementation of a type-safe deserialisation function.

\subsection{A Motivating Example}
Algebraic datatypes are very suitable for capturing structured data, in particular trees.  However general tree operations need to be defined for each specific tree type, resulting in repetitive code.

Consider the height of a tree, which is the length of the longest path from the root to a leaf.
We will define a different height function on lists, binary trees and rose trees.

For lists, the height corresponds to the length of the list.
\begin{hscode}\SaveRestoreHook
\column{B}{@{}>{\hspre}l<{\hspost}@{}}%
\column{28}{@{}>{\hspre}l<{\hspost}@{}}%
\column{41}{@{}>{\hspre}l<{\hspost}@{}}%
\column{E}{@{}>{\hspre}l<{\hspost}@{}}%
\>[B]{}\Keyword{let}\;\Keyword{rec}\;\Varid{length}\mathrel{=}\Keyword{function}{}\<[28]%
\>[28]{}\mid [\mskip1.5mu \mskip1.5mu]{}\<[41]%
\>[41]{}\to \mathrm{0}{}\<[E]%
\\
\>[28]{}\mid \anonymous \mathbin{::}\Varid{tail}{}\<[41]%
\>[41]{}\to \mathrm{1}\mathbin{+}\Varid{length}\;\Varid{tail}{}\<[E]%
\ColumnHook
\end{hscode}\resethooks
For binary trees the definition is very similar, but in the inductive step we must now take the maximum of the heights of the children.
\begin{hscode}\SaveRestoreHook
\column{B}{@{}>{\hspre}l<{\hspost}@{}}%
\column{10}{@{}>{\hspre}l<{\hspost}@{}}%
\column{20}{@{}>{\hspre}l<{\hspost}@{}}%
\column{32}{@{}>{\hspre}l<{\hspost}@{}}%
\column{50}{@{}>{\hspre}l<{\hspost}@{}}%
\column{E}{@{}>{\hspre}l<{\hspost}@{}}%
\>[B]{}\mathbf{type\;}\;{}\<[10]%
\>[10]{}\Tyvarid{a}\;\Varid{btree}{}\<[20]%
\>[20]{}\mathrel{=}\Conid{Empty}\mid \Conid{Node}\;\mathbf{of\;}\;\Tyvarid{a}\;\Varid{btree}\mathbin{\times}\Tyvarid{a}\mathbin{\times}\Tyvarid{a}\;\Varid{btree}{}\<[E]%
\\
\>[B]{}\Keyword{let}\;\Keyword{rec}\;{}\<[10]%
\>[10]{}\Varid{bheight}{}\<[20]%
\>[20]{}\mathrel{=}\Keyword{function}{}\<[32]%
\>[32]{}\mid \Conid{Empty}{}\<[50]%
\>[50]{}\to \mathrm{0}{}\<[E]%
\\
\>[32]{}\mid \Conid{Node}\;(\Varid{l},\,\anonymous ,\,\Varid{r}){}\<[50]%
\>[50]{}\to \mathrm{1}\mathbin{+}\Varid{max}\;(\Varid{bheight}\;\Varid{l})\;(\Varid{bheight}\;\Varid{r}){}\<[E]%
\ColumnHook
\end{hscode}\resethooks
Rose trees have nodes of variable arity, we define them as records with two fields: \ensuremath{\Varid{attr}} is the data associated to a node, and \ensuremath{\Varid{children}} is the list of its immediate subtrees.
\begin{hscode}\SaveRestoreHook
\column{B}{@{}>{\hspre}l<{\hspost}@{}}%
\column{10}{@{}>{\hspre}l<{\hspost}@{}}%
\column{25}{@{}>{\hspre}c<{\hspost}@{}}%
\column{25E}{@{}l@{}}%
\column{28}{@{}>{\hspre}l<{\hspost}@{}}%
\column{39}{@{}>{\hspre}l<{\hspost}@{}}%
\column{E}{@{}>{\hspre}l<{\hspost}@{}}%
\>[B]{}\mathbf{type\;}\;{}\<[10]%
\>[10]{}\Tyvarid{a}\;\Varid{rtree}{}\<[25]%
\>[25]{}\mathrel{=}{}\<[25E]%
\>[28]{}\{\mskip1.5mu \Varid{attr}\mathrel{\;:\;}\Tyvarid{a}\mathbin{;\;}\quad\Varid{children}\mathrel{\;:\;}\Tyvarid{a}\;\Varid{rtree}\;\Varid{list}\mskip1.5mu\}{}\<[E]%
\\
\>[B]{}\Keyword{let}\;\Keyword{rec}\;{}\<[10]%
\>[10]{}\Varid{rheight}\;\Varid{rtree}{}\<[25]%
\>[25]{}\mathrel{=}{}\<[25E]%
\>[28]{}\Keyword{match}\;\Varid{\Conid{List}.map}\;\Varid{rheight}\;\Varid{rtree}.\Varid{children}\;\Keyword{with}{}\<[E]%
\\
\>[28]{}\mid [\mskip1.5mu \mskip1.5mu]{}\<[39]%
\>[39]{}\to \mathrm{0}\mbox{\commentbegin  The height of a leaf is zero.  \commentend}{}\<[E]%
\\
\>[28]{}\mid \Varid{h}\mathbin{::}\Varid{hs}{}\<[39]%
\>[39]{}\to \mathrm{1}\mathbin{+}\Varid{\Conid{List}.fold\char95 left}\;\Varid{max}\;\Varid{h}\;\Varid{hs}{}\<[E]%
\ColumnHook
\end{hscode}\resethooks
The reader can see how the definition of new types of trees would require the implementation of their own specialised height function.
Yet we can see a common pattern emerging. Is it possible to factorise the common behaviour?
Yes! thanks to parametric polymorphism and higher-order functions, we may abstract over the notion of children: the function \ensuremath{\Varid{gheight}} below takes an argument function  \ensuremath{\Varid{children}} that computes the list of children of a node.

\begin{hscode}\SaveRestoreHook
\column{B}{@{}>{\hspre}l<{\hspost}@{}}%
\column{10}{@{}>{\hspre}l<{\hspost}@{}}%
\column{33}{@{}>{\hspre}c<{\hspost}@{}}%
\column{33E}{@{}l@{}}%
\column{36}{@{}>{\hspre}l<{\hspost}@{}}%
\column{40}{@{}>{\hspre}l<{\hspost}@{}}%
\column{51}{@{}>{\hspre}l<{\hspost}@{}}%
\column{E}{@{}>{\hspre}l<{\hspost}@{}}%
\>[B]{}\Keyword{val}\;{}\<[10]%
\>[10]{}\Varid{gheight}{}\<[33]%
\>[33]{}\mathrel{\;:\;}{}\<[33E]%
\>[36]{}(\Tyvarid{a}\to \Tyvarid{a}\;\Varid{list})\to \Tyvarid{a}\to \Varid{int}{}\<[E]%
\\
\>[B]{}\Keyword{let}\;\Keyword{rec}\;{}\<[10]%
\>[10]{}\Varid{gheight}\;\Varid{children}\;\Varid{tree}{}\<[33]%
\>[33]{}\mathrel{=}{}\<[33E]%
\>[36]{}\Keyword{let}\;\Varid{subtrees}\mathrel{=}\Varid{children}\;\Varid{tree}{}\<[E]%
\\
\>[36]{}\Keyword{in}\;{}\<[40]%
\>[40]{}\Keyword{match}\;\Varid{\Conid{List}.map}\;(\Varid{gheight}\;\Varid{children})\;\Varid{subtrees}\;\Keyword{with}{}\<[E]%
\\
\>[40]{}\mid [\mskip1.5mu \mskip1.5mu]{}\<[51]%
\>[51]{}\to \mathrm{0}{}\<[E]%
\\
\>[40]{}\mid \Varid{h}\mathbin{::}\Varid{hs}{}\<[51]%
\>[51]{}\to \mathrm{1}\mathbin{+}\Varid{\Conid{List}.fold\char95 left}\;\Varid{max}\;\Varid{h}\;\Varid{hs}{}\<[E]%
\ColumnHook
\end{hscode}\resethooks
Then each particular case above can be implemented using \ensuremath{\Varid{gheight}} by providing the appropriate implementation of \ensuremath{\Varid{children}}:
\begin{hscode}\SaveRestoreHook
\column{B}{@{}>{\hspre}l<{\hspost}@{}}%
\column{15}{@{}>{\hspre}l<{\hspost}@{}}%
\column{18}{@{}>{\hspre}c<{\hspost}@{}}%
\column{18E}{@{}l@{}}%
\column{21}{@{}>{\hspre}l<{\hspost}@{}}%
\column{46}{@{}>{\hspre}l<{\hspost}@{}}%
\column{53}{@{}>{\hspre}l<{\hspost}@{}}%
\column{71}{@{}>{\hspre}l<{\hspost}@{}}%
\column{83}{@{}>{\hspre}l<{\hspost}@{}}%
\column{E}{@{}>{\hspre}l<{\hspost}@{}}%
\>[B]{}\Keyword{let}\;\Varid{length'}\;{}\<[15]%
\>[15]{}\Varid{x}{}\<[18]%
\>[18]{}\mathrel{=}{}\<[18E]%
\>[21]{}\Varid{gheight}\;(\Keyword{function}\;[\mskip1.5mu \mskip1.5mu]{}\<[46]%
\>[46]{}\to [\mskip1.5mu \mskip1.5mu]{}\<[53]%
\>[53]{}\mid \anonymous \mathbin{::}\Varid{tail}{}\<[71]%
\>[71]{}\to [\mskip1.5mu \Varid{tail}\mskip1.5mu])\;{}\<[83]%
\>[83]{}\Varid{x}{}\<[E]%
\\
\>[B]{}\Keyword{let}\;\Varid{bheight'}\;{}\<[15]%
\>[15]{}\Varid{x}{}\<[18]%
\>[18]{}\mathrel{=}{}\<[18E]%
\>[21]{}\Varid{gheight}\;(\Keyword{function}\;\Conid{Empty}{}\<[46]%
\>[46]{}\to [\mskip1.5mu \mskip1.5mu]{}\<[53]%
\>[53]{}\mid \Conid{Node}\;(\Varid{l},\,\anonymous ,\,\Varid{r}){}\<[71]%
\>[71]{}\to [\mskip1.5mu \Varid{l}\mathbin{;\;}\Varid{r}\mskip1.5mu])\;{}\<[83]%
\>[83]{}\Varid{x}{}\<[E]%
\\
\>[B]{}\Keyword{let}\;\Varid{rheight'}\;{}\<[15]%
\>[15]{}\Varid{x}{}\<[18]%
\>[18]{}\mathrel{=}{}\<[18E]%
\>[21]{}\Varid{gheight}\;(\Keyword{fun}\;\Varid{x}\to \Varid{x}.\Varid{children})\;\Varid{x}{}\<[E]%
\ColumnHook
\end{hscode}\resethooks
Having factored the functionality of \ensuremath{\Varid{height}}, we are left with the task of implementing \ensuremath{\Varid{children}} for each datatype.
This task follows systematically from the definition of a type and this time the pattern cannot be abstracted.
This is when generic programming comes into play!
With generic programming, we can write a single \ensuremath{\Varid{children}} function working over all types. It is indexed by the type representation of its tree argument: a value of type \ensuremath{\Tyvarid{a}\;\Varid{ty}} is the value-level representation of the type \ensuremath{\Tyvarid{a}}.

\savecolumns
\begin{hscode}\SaveRestoreHook
\column{B}{@{}>{\hspre}l<{\hspost}@{}}%
\column{16}{@{}>{\hspre}c<{\hspost}@{}}%
\column{16E}{@{}l@{}}%
\column{19}{@{}>{\hspre}l<{\hspost}@{}}%
\column{E}{@{}>{\hspre}l<{\hspost}@{}}%
\>[B]{}\Keyword{val}\;\Varid{children}{}\<[16]%
\>[16]{}\mathrel{\;:\;}{}\<[16E]%
\>[19]{}\Tyvarid{a}\;\Varid{ty}\to \Tyvarid{a}\to \Tyvarid{a}\;\Varid{list}{}\<[E]%
\ColumnHook
\end{hscode}\resethooks
The type-indexed version of \ensuremath{\Varid{height}} is obtained by composing \ensuremath{\Varid{gheight}} and \ensuremath{\Varid{children}}:
\restorecolumns
\begin{hscode}\SaveRestoreHook
\column{B}{@{}>{\hspre}l<{\hspost}@{}}%
\column{6}{@{}>{\hspre}l<{\hspost}@{}}%
\column{16}{@{}>{\hspre}c<{\hspost}@{}}%
\column{16E}{@{}l@{}}%
\column{19}{@{}>{\hspre}l<{\hspost}@{}}%
\column{E}{@{}>{\hspre}l<{\hspost}@{}}%
\>[B]{}\Keyword{val}\;{}\<[6]%
\>[6]{}\Varid{height}{}\<[16]%
\>[16]{}\mathrel{\;:\;}{}\<[16E]%
\>[19]{}\Tyvarid{a}\;\Varid{ty}\to \Tyvarid{a}\to \Varid{int}{}\<[E]%
\\
\>[B]{}\Keyword{let}\;{}\<[6]%
\>[6]{}\Varid{height}\;\Varid{t}{}\<[16]%
\>[16]{}\mathrel{=}{}\<[16E]%
\>[19]{}\Varid{gheight}\;(\Varid{children}\;\Varid{t}){}\<[E]%
\ColumnHook
\end{hscode}\resethooks
The implementation of \ensuremath{\Varid{children}} will be explained in section~\ref{sec:children}.

Note: the type witness \ensuremath{\Tyvarid{a}\;\Varid{ty}} is explicitly given to a generic function, for instance if \ensuremath{\Varid{x}\mathrel{\;:\;}\Tyvarid{a}\;\Varid{list}} we might call \ensuremath{\Varid{height}\;\Varid{list}\;\Varid{x}} where \ensuremath{\Varid{list}} is a suitable value of type \ensuremath{\Tyvarid{a}\;\Varid{list}\;\Varid{ty}}.
It is theoretically possible to infer the type witness since there is a one to one correspondence between the witnesses and types. The work on \emph{modular implicits}~\cite{modular-implicits} promises to offer this functionality.

\subsection{A Case for Generic Programming}
\paragraph{Generic Traversals}
Some common operations on structured data, eg. abstract syntax trees require a lot of boilerplate code to traverse the data structure and modify it recursively, or extract some result.
This boilerplate code needs to be adapted to each new datatype.

When writing specific traversals over an AST using pattern matching over the constructors, it often happens that only a few cases carry the meaningful computation, others being default cases. Boilerplate removal allows us to write such function by only giving the meaningful cases. In addition to conciseness, this has the benefit of making the code robust to changes in the AST type: since the same function would treat additional constructors using the default case.
Generic traversals is the focus of section~\ref{sec:traversals}.

\paragraph{Ad-hoc Polymorphism}
\ocaml{} lacks an overloading mechanism such as Haskell type classes.
The generic library implements similar mechanisms, through explicit type representation and dynamic dispatch.
This feature is illustrated in the generic traversals of section~\ref{sec:traversals} in which we adapted some Haskell libraries that rely heavily on type classes.

\paragraph{Safer Alternatives to the Built-in Generic Functions in \ocaml{}}
The \ocaml{} standard library provides a few functions that perform black magic.
Such functions are defined over the concrete memory model of the \ocaml{} value runtime representation.
In fact one of them is actually called \ensuremath{\Varid{magic}\mathrel{\;:\;}\Tyvarid{a}\to \Tyvarid{b}} and does what its type suggests: casting a value to an arbitrary type, which is unsafe.
Deserialisation, as implemented by \ensuremath{\Varid{\Conid{Marshal}.from\char95 string}} is also unsafe.
Such operations can provoke segmentation faults if used unwisely.
Other magical operations---such as polymorphic comparisons and the polymorphic hash function---break the abstraction provided by abstract types: such types are often defined as quotients over an equivalence relation, yet the structural comparisons work on their concrete implementation instead of the equivalence classes.

With a generic programming library, the user can define alternatives to the built-in functions that are well-behaved regarding both type safety and abstraction.

\subsection{Overview of the article}

Section \ref{sec:GP} explains the three elements of a generic programming library: means to reflect types, to define type-dependent functions
and to represent the structure of types.
Section \ref{sec:traversals} shows how generic traversals can be defined on top of the library.
Section \ref{sec:unmarshal} covers a complex generic program implementing safe deserialisation.
Section \ref{sec:comparison} gives some context to our approach, which is compared with other implementations. We also discuss genericity within other type systems.
Section \ref{sec:conclusion} sums up the main points of the article.

\section{The Three Elements of Generic Programming}
\label{sec:GP}

Following the \emph{Generic Programming in 3D} approach~\cite{generic-programming-in-3d},
we identify three orthogonal dimensions in the design of generic programming libraries:
\begin{description}
\item[A reflection of types at the value level]
over which our generic functions are defined.
\item[A mechanism for overloading]
that enables us to define and call generic functions over different types.
\item[A generic view of types]
  that provides a uniform representation of types on top of which generic functions are recursively defined.
\end{description}

We describe in this section each dimension in turn, and give some examples of generic programs in section \ref{sec:traversals}.

\subsection{Type Reflection}
\label{sec:ty}
Generalised algebraic datatypes (GADT), introduced in \ocaml{} version~4, are \emph{type indexed} families of types.
Using GADTs, we can define singleton types where each index of the family is associated with a single data constructor.
The one to one correspondence between type indices and data constructors allows us to reflect types as values.

The syntax of GADT extends the syntax of variants by allowing the return type to be specified, where the indices may be instantiated to concrete types. Hence, we may reflect types as follows:
\begin{hscode}\SaveRestoreHook
\column{B}{@{}>{\hspre}l<{\hspost}@{}}%
\column{3}{@{}>{\hspre}l<{\hspost}@{}}%
\column{13}{@{}>{\hspre}c<{\hspost}@{}}%
\column{13E}{@{}l@{}}%
\column{16}{@{}>{\hspre}l<{\hspost}@{}}%
\column{23}{@{}>{\hspre}l<{\hspost}@{}}%
\column{32}{@{}>{\hspre}c<{\hspost}@{}}%
\column{32E}{@{}l@{}}%
\column{36}{@{}>{\hspre}c<{\hspost}@{}}%
\column{36E}{@{}l@{}}%
\column{39}{@{}>{\hspre}l<{\hspost}@{}}%
\column{43}{@{}>{\hspre}c<{\hspost}@{}}%
\column{43E}{@{}l@{}}%
\column{44}{@{}>{\hspre}l<{\hspost}@{}}%
\column{47}{@{}>{\hspre}l<{\hspost}@{}}%
\column{52}{@{}>{\hspre}l<{\hspost}@{}}%
\column{E}{@{}>{\hspre}l<{\hspost}@{}}%
\>[B]{}\mathbf{type\;}\;\anonymous \;\Varid{ty}\mathrel{=}{}\<[E]%
\\
\>[B]{}\hsindent{3}{}\<[3]%
\>[3]{}\mid \Conid{Int}{}\<[13]%
\>[13]{}\mathrel{\;:\;}{}\<[13E]%
\>[44]{}\Varid{int}\;{}\<[52]%
\>[52]{}\Varid{ty}{}\<[E]%
\\
\>[B]{}\hsindent{3}{}\<[3]%
\>[3]{}\mid \Conid{String}{}\<[13]%
\>[13]{}\mathrel{\;:\;}{}\<[13E]%
\>[44]{}\Varid{string}\;{}\<[52]%
\>[52]{}\Varid{ty}{}\<[E]%
\\
\>[B]{}\hsindent{3}{}\<[3]%
\>[3]{}\mid \Conid{List}{}\<[13]%
\>[13]{}\mathrel{\;:\;}{}\<[13E]%
\>[23]{}\Tyvarid{a}\;\Varid{ty}{}\<[32]%
\>[32]{}\to {}\<[32E]%
\>[39]{}\Tyvarid{a}\;{}\<[44]%
\>[44]{}\Varid{list}\;{}\<[52]%
\>[52]{}\Varid{ty}{}\<[E]%
\\
\>[B]{}\hsindent{3}{}\<[3]%
\>[3]{}\mid \Conid{Pair}{}\<[13]%
\>[13]{}\mathrel{\;:\;}{}\<[13E]%
\>[16]{}\Tyvarid{a}\;\Varid{ty}{}\<[23]%
\>[23]{}\mathbin{\times}\Tyvarid{b}\;\Varid{ty}{}\<[32]%
\>[32]{}\to {}\<[32E]%
\>[36]{}({}\<[36E]%
\>[39]{}\Tyvarid{a}{}\<[43]%
\>[43]{}\mathbin{\times}{}\<[43E]%
\>[47]{}\Tyvarid{b})\;{}\<[52]%
\>[52]{}\Varid{ty}{}\<[E]%
\\
\>[B]{}\hsindent{3}{}\<[3]%
\>[3]{}\mid \Conid{Fun}{}\<[13]%
\>[13]{}\mathrel{\;:\;}{}\<[13E]%
\>[16]{}\Tyvarid{a}\;\Varid{ty}{}\<[23]%
\>[23]{}\mathbin{\times}\Tyvarid{b}\;\Varid{ty}{}\<[32]%
\>[32]{}\to {}\<[32E]%
\>[36]{}({}\<[36E]%
\>[39]{}\Tyvarid{a}{}\<[43]%
\>[43]{}\to {}\<[43E]%
\>[47]{}\Tyvarid{b})\;{}\<[52]%
\>[52]{}\Varid{ty}{}\<[E]%
\ColumnHook
\end{hscode}\resethooks
Notice how we reflected type formers as value constructors of the same arity with type witnesses as arguments.
A complex type is reflected straightforwardly:
\begin{hscode}\SaveRestoreHook
\column{B}{@{}>{\hspre}l<{\hspost}@{}}%
\column{3}{@{}>{\hspre}c<{\hspost}@{}}%
\column{3E}{@{}l@{}}%
\column{6}{@{}>{\hspre}l<{\hspost}@{}}%
\column{27}{@{}>{\hspre}l<{\hspost}@{}}%
\column{41}{@{}>{\hspre}l<{\hspost}@{}}%
\column{51}{@{}>{\hspre}l<{\hspost}@{}}%
\column{E}{@{}>{\hspre}l<{\hspost}@{}}%
\>[B]{}\Conid{Fun}\;{}\<[6]%
\>[6]{}(\Conid{List}\;\Conid{String},\,\Conid{Fun}\;({}\<[27]%
\>[27]{}\Conid{Int},\,(\Conid{Pair}\;({}\<[41]%
\>[41]{}\Conid{String},\,{}\<[51]%
\>[51]{}\Conid{Int})))){}\<[E]%
\\
\>[B]{}\hsindent{3}{}\<[3]%
\>[3]{}\mathrel{\;:\;}{}\<[3E]%
\>[6]{}(\Varid{string}\;\Varid{list}\to {}\<[27]%
\>[27]{}\Varid{int}\to {}\<[41]%
\>[41]{}\Varid{string}\mathbin{\times}{}\<[51]%
\>[51]{}\Varid{int})\;\Varid{ty}{}\<[E]%
\ColumnHook
\end{hscode}\resethooks
\subsubsection{Open Types}
Introduced in version~4.02, open types allow us to extend \ensuremath{\Varid{ty}} with new cases reflecting newly introduced user types.
We declare an extensible type with:
\savecolumns
\begin{hscode}\SaveRestoreHook
\column{B}{@{}>{\hspre}l<{\hspost}@{}}%
\column{12}{@{}>{\hspre}c<{\hspost}@{}}%
\column{12E}{@{}l@{}}%
\column{16}{@{}>{\hspre}c<{\hspost}@{}}%
\column{16E}{@{}l@{}}%
\column{E}{@{}>{\hspre}l<{\hspost}@{}}%
\>[B]{}\mathbf{type\;}\;\anonymous \;\Varid{ty}{}\<[12]%
\>[12]{}\mathrel{=}{}\<[12E]%
\>[16]{}\mathinner{\ldotp\ldotp}{}\<[16E]%
\ColumnHook
\end{hscode}\resethooks
New cases are added with the syntax:
\restorecolumns
\begin{hscode}\SaveRestoreHook
\column{B}{@{}>{\hspre}l<{\hspost}@{}}%
\column{12}{@{}>{\hspre}c<{\hspost}@{}}%
\column{12E}{@{}l@{}}%
\column{16}{@{}>{\hspre}l<{\hspost}@{}}%
\column{23}{@{}>{\hspre}l<{\hspost}@{}}%
\column{35}{@{}>{\hspre}l<{\hspost}@{}}%
\column{39}{@{}>{\hspre}l<{\hspost}@{}}%
\column{46}{@{}>{\hspre}l<{\hspost}@{}}%
\column{E}{@{}>{\hspre}l<{\hspost}@{}}%
\>[B]{}\mathbf{type\;}\;\anonymous \;\Varid{ty}{}\<[12]%
\>[12]{}\mathrel{+\!\!\!=}{}\<[12E]%
\>[16]{}\Conid{Float}{}\<[23]%
\>[23]{}\mathrel{\;:\;}{}\<[39]%
\>[39]{}\Varid{float}\;{}\<[46]%
\>[46]{}\Varid{ty}{}\<[E]%
\\
\>[B]{}\mathbf{type\;}\;\anonymous \;\Varid{ty}{}\<[12]%
\>[12]{}\mathrel{+\!\!\!=}{}\<[12E]%
\>[16]{}\Conid{Btree}{}\<[23]%
\>[23]{}\mathrel{\;:\;}\Tyvarid{a}\;\Varid{ty}\to {}\<[35]%
\>[35]{}\Tyvarid{a}\;{}\<[39]%
\>[39]{}\Varid{btree}\;{}\<[46]%
\>[46]{}\Varid{ty}{}\<[E]%
\ColumnHook
\end{hscode}\resethooks
See \cite{ocaml-manual}, section ``Extensible variant types'' for more information about OCaml extensible types.

\paragraph{Objects and Polymorphic Variants}
Objects of anonymous classes and polymorphic variants are special amongst \ocaml{} types in that they are not nominal types. Therefore, they do not fit nicely with the nominal type witnesses. One possibility to support them indirectly is to give them a name. Another possibility is to break the general scheme of type witnesses, and provide two special constructors for objects and polymorphic variants:
\restorecolumns
\begin{hscode}\SaveRestoreHook
\column{B}{@{}>{\hspre}l<{\hspost}@{}}%
\column{12}{@{}>{\hspre}c<{\hspost}@{}}%
\column{12E}{@{}l@{}}%
\column{16}{@{}>{\hspre}l<{\hspost}@{}}%
\column{29}{@{}>{\hspre}c<{\hspost}@{}}%
\column{29E}{@{}l@{}}%
\column{32}{@{}>{\hspre}l<{\hspost}@{}}%
\column{53}{@{}>{\hspre}l<{\hspost}@{}}%
\column{E}{@{}>{\hspre}l<{\hspost}@{}}%
\>[B]{}\mathbf{type\;}\;\anonymous \;\Varid{ty}{}\<[12]%
\>[12]{}\mathrel{+\!\!\!=}{}\<[12E]%
\>[16]{}\Conid{Object}{}\<[29]%
\>[29]{}\mathrel{\;:\;}{}\<[29E]%
\>[32]{}\Tyvarid{a}\;\Varid{object\char95 desc}{}\<[53]%
\>[53]{}\to \Tyvarid{a}\;\Varid{ty}{}\<[E]%
\\
\>[B]{}\mathbf{type\;}\;\anonymous \;\Varid{ty}{}\<[12]%
\>[12]{}\mathrel{+\!\!\!=}{}\<[12E]%
\>[16]{}\Conid{PolyVariant}{}\<[29]%
\>[29]{}\mathrel{\;:\;}{}\<[29E]%
\>[32]{}\Tyvarid{a}\;\Varid{polyvariant\char95 desc}{}\<[53]%
\>[53]{}\to \Tyvarid{a}\;\Varid{ty}{}\<[E]%
\ColumnHook
\end{hscode}\resethooks
with suitable generic views \ensuremath{\Varid{object\char95 desc}} and \ensuremath{\Varid{polyvariant\char95 desc}} described in section~\ref{sec:view/objects-polyvariants}.

\subsection{Type-Indexed Functions}
\label{sec:type-indexed}
With type reflection we can write type-indexed functions, for instance a pretty printer has the following type:
\begin{hscode}\SaveRestoreHook
\column{B}{@{}>{\hspre}l<{\hspost}@{}}%
\column{E}{@{}>{\hspre}l<{\hspost}@{}}%
\>[B]{}\Keyword{val}\;\Varid{show}\mathrel{\;:\;}\Tyvarid{a}\;\Varid{ty}\to \Tyvarid{a}\to \Varid{string}{}\<[E]%
\ColumnHook
\end{hscode}\resethooks
Note how the reflected type is also used as a parameter of the function.

To implement \ensuremath{\Varid{show}} we need another extension to \ocaml{} type system introduced in version~4.00: \emph{locally abstract types}.
This type annotation is necessary to help the type checker while pattern matching over a GADT, since the type indices of a GADT may be instantiated to different concrete types depending on the constructor case, which is not possible with the classical Hindley-Milner algorithm.
In addition, our function uses polymorphic recursion: a call \ensuremath{\Varid{show}\;(\Conid{List}\;\Varid{a})} recurses on the type of the list elements: \ensuremath{\Varid{show}\;\Varid{a}}, which requires the explicit polymorphic quantification of the locally abstract type \ensuremath{\Varid{a}}.

\begin{hscode}\SaveRestoreHook
\column{B}{@{}>{\hspre}l<{\hspost}@{}}%
\column{15}{@{}>{\hspre}c<{\hspost}@{}}%
\column{15E}{@{}l@{}}%
\column{18}{@{}>{\hspre}l<{\hspost}@{}}%
\column{30}{@{}>{\hspre}l<{\hspost}@{}}%
\column{45}{@{}>{\hspre}c<{\hspost}@{}}%
\column{45E}{@{}l@{}}%
\column{49}{@{}>{\hspre}l<{\hspost}@{}}%
\column{E}{@{}>{\hspre}l<{\hspost}@{}}%
\>[B]{}\Keyword{let}\;\Keyword{rec}\;\Varid{show}{}\<[15]%
\>[15]{}\mathrel{\;:\;}{}\<[15E]%
\>[18]{}\mathbf{type\;}\;\Varid{a}\mathbin{°.°}\Varid{a}\;\Varid{ty}\to \Varid{a}\to \Varid{string}{}\<[E]%
\\
\>[15]{}\mathrel{=}{}\<[15E]%
\>[18]{}\Keyword{fun}\;\Varid{t}\;\Varid{x}\to {}\<[30]%
\>[30]{}\Keyword{match}\;\Varid{t}\;\Keyword{with}{}\<[E]%
\\
\>[30]{}\mid \Conid{Int}{}\<[45]%
\>[45]{}\to {}\<[45E]%
\>[49]{}\Varid{string\char95 of\char95 int}\;\Varid{x}{}\<[E]%
\\
\>[30]{}\mid \Conid{String}{}\<[45]%
\>[45]{}\to {}\<[45E]%
\>[49]{}\text{\tt \char34 \char92 \char34 \char34}\mathbin{\mbox{}^\wedge}\Varid{x}\mathbin{\mbox{}^\wedge}\text{\tt \char34 \char92 \char34 \char34}{}\<[E]%
\\
\>[30]{}\mid \Conid{List}\;\Varid{a}{}\<[45]%
\>[45]{}\to {}\<[45E]%
\>[49]{}\text{\tt \char34 [\char34}\mathbin{\mbox{}^\wedge}\Varid{\Conid{String}.concat}\;\text{\tt \char34 ;~\char34}\;(\Varid{\Conid{List}.map}\;(\Varid{show}\;\Varid{a})\;\Varid{x})\mathbin{\mbox{}^\wedge}\text{\tt \char34 ]\char34}{}\<[E]%
\\
\>[30]{}\mid \Conid{Pair}\;(\Varid{a},\,\Varid{b}){}\<[45]%
\>[45]{}\to {}\<[45E]%
\>[49]{}\text{\tt \char34 (\char34}\mathbin{\mbox{}^\wedge}\Varid{show}\;\Varid{a}\;(\Varid{fst}\;\Varid{x})\mathbin{\mbox{}^\wedge}\text{\tt \char34 ,~\char34}\mathbin{\mbox{}^\wedge}\Varid{show}\;\Varid{b}\;(\Varid{snd}\;\Varid{x})\mathbin{\mbox{}^\wedge}\text{\tt \char34 )\char34}{}\<[E]%
\\
\>[30]{}\mid \Conid{Fun}\;(\Varid{a},\,\Varid{b}){}\<[45]%
\>[45]{}\to {}\<[45E]%
\>[49]{}\text{\tt \char34 <fun>\char34}{}\<[E]%
\ColumnHook
\end{hscode}\resethooks
Such a definition by pattern matching is suitable for a closed type universe where the cases may be given exhaustively. However we want our universe to be extensible, so that we may add new types. Consequently the type indexed functions must also be extensible: as new type witnesses are added to \ensuremath{\Varid{ty}}, new cases must be added to the type indexed functions.

This problem of extending a datatype and a function on that datatype is known as Wadler's expression problem~\cite{wadler:98:expression-problem} and is indicative of the modularity of the language. Solutions in Haskell have been given involving type classes \cite{swierstra:08:alacarte,bahr:11:compositional-data-types} and cannot easily be adapted to \ocaml{}. However, \ocaml{} version 4.02 introduced extensible variant types. The only missing ingredient is extensible functions. The rest of the section explains how they are implemented in the library.

\subsubsection{Extensible Functions}
To define extensible functions, we will use the imperative features of \ocaml{}.
The idea is to keep a reference to a function which will be updated when a new case is added.
The reference is kept private while the public interface offers the means to
add a new case: 
\savecolumns
\begin{hscode}\SaveRestoreHook
\column{B}{@{}>{\hspre}l<{\hspost}@{}}%
\column{7}{@{}>{\hspre}l<{\hspost}@{}}%
\column{17}{@{}>{\hspre}c<{\hspost}@{}}%
\column{17E}{@{}l@{}}%
\column{20}{@{}>{\hspre}l<{\hspost}@{}}%
\column{E}{@{}>{\hspre}l<{\hspost}@{}}%
\>[B]{}\Keyword{val}\;{}\<[7]%
\>[7]{}\Varid{show\char95 ext}{}\<[17]%
\>[17]{}\mathrel{\;:\;}{}\<[17E]%
\>[20]{}(\forall\!\;\Tyvarid{a}\mathbin{°.°}\Tyvarid{a}\;\Varid{ty}\to \Tyvarid{a}\to \Varid{string})\to \Varid{unit}{}\<[E]%
\ColumnHook
\end{hscode}\resethooks
However this type is not correct in \ocaml{} because a polymorphic function is not allowed as an argument. Fortunately, we are allowed polymorphic record fields, thus we define:
\restorecolumns
\begin{hscode}\SaveRestoreHook
\column{B}{@{}>{\hspre}l<{\hspost}@{}}%
\column{7}{@{}>{\hspre}l<{\hspost}@{}}%
\column{17}{@{}>{\hspre}c<{\hspost}@{}}%
\column{17E}{@{}l@{}}%
\column{20}{@{}>{\hspre}l<{\hspost}@{}}%
\column{E}{@{}>{\hspre}l<{\hspost}@{}}%
\>[B]{}\mathbf{type\;}\;{}\<[7]%
\>[7]{}\Varid{show\char95 fun}{}\<[17]%
\>[17]{}\mathrel{=}{}\<[17E]%
\>[20]{}\{\mskip1.5mu \Varid{apply}\mathrel{\;:\;}\forall\!\;\Tyvarid{a}\mathbin{°.°}\Tyvarid{a}\;\Varid{ty}\to \Tyvarid{a}\to \Varid{string}\mskip1.5mu\}{}\<[E]%
\\
\>[B]{}\Keyword{val}\;{}\<[7]%
\>[7]{}\Varid{show\char95 ext}{}\<[17]%
\>[17]{}\mathrel{\;:\;}{}\<[17E]%
\>[20]{}\Varid{show\char95 fun}\to \Varid{unit}{}\<[E]%
\ColumnHook
\end{hscode}\resethooks
We may already use this public interface to define the cases above.
Once again the GADT forces us to provide type annotations.
Compiler warnings also encourage us to explicitly raise an exception for the cases that do not concern us.
Note that we use \ensuremath{\Varid{show}} for the recursive calls.
\begin{hscode}\SaveRestoreHook
\column{B}{@{}>{\hspre}l<{\hspost}@{}}%
\column{3}{@{}>{\hspre}l<{\hspost}@{}}%
\column{23}{@{}>{\hspre}l<{\hspost}@{}}%
\column{30}{@{}>{\hspre}l<{\hspost}@{}}%
\column{33}{@{}>{\hspre}l<{\hspost}@{}}%
\column{48}{@{}>{\hspre}l<{\hspost}@{}}%
\column{E}{@{}>{\hspre}l<{\hspost}@{}}%
\>[B]{}\Keyword{let}\;()\mathrel{=}\Keyword{begin}{}\<[E]%
\\
\>[B]{}\hsindent{3}{}\<[3]%
\>[3]{}\Varid{show\char95 ext}\;\{\mskip1.5mu \Varid{apply}\mathrel{=}{}\<[23]%
\>[23]{}\Keyword{fun}\;(\mathbf{type\;}\;\Varid{a})\;(\Varid{t}\mathrel{\;:\;}\Varid{a}\;\Varid{ty})\;(\Varid{x}\mathrel{\;:\;}\Varid{a})\to \Keyword{match}\;\Varid{t}\;\Keyword{with}{}\<[E]%
\\
\>[23]{}\mid \Conid{Int}{}\<[30]%
\>[30]{}\to \Varid{string\char95 of\char95 int}\;\Varid{x}{}\<[E]%
\\
\>[23]{}\mid \anonymous {}\<[30]%
\>[30]{}\to \Keyword{raise}\;\Conid{Not\char95 found}\mskip1.5mu\}\mathbin{;\;}{}\<[E]%
\\
\>[B]{}\hsindent{3}{}\<[3]%
\>[3]{}\Varid{show\char95 ext}\;\{\mskip1.5mu \Varid{apply}\mathrel{=}{}\<[23]%
\>[23]{}\Keyword{fun}\;(\mathbf{type\;}\;\Varid{a})\;(\Varid{t}\mathrel{\;:\;}\Varid{a}\;\Varid{ty})\;(\Varid{x}\mathrel{\;:\;}\Varid{a})\to \Keyword{match}\;\Varid{t}\;\Keyword{with}{}\<[E]%
\\
\>[23]{}\mid \Conid{List}\;\Varid{a}{}\<[33]%
\>[33]{}\to \text{\tt \char34 [\char34}\mathbin{\mbox{}^\wedge}\Varid{\Conid{String}.concat}\;\text{\tt \char34 ;~\char34}\;(\Varid{\Conid{List}.map}\;(\Varid{show}\;\Varid{a})\;\Varid{x})\mathbin{\mbox{}^\wedge}\text{\tt \char34 ]\char34}{}\<[E]%
\\
\>[23]{}\mid \anonymous {}\<[33]%
\>[33]{}\to \Keyword{raise}\;\Conid{Not\char95 found}\mskip1.5mu\}\mathbin{;\;}{}\<[E]%
\\
\>[B]{}\hsindent{3}{}\<[3]%
\>[3]{}\Varid{show\char95 ext}\;\{\mskip1.5mu \Varid{apply}\mathrel{=}{}\<[23]%
\>[23]{}\Keyword{fun}\;(\mathbf{type\;}\;\Varid{a})\;(\Varid{t}\mathrel{\;:\;}\Varid{a}\;\Varid{ty})\;(\Varid{x}\mathrel{\;:\;}\Varid{a})\to \Keyword{match}\;\Varid{t},\,\Varid{x}\;\Keyword{with}{}\<[E]%
\\
\>[23]{}\mid (\Conid{Pair}\;(\Varid{a},\,\Varid{b})),\,(\Varid{x},\,\Varid{y}){}\<[48]%
\>[48]{}\to \text{\tt \char34 (\char34}\mathbin{\mbox{}^\wedge}\Varid{show}\;\Varid{a}\;\Varid{x}\mathbin{\mbox{}^\wedge}\text{\tt \char34 ,~\char34}\mathbin{\mbox{}^\wedge}\Varid{show}\;\Varid{b}\;\Varid{y}\mathbin{\mbox{}^\wedge}\text{\tt \char34 )\char34}{}\<[E]%
\\
\>[23]{}\mid \anonymous {}\<[48]%
\>[48]{}\to \Keyword{raise}\;\Conid{Not\char95 found}\mskip1.5mu\}\mathbin{;\;}{}\<[E]%
\\
\>[B]{}\Keyword{end}{}\<[E]%
\ColumnHook
\end{hscode}\resethooks
All this syntactic noise could be avoided by the use of a PPX (see \cite{ocaml-manual}, section ``Extension nodes'') providing the following syntactic sugar:
\begin{hscode}\SaveRestoreHook
\column{B}{@{}>{\hspre}l<{\hspost}@{}}%
\column{17}{@{}>{\hspre}l<{\hspost}@{}}%
\column{32}{@{}>{\hspre}l<{\hspost}@{}}%
\column{40}{@{}>{\hspre}c<{\hspost}@{}}%
\column{40E}{@{}l@{}}%
\column{43}{@{}>{\hspre}l<{\hspost}@{}}%
\column{E}{@{}>{\hspre}l<{\hspost}@{}}%
\>[B]{}[\!\mathbin{\%}\Conid{Extend}]\;\Varid{show}\;{}\<[17]%
\>[17]{}\Conid{Int}\;{}\<[32]%
\>[32]{}\Varid{x}{}\<[40]%
\>[40]{}\mathrel{=}{}\<[40E]%
\>[43]{}\Varid{string\char95 of\char95 int}\;\Varid{x}{}\<[E]%
\\
\>[B]{}[\!\mathbin{\%}\Conid{Extend}]\;\Varid{show}\;{}\<[17]%
\>[17]{}(\Conid{List}\;\Varid{a})\;{}\<[32]%
\>[32]{}\Varid{x}{}\<[40]%
\>[40]{}\mathrel{=}{}\<[40E]%
\>[43]{}\text{\tt \char34 [\char34}\mathbin{\mbox{}^\wedge}\Varid{\Conid{String}.concat}\;\text{\tt \char34 ;~\char34}\;(\Varid{\Conid{List}.map}\;(\Varid{show}\;\Varid{a})\;\Varid{x})\mathbin{\mbox{}^\wedge}\text{\tt \char34 ]\char34}{}\<[E]%
\\
\>[B]{}[\!\mathbin{\%}\Conid{Extend}]\;\Varid{show}\;{}\<[17]%
\>[17]{}(\Conid{Pair}\;(\Varid{a},\,\Varid{b}))\;{}\<[32]%
\>[32]{}(\Varid{x},\,\Varid{y}){}\<[40]%
\>[40]{}\mathrel{=}{}\<[40E]%
\>[43]{}\text{\tt \char34 (\char34}\mathbin{\mbox{}^\wedge}\Varid{show}\;\Varid{a}\;\Varid{x}\mathbin{\mbox{}^\wedge}\text{\tt \char34 ,~\char34}\mathbin{\mbox{}^\wedge}\Varid{show}\;\Varid{b}\;\Varid{y}\mathbin{\mbox{}^\wedge}\text{\tt \char34 )\char34}{}\<[E]%
\ColumnHook
\end{hscode}\resethooks

\subsubsection{A Simple Implementation of Extensible Functions}
\label{sec:gp/ext}
How can we implement \ensuremath{\Varid{show\char95 ext}}? First we need to define a reference to a \ensuremath{\Varid{show\char95 fun}} record.
\savecolumns
\begin{hscode}\SaveRestoreHook
\column{B}{@{}>{\hspre}l<{\hspost}@{}}%
\column{6}{@{}>{\hspre}l<{\hspost}@{}}%
\column{16}{@{}>{\hspre}c<{\hspost}@{}}%
\column{16E}{@{}l@{}}%
\column{19}{@{}>{\hspre}l<{\hspost}@{}}%
\column{E}{@{}>{\hspre}l<{\hspost}@{}}%
\>[B]{}\Keyword{val}\;{}\<[6]%
\>[6]{}\Varid{show\char95 ref}{}\<[16]%
\>[16]{}\mathrel{\;:\;}{}\<[16E]%
\>[19]{}\Varid{show\char95 fun}\;\Varid{ref}{}\<[E]%
\ColumnHook
\end{hscode}\resethooks
The reference is initialised to a function that always fails.
\restorecolumns
\begin{hscode}\SaveRestoreHook
\column{B}{@{}>{\hspre}l<{\hspost}@{}}%
\column{6}{@{}>{\hspre}l<{\hspost}@{}}%
\column{16}{@{}>{\hspre}c<{\hspost}@{}}%
\column{16E}{@{}l@{}}%
\column{19}{@{}>{\hspre}l<{\hspost}@{}}%
\column{E}{@{}>{\hspre}l<{\hspost}@{}}%
\>[B]{}\Keyword{let}\;{}\<[6]%
\>[6]{}\Varid{show\char95 ref}{}\<[16]%
\>[16]{}\mathrel{=}{}\<[16E]%
\>[19]{}\Varid{ref}\;\{\mskip1.5mu \Varid{apply}\mathrel{=}\Keyword{fun}\;\Varid{t}\;\Varid{x}\to \Keyword{failwith}\;\text{\tt \char34 show:~type~not~supported~yet\char34}\mskip1.5mu\}{}\<[E]%
\ColumnHook
\end{hscode}\resethooks
This reference is private, we define two public functions: \ensuremath{\Varid{show}} to call the function in the reference,
and \ensuremath{\Varid{show\char95 ext}} to update it.
\restorecolumns
\begin{hscode}\SaveRestoreHook
\column{B}{@{}>{\hspre}l<{\hspost}@{}}%
\column{6}{@{}>{\hspre}l<{\hspost}@{}}%
\column{16}{@{}>{\hspre}c<{\hspost}@{}}%
\column{16E}{@{}l@{}}%
\column{19}{@{}>{\hspre}l<{\hspost}@{}}%
\column{E}{@{}>{\hspre}l<{\hspost}@{}}%
\>[B]{}\Keyword{let}\;{}\<[6]%
\>[6]{}\Varid{show}\;\Varid{t}\;\Varid{x}{}\<[16]%
\>[16]{}\mathrel{=}{}\<[16E]%
\>[19]{}\mathbin{!}\Varid{show\char95 ref}.\Varid{apply}\;\Varid{t}\;\Varid{x}{}\<[E]%
\ColumnHook
\end{hscode}\resethooks
To update the function, we simply try the new case, and resort to the previous version if it raises a \ensuremath{\Conid{Not\char95 found}} exception.
\begin{hscode}\SaveRestoreHook
\column{B}{@{}>{\hspre}l<{\hspost}@{}}%
\column{24}{@{}>{\hspre}c<{\hspost}@{}}%
\column{24E}{@{}l@{}}%
\column{27}{@{}>{\hspre}l<{\hspost}@{}}%
\column{31}{@{}>{\hspre}l<{\hspost}@{}}%
\column{64}{@{}>{\hspre}l<{\hspost}@{}}%
\column{E}{@{}>{\hspre}l<{\hspost}@{}}%
\>[B]{}\Keyword{let}\;\Varid{show\char95 ext}\;\Varid{new\char95 case}{}\<[24]%
\>[24]{}\mathrel{=}{}\<[24E]%
\>[27]{}\Keyword{let}\;\Varid{old\char95 show}\mathrel{=}\mathbin{!}\Varid{show\char95 ref}{}\<[E]%
\\
\>[27]{}\Keyword{in}\;{}\<[31]%
\>[31]{}\Varid{show\char95 ref}\mathrel{:=}\{\mskip1.5mu \Varid{apply}\mathrel{=}\Keyword{fun}\;\Varid{t}\;\Varid{x}\to {}\<[64]%
\>[64]{}\Keyword{try}\;\Varid{new\char95 case}.\Varid{apply}\;\Varid{t}\;\Varid{x}{}\<[E]%
\\
\>[64]{}\Keyword{with}\;\Conid{Not\char95 found}\to \Varid{old\char95 show}.\Varid{apply}\;\Varid{t}\;\Varid{x}\mskip1.5mu\}{}\<[E]%
\ColumnHook
\end{hscode}\resethooks
\paragraph{Semantics}
The semantics of \ensuremath{\Varid{show}} depends on the order of the calls to \ensuremath{\Varid{show\char95 ext}}, since the most recent extension is tried before the previous ones. When some patterns overlap between two extensions, it is the most recent extension that succeeds. This semantics is fragile since it depends on the order in which top level modules are linked.

\subsubsection{A Generic, Efficient and Robust Implementation of Extensible Functions}
\label{sec:extensible}
The previous implementation of extensible functions has a couple of issues: (1) it is not general, the same boilerplate we wrote for \ensuremath{\Varid{show}} would need to be written for new functions, (2) it is not very efficient because a list of cases is tried until one match succeeds and (3) it is fragile as the semantics depends on the order in which the cases are added at runtime.

The implementation we give now solves those problems by (1) using an encoding of higher-kinded polymorphism, (2) using a hash-table indexed by the constructors, and (3) using a partial order on patterns.

\paragraph{Higher Kinded Polymorphism}
To state the obvious, the type of a type indexed function may depend on its type index. For instance
\begin{hscode}\SaveRestoreHook
\column{B}{@{}>{\hspre}l<{\hspost}@{}}%
\column{12}{@{}>{\hspre}c<{\hspost}@{}}%
\column{12E}{@{}l@{}}%
\column{15}{@{}>{\hspre}l<{\hspost}@{}}%
\column{E}{@{}>{\hspre}l<{\hspost}@{}}%
\>[B]{}\Varid{show}{}\<[12]%
\>[12]{}\mathrel{\;:\;}{}\<[12E]%
\>[15]{}\Tyvarid{a}\;\Varid{ty}\to \Tyvarid{a}\to \Varid{string}{}\<[E]%
\\
\>[B]{}\Varid{read}{}\<[12]%
\>[12]{}\mathrel{\;:\;}{}\<[12E]%
\>[15]{}\Tyvarid{a}\;\Varid{ty}\to \Varid{string}\to \Tyvarid{a}{}\<[E]%
\\
\>[B]{}\Varid{enumerate}{}\<[12]%
\>[12]{}\mathrel{\;:\;}{}\<[12E]%
\>[15]{}\Tyvarid{a}\;\Varid{ty}\to \Tyvarid{a}\;\Varid{list}{}\<[E]%
\\
\>[B]{}\Varid{equal}{}\<[12]%
\>[12]{}\mathrel{\;:\;}{}\<[12E]%
\>[15]{}\Tyvarid{a}\;\Varid{ty}\to \Tyvarid{a}\to \Tyvarid{a}\to \Varid{bool}{}\<[E]%
\ColumnHook
\end{hscode}\resethooks
How can we define a general type for type-indexed functions?
They are all instances of the same scheme, which would be expressed like this if \ocaml{} allowed it:
\begin{hscode}\SaveRestoreHook
\column{B}{@{}>{\hspre}l<{\hspost}@{}}%
\column{E}{@{}>{\hspre}l<{\hspost}@{}}%
\>[B]{}\mathbf{type\;}\;\Tyvarid{f}\;\Varid{ty\char95 fun}\mathrel{=}\forall\!\;\Tyvarid{a}\mathbin{°.°}\Tyvarid{a}\;\Varid{ty}\to \Tyvarid{a}\;\Tyvarid{f}{}\<[E]%
\ColumnHook
\end{hscode}\resethooks
For instance \ensuremath{\Varid{show}\mathrel{\;:\;}\Varid{show\char95 t}\;\Varid{ty\char95 fun}}
where \ensuremath{\mathbf{type\;}\;\Tyvarid{a}\;\Varid{show\char95 t}\mathrel{=}\Tyvarid{a}\to \Varid{string}}.
Unfortunately \ocaml{} does not allow the use of type parameters of higher kinds (partially applied types like \ensuremath{\Varid{show\char95 t}} above). However, we may use a defunctionalisation method to emulate them \cite{higher-kinded-polymorphism}.
The idea is that \ensuremath{(\Tyvarid{a},\,\Tyvarid{f})\;\Varid{app}} represents the type application of \ensuremath{\Tyvarid{f}} to the type \ensuremath{\Tyvarid{a}}.\label{app}
We make \ensuremath{\Varid{app}} an extensible variant so that new constructors may be added as we need them.
\savecolumns
\begin{hscode}\SaveRestoreHook
\column{B}{@{}>{\hspre}l<{\hspost}@{}}%
\column{19}{@{}>{\hspre}c<{\hspost}@{}}%
\column{19E}{@{}l@{}}%
\column{23}{@{}>{\hspre}c<{\hspost}@{}}%
\column{23E}{@{}l@{}}%
\column{E}{@{}>{\hspre}l<{\hspost}@{}}%
\>[B]{}\mathbf{type\;}\;(\Tyvarid{a},\,\Tyvarid{f})\;\Varid{app}{}\<[19]%
\>[19]{}\mathrel{=}{}\<[19E]%
\>[23]{}\mathinner{\ldotp\ldotp}{}\<[23E]%
\ColumnHook
\end{hscode}\resethooks
Concretely, each type abstraction \ensuremath{\Lambda\!\;\Tyvarid{a}.\Varid{f}\;(\Tyvarid{a})} is represented by an empty type \ensuremath{\Varid{f'}} and we add a new constructor \ensuremath{\Conid{F}} such that \ensuremath{(\Tyvarid{a},\,\Varid{f'})\;\Varid{app}} is isomorphic to \ensuremath{\Varid{f}\;(\Tyvarid{a})}
\restorecolumns
\begin{hscode}\SaveRestoreHook
\column{B}{@{}>{\hspre}l<{\hspost}@{}}%
\column{19}{@{}>{\hspre}c<{\hspost}@{}}%
\column{19E}{@{}l@{}}%
\column{23}{@{}>{\hspre}l<{\hspost}@{}}%
\column{E}{@{}>{\hspre}l<{\hspost}@{}}%
\>[B]{}\mathbf{type\;}\;(\anonymous ,\,\anonymous )\;\Varid{app}{}\<[19]%
\>[19]{}\mathrel{+\!\!\!=}{}\<[19E]%
\>[23]{}\Conid{F}\mathrel{\;:\;}\Varid{f}\;(\Tyvarid{a})\to (\Tyvarid{a},\,\Varid{f'})\;\Varid{app}{}\<[E]%
\ColumnHook
\end{hscode}\resethooks
For instance, the list type former is represented by a type \ensuremath{\Varid{list'}} (with no parameters), and its semantics is given by:
\restorecolumns
\begin{hscode}\SaveRestoreHook
\column{B}{@{}>{\hspre}l<{\hspost}@{}}%
\column{19}{@{}>{\hspre}c<{\hspost}@{}}%
\column{19E}{@{}l@{}}%
\column{23}{@{}>{\hspre}l<{\hspost}@{}}%
\column{E}{@{}>{\hspre}l<{\hspost}@{}}%
\>[B]{}\mathbf{type\;}\;(\anonymous ,\,\anonymous )\;\Varid{app}{}\<[19]%
\>[19]{}\mathrel{+\!\!\!=}{}\<[19E]%
\>[23]{}\Conid{List}\mathrel{\;:\;}\Tyvarid{a}\;\Varid{list}\to (\Tyvarid{a},\,\Varid{list'})\;\Varid{app}{}\<[E]%
\ColumnHook
\end{hscode}\resethooks
\paragraph{Type Indexed Functions}
Using \ensuremath{\Varid{app}} we can give a general type for type-indexed functions:
\begin{hscode}\SaveRestoreHook
\column{B}{@{}>{\hspre}l<{\hspost}@{}}%
\column{E}{@{}>{\hspre}l<{\hspost}@{}}%
\>[B]{}\mathbf{type\;}\;\Tyvarid{f}\;\Varid{ty\char95 fun}\mathrel{=}\{\mskip1.5mu \Varid{f}\mathrel{\;:\;}\forall\!\;\Tyvarid{a}\mathbin{°.°}\Tyvarid{a}\;\Varid{ty}\to (\Tyvarid{a},\,\Tyvarid{f})\;\Varid{app}\mskip1.5mu\}{}\<[E]%
\ColumnHook
\end{hscode}\resethooks
For instance, \ensuremath{\Varid{show}} may be defined using the type abstraction \ensuremath{\Lambda\!\;\Tyvarid{a}.\Tyvarid{a}\to \Varid{string}}, represented with the abstract type \ensuremath{\Varid{show'}} and the \ensuremath{\Varid{app}} constructor:
\begin{hscode}\SaveRestoreHook
\column{B}{@{}>{\hspre}l<{\hspost}@{}}%
\column{E}{@{}>{\hspre}l<{\hspost}@{}}%
\>[B]{}\mathbf{type\;}\;(\anonymous ,\,\anonymous )\;\Varid{app}\mathrel{+\!\!\!=}\Conid{Show}\mathrel{\;:\;}(\Tyvarid{a}\to \Varid{string})\to (\Tyvarid{a},\,\Varid{show'})\;\Varid{app}{}\<[E]%
\ColumnHook
\end{hscode}\resethooks
So that \ensuremath{\Varid{show'}\;\Varid{ty\char95 fun}} is isomorphic with \ensuremath{\forall\!\;\Tyvarid{a}\mathbin{°.°}\Tyvarid{a}\;\Varid{ty}\to \Tyvarid{a}\to \Varid{string}}.

An extensible function is a collection of \ensuremath{\Varid{ty\char95 fun}}s.
A collection is created with a function \ensuremath{\Varid{create}}, that takes a string that will be used as documentation, and returns
a record \ensuremath{\Tyvarid{f}\;\Varid{closure}} whose field \ensuremath{\Varid{f}} is
the extensible function, and field \ensuremath{\Varid{ext}} allows us to extend it
with a new case by providing a type pattern and a \ensuremath{\Varid{ty\char95 fun}}
matching that type pattern.
\begin{hscode}\SaveRestoreHook
\column{B}{@{}>{\hspre}l<{\hspost}@{}}%
\column{20}{@{}>{\hspre}c<{\hspost}@{}}%
\column{20E}{@{}l@{}}%
\column{23}{@{}>{\hspre}l<{\hspost}@{}}%
\column{28}{@{}>{\hspre}l<{\hspost}@{}}%
\column{E}{@{}>{\hspre}l<{\hspost}@{}}%
\>[B]{}\Keyword{val}\;\Varid{create}\mathrel{\;:\;}\Varid{string}\to \Tyvarid{f}\;\Varid{closure}{}\<[E]%
\\
\>[B]{}\mathbf{type\;}\;\Tyvarid{f}\;\Varid{closure}\mathrel{=}{}\<[20]%
\>[20]{}\{\mskip1.5mu {}\<[20E]%
\>[23]{}\Varid{f}{}\<[28]%
\>[28]{}\mathrel{\;:\;}\forall\!\;\Tyvarid{a}\mathbin{°.°}\Tyvarid{a}\;\Varid{ty}\to (\Tyvarid{a},\,\Tyvarid{f})\;\Varid{app}{}\<[E]%
\\
\>[20]{}\mathbin{;\;}{}\<[20E]%
\>[23]{}\Varid{ext}{}\<[28]%
\>[28]{}\mathrel{\;:\;}\forall\!\;\Tyvarid{a}\mathbin{°.°}\Tyvarid{a}\;\Varid{pat}\to \Tyvarid{f}\;\Varid{ty\char95 fun}\to \Varid{unit}\mskip1.5mu\}{}\<[E]%
\ColumnHook
\end{hscode}\resethooks
Type patterns are inductively defined using the type constructors of \ensuremath{\Tyvarid{a}\;\Varid{ty}} plus a universal pattern \ensuremath{\Conid{Any}} that acts as a wild-card, matching any type.
For simplicity, type patterns are defined as synonyms of type witnesses.
\begin{hscode}\SaveRestoreHook
\column{B}{@{}>{\hspre}l<{\hspost}@{}}%
\column{E}{@{}>{\hspre}l<{\hspost}@{}}%
\>[B]{}\mathbf{type\;}\;\Tyvarid{a}\;\Varid{pat}\mathrel{=}\Tyvarid{a}\;\Varid{ty}{}\<[E]%
\ColumnHook
\end{hscode}\resethooks
The constructor \ensuremath{\Conid{Any}\mathrel{\;:\;}\forall\!\;\Tyvarid{a}\mathbin{°.°}\Tyvarid{a}\;\Varid{ty}}, may only be used in a context where a pattern is expected. For instance, we may extend our \ensuremath{\Varid{show}} function to lists with the statement:
\begin{hscode}\SaveRestoreHook
\column{B}{@{}>{\hspre}l<{\hspost}@{}}%
\column{E}{@{}>{\hspre}l<{\hspost}@{}}%
\>[B]{}\Varid{show}.\Varid{ext}\;(\Conid{List}\;\Conid{Any})\;\{\mskip1.5mu \Varid{f}\mathrel{=}\Varid{show\char95 list}\mskip1.5mu\}{}\<[E]%
\ColumnHook
\end{hscode}\resethooks
\ensuremath{\Varid{show\char95 list}} is a type indexed function that expects a type witness of the form \ensuremath{\Conid{List}\;\Varid{a}}.
\begin{hscode}\SaveRestoreHook
\column{B}{@{}>{\hspre}l<{\hspost}@{}}%
\column{19}{@{}>{\hspre}c<{\hspost}@{}}%
\column{19E}{@{}l@{}}%
\column{22}{@{}>{\hspre}l<{\hspost}@{}}%
\column{32}{@{}>{\hspre}l<{\hspost}@{}}%
\column{42}{@{}>{\hspre}l<{\hspost}@{}}%
\column{E}{@{}>{\hspre}l<{\hspost}@{}}%
\>[B]{}\Keyword{let}\;\Varid{show\char95 list}{}\<[19]%
\>[19]{}\mathrel{\;:\;}{}\<[19E]%
\>[22]{}\mathbf{type\;}\;\Varid{a}\mathbin{°.°}\Varid{a}\;\Varid{ty}\to (\Varid{a},\,\Varid{show'})\;\Varid{app}{}\<[E]%
\\
\>[19]{}\mathrel{=}{}\<[19E]%
\>[22]{}\Keyword{function}{}\<[32]%
\>[32]{}\mid \Conid{List}\;\Varid{a}{}\<[42]%
\>[42]{}\to \Conid{Show}\;(\Varid{show\char95 list\char95 of}\;\Varid{a}){}\<[E]%
\\
\>[32]{}\mid \anonymous {}\<[42]%
\>[42]{}\to \Keyword{invalid\_arg}\;\text{\tt \char34 show\char95 list:~expected~a~list\char34}{}\<[E]%
\ColumnHook
\end{hscode}\resethooks
where \ensuremath{\Varid{show\char95 list\char95 of}} can be defined as:
\begin{hscode}\SaveRestoreHook
\column{B}{@{}>{\hspre}l<{\hspost}@{}}%
\column{E}{@{}>{\hspre}l<{\hspost}@{}}%
\>[B]{}\Keyword{let}\;\Varid{show\char95 list\char95 of}\;\Varid{a}\;\Varid{xs}\mathrel{=}\text{\tt \char34 [\char34}\mathbin{\mbox{}^\wedge}\Varid{\Conid{String}.concat}\;\text{\tt \char34 ;~\char34}\;(\Varid{\Conid{List}.map}\;(\Varid{show}\;\Varid{a})\;\Varid{xs})\mathbin{\mbox{}^\wedge}\text{\tt \char34 ]\char34}{}\<[E]%
\\
\>[B]{}\Keyword{val}\;\Varid{show\char95 list\char95 of}\mathrel{\;:\;}\Tyvarid{a}\;\Varid{ty}\to \Tyvarid{a}\;\Varid{list}\to \Varid{string}{}\<[E]%
\ColumnHook
\end{hscode}\resethooks
\paragraph{Extensibility}


For fast application, we store the \ensuremath{\Varid{ty\char95 fun}}'s in a hash-table indexed by a type pattern where all the parameters of a type constructor are set to \ensuremath{\Conid{Any}}. For instance, if an indexed function has independent cases for \ensuremath{\Conid{Pair}\;(\Conid{Int},\,\Conid{Bool})}, \ensuremath{\Conid{Pair}\;(\Conid{List}\;\Conid{String},\,\Conid{Bool})}, \ensuremath{\Conid{Pair}\;(\Conid{Any},\,\Conid{Bool})}, \ensuremath{\Conid{Pair}\;(\Conid{Float},\,\Conid{Any})}, they will all be associated to the same entry in the hash-table, with key \ensuremath{\Conid{Pair}\;(\Conid{Any},\,\Conid{Any})}.



For each constructor pattern we store a list of functions ordered by their type pattern so that when applying the extensible function to some given type, the more general patterns are tried after all the more specific ones have failed to match the type.
This mechanism ensures that the behaviour of an extensible function does not depend on the order in which the cases are given. In the example above,
\ensuremath{\Conid{Pair}\;(\Conid{Int},\,\Conid{Bool})} and \ensuremath{\Conid{Pair}\;(\Conid{List}\;\Conid{String},\,\Conid{Bool})} will be tried before \ensuremath{\Conid{Pair}\;(\Conid{Any},\,\Conid{Bool})}
and \ensuremath{\Conid{Pair}\;(\Conid{Float},\,\Conid{Any})} will be tried before \ensuremath{\Conid{Pair}\;(\Conid{Any},\,\Conid{Any})}.

Using a lexicographic order, \ensuremath{\Conid{Pair}\;(\Conid{Int},\,\Conid{Any})} matches before \ensuremath{\Conid{Pair}\;(\Conid{Any},\,\Conid{Int})}; \ensuremath{\Conid{Pair}\;(\Conid{Int},\,\Conid{Int})} matches before both of them, and \ensuremath{\Conid{Pair}\;(\Conid{Any},\,\Conid{Any})} matches after all of them. \ensuremath{\Conid{Any}} is the most general pattern and matches when all the other patterns fail to match.

This approach combines both \emph{efficient} application in the most frequent case in which there will be only one definition per constructor, and \emph{flexibility} as it allows nested pattern matching and the order in which the function is extended does not matter.

\subsubsection{Type Equality and Safe Coercion}
\label{sec:coercion}
\label{sec:type-equality}
GADTs allow us to define type indexed types, which can also be seen as type predicates (or relations), and their values can be seen as the proof of the predicates.
Hence we may define a binary type predicate for type equality, with a single constructor for the proof of reflexivity.
\begin{hscode}\SaveRestoreHook
\column{B}{@{}>{\hspre}l<{\hspost}@{}}%
\column{E}{@{}>{\hspre}l<{\hspost}@{}}%
\>[B]{}\mathbf{type\;}\;(\anonymous ,\,\anonymous )\;\Varid{equal}\mathrel{=}\Conid{Refl}\mathrel{\;:\;}(\Tyvarid{a},\,\Tyvarid{a})\;\Varid{equal}{}\<[E]%
\ColumnHook
\end{hscode}\resethooks
Pattern matching on the \ensuremath{\Conid{Refl}} constructor forces the type-checker to unify the type parameters of \ensuremath{\Varid{equal}}. This allows us to define a safe coercion function.
\begin{hscode}\SaveRestoreHook
\column{B}{@{}>{\hspre}l<{\hspost}@{}}%
\column{24}{@{}>{\hspre}c<{\hspost}@{}}%
\column{24E}{@{}l@{}}%
\column{27}{@{}>{\hspre}l<{\hspost}@{}}%
\column{E}{@{}>{\hspre}l<{\hspost}@{}}%
\>[B]{}\Keyword{let}\;\Varid{coerce\char95 from\char95 equal}{}\<[24]%
\>[24]{}\mathrel{\;:\;}{}\<[24E]%
\>[27]{}\mathbf{type\;}\;\Varid{a}\;\Varid{b}\mathbin{°.°}(\Varid{a},\,\Varid{b})\;\Varid{equal}\to \Varid{a}\to \Varid{b}{}\<[E]%
\\
\>[24]{}\mathrel{=}{}\<[24E]%
\>[27]{}\Keyword{function}\;\Conid{Refl}\to \Keyword{fun}\;\Varid{x}\to \Varid{x}{}\<[E]%
\ColumnHook
\end{hscode}\resethooks
That definition is possible because by the time we match \ensuremath{\Conid{Refl}}, the type \ensuremath{\Varid{b}} is unified with \ensuremath{\Varid{a}} hence, the variable \ensuremath{\Varid{x}} may be used both with type \ensuremath{\Varid{a}} and \ensuremath{\Varid{b}}.

We define an extensible type-indexed function \ensuremath{\Varid{ty\char95 equal}}.
\begin{hscode}\SaveRestoreHook
\column{B}{@{}>{\hspre}l<{\hspost}@{}}%
\column{E}{@{}>{\hspre}l<{\hspost}@{}}%
\>[B]{}\Keyword{val}\;\Varid{ty\char95 equal}\mathrel{\;:\;}\Tyvarid{a}\;\Varid{ty}\to \Tyvarid{b}\;\Varid{ty}\to (\Tyvarid{a},\,\Tyvarid{b})\;\Varid{equal}\;\Varid{option}{}\<[E]%
\ColumnHook
\end{hscode}\resethooks
Adding new cases to \ensuremath{\Varid{ty\char95 equal}} is very systematic. They may be automatically derived by a PPX extension described in section~\ref{sec:ppx}.
\begin{hscode}\SaveRestoreHook
\column{B}{@{}>{\hspre}l<{\hspost}@{}}%
\column{26}{@{}>{\hspre}l<{\hspost}@{}}%
\column{30}{@{}>{\hspre}l<{\hspost}@{}}%
\column{39}{@{}>{\hspre}l<{\hspost}@{}}%
\column{45}{@{}>{\hspre}l<{\hspost}@{}}%
\column{E}{@{}>{\hspre}l<{\hspost}@{}}%
\>[B]{}\Varid{ty\char95 equal\char95 ext}\;\Conid{Char}\;\{\mskip1.5mu \Varid{f}\mathrel{=}{}\<[26]%
\>[26]{}\Keyword{fun}\;(\mathbf{type\;}\;\Varid{a})\;(\mathbf{type\;}\;\Varid{b})\;(\Varid{a}\mathrel{\;:\;}\Varid{a}\;\Varid{ty})\;(\Varid{b}\mathrel{\;:\;}\Varid{b}\;\Varid{ty}){}\<[E]%
\\
\>[26]{}\to {}\<[30]%
\>[30]{}\Keyword{match}\;\Varid{a},\,\Varid{b}\;\Keyword{with}{}\<[E]%
\\
\>[30]{}\mid \Conid{Char},\,{}\<[39]%
\>[39]{}\Conid{Char}{}\<[45]%
\>[45]{}\to \Conid{Some}\;(\Conid{Refl}\mathrel{\;:\;}(\Varid{a},\,\Varid{b})\;\Varid{equal}){}\<[E]%
\\
\>[30]{}\mid \anonymous ,\,{}\<[39]%
\>[39]{}\anonymous {}\<[45]%
\>[45]{}\to \Conid{None}\mskip1.5mu\}\mathbin{;\;}{}\<[E]%
\ColumnHook
\end{hscode}\resethooks
In the case of parametric types, all type parameters must be recursively checked for equality.
\begin{hscode}\SaveRestoreHook
\column{B}{@{}>{\hspre}l<{\hspost}@{}}%
\column{32}{@{}>{\hspre}l<{\hspost}@{}}%
\column{36}{@{}>{\hspre}l<{\hspost}@{}}%
\column{47}{@{}>{\hspre}l<{\hspost}@{}}%
\column{55}{@{}>{\hspre}l<{\hspost}@{}}%
\column{61}{@{}>{\hspre}l<{\hspost}@{}}%
\column{74}{@{}>{\hspre}l<{\hspost}@{}}%
\column{E}{@{}>{\hspre}l<{\hspost}@{}}%
\>[B]{}\Varid{ty\char95 equal\char95 ext}\;(\Conid{List}\;\Conid{Any})\;\{\mskip1.5mu \Varid{f}\mathrel{=}{}\<[32]%
\>[32]{}\Keyword{fun}\;(\mathbf{type\;}\;\Varid{a})\;(\mathbf{type\;}\;\Varid{b})\;(\Varid{a}\mathrel{\;:\;}\Varid{a}\;\Varid{ty})\;(\Varid{b}\mathrel{\;:\;}\Varid{b}\;\Varid{ty}){}\<[E]%
\\
\>[32]{}\to {}\<[36]%
\>[36]{}\Keyword{match}\;\Varid{a},\,\Varid{b}\;\Keyword{with}{}\<[E]%
\\
\>[36]{}\mid \Conid{List}\;\Varid{x},\,{}\<[47]%
\>[47]{}\Conid{List}\;\Varid{y}{}\<[55]%
\>[55]{}\to ({}\<[61]%
\>[61]{}\Keyword{match}\;\Varid{ty\char95 equal}\;\Varid{x}\;\Varid{y}\;\Keyword{with}{}\<[E]%
\\
\>[61]{}\mid \Conid{Some}\;\Conid{Refl}{}\<[74]%
\>[74]{}\to \Conid{Some}\;(\Conid{Refl}\mathrel{\;:\;}(\Varid{a},\,\Varid{b})\;\Varid{equal}){}\<[E]%
\\
\>[61]{}\mid \Conid{None}{}\<[74]%
\>[74]{}\to \Conid{None}){}\<[E]%
\\
\>[36]{}\mid \anonymous ,\,{}\<[47]%
\>[47]{}\anonymous {}\<[55]%
\>[55]{}\to \Conid{None}\mskip1.5mu\}\mathbin{;\;}{}\<[E]%
\ColumnHook
\end{hscode}\resethooks
By composing the previous functions we may derive a type-indexed safe coercion function:
\label{prog:coerce}
\begin{hscode}\SaveRestoreHook
\column{B}{@{}>{\hspre}l<{\hspost}@{}}%
\column{13}{@{}>{\hspre}l<{\hspost}@{}}%
\column{16}{@{}>{\hspre}l<{\hspost}@{}}%
\column{E}{@{}>{\hspre}l<{\hspost}@{}}%
\>[B]{}\Keyword{let}\;\Varid{coerce}{}\<[13]%
\>[13]{}\mathrel{\;:\;}{}\<[16]%
\>[16]{}\mathbf{type\;}\;\Varid{a}\;\Varid{b}\mathbin{°.°}\Varid{a}\;\Varid{ty}\to \Varid{b}\;\Varid{ty}\to (\Varid{a}\to \Varid{b})\;\Varid{option}{}\<[E]%
\\
\>[13]{}\mathrel{=}\Keyword{fun}\;\Varid{a}\;\Varid{b}\to \Varid{\Conid{Option}.map}\;\Varid{coerce\char95 from\char95 equal}\;(\Varid{ty\char95 equal}\;\Varid{a}\;\Varid{b}){}\<[E]%
\ColumnHook
\end{hscode}\resethooks
Where \ensuremath{\Keyword{val}\;\Varid{\Conid{Option}.map}\mathrel{\;:\;}(\Tyvarid{a}\to \Tyvarid{b})\to \Tyvarid{a}\;\Varid{option}\to \Tyvarid{b}\;\Varid{option}}.

The \ensuremath{\Varid{coerce}} function is used as the basis for implementing the function \ensuremath{\Varid{children}}, as well as the Uniplate and Multiplate \ensuremath{\Varid{scrap}} functions (sections \ref{sec:Uniplate} and \ref{sec:Multiplate}).

\subsection{Generic Views}
\label{sec:gp/views}
\label{sec:children}
A generic view is a uniform representation of the structure of types.
Most libraries are built around a single view. In our design, we allow the user to choose among many views and even to define his own.

Depending on the task, some views might be more appropriate than others. For instance, to implement safe deserialisation in section \ref{sec:unmarshal} we need a low level view reflecting the specifics of \ocaml{} types.
When such details can be ignored, a higher level view is more adequate and easier to work with.
For instance, the \emph{Multiplate} library is written on top of the list-of-constructors view.

The low level view \ensuremath{\Varid{desc}} is special in the library in that it is the primitive view and is automatically derived from type definitions using the PPX attribute \ensuremath{\Varid{reify}} described in section~\ref{sec:ppx}.
All the high level views are defined as a generic function by using the low level view.
We show some examples in section  \ref{sec:traversals}.

\subsubsection{What is a View?}
A generic view is given by a datatype \ensuremath{\Tyvarid{a}\;\Varid{view}} together with a type indexed function \ensuremath{\Varid{view}\mathrel{\;:\;}\Tyvarid{a}\;\Varid{ty}\to \Tyvarid{a}\;\Varid{view}}.
It maps a type witness to a value giving the structure of the type.

The rest of the section describes some common views that are available in the library. For each view, we show how binary trees are represented and we give an implementation of the function \ensuremath{\Varid{children}\mathrel{\;:\;}\Tyvarid{a}\;\Varid{ty}\to \Tyvarid{a}\to \Tyvarid{a}\;\Varid{list}} seen in the introduction.

Note that abstract types may have a public representation associated to them: for instance when abstract values have to be exported or imported to/from a file, one may prefer to convert them to/from an external, public, representation.  Their generic view gives the structure of that public representation.

\subsubsection{Sum of Products}
\label{sec:sumprod}

The sum of products view represents algebraic datatypes using finite products and finite sums.
The implementation follows closely that of the Haskell LIGD library~\cite{Cheney:02:LIGD}.

We must define an empty type and a binary sum type from which all finite sums may be constructed:
\begin{hscode}\SaveRestoreHook
\column{B}{@{}>{\hspre}l<{\hspost}@{}}%
\column{E}{@{}>{\hspre}l<{\hspost}@{}}%
\>[B]{}\mathbf{type\;}\;\Varid{empty}{}\<[E]%
\\
\>[B]{}\mathbf{type\;}\;(\Tyvarid{a},\,\Tyvarid{b})\;\Varid{sum}\mathrel{=}\Conid{Left}\;\mathbf{of\;}\;\Tyvarid{a}\mid \Conid{Right}\;\mathbf{of\;}\;\Tyvarid{b}{}\<[E]%
\ColumnHook
\end{hscode}\resethooks
The sum of products representation is given by an indexed type \ensuremath{\Tyvarid{a}\;\Varid{sp}} whose index \ensuremath{\Tyvarid{a}} is the type being represented. We first give a representation of finite sums and products by reifying the index:
\savecolumns
\begin{hscode}\SaveRestoreHook
\column{B}{@{}>{\hspre}l<{\hspost}@{}}%
\column{13}{@{}>{\hspre}c<{\hspost}@{}}%
\column{13E}{@{}l@{}}%
\column{16}{@{}>{\hspre}l<{\hspost}@{}}%
\column{23}{@{}>{\hspre}c<{\hspost}@{}}%
\column{23E}{@{}l@{}}%
\column{26}{@{}>{\hspre}l<{\hspost}@{}}%
\column{E}{@{}>{\hspre}l<{\hspost}@{}}%
\>[B]{}\mathbf{type\;}\;\Tyvarid{a}\;\Varid{sp}{}\<[13]%
\>[13]{}\mathrel{=}{}\<[13E]%
\>[16]{}\Conid{Empty}{}\<[23]%
\>[23]{}\mathrel{\;:\;}{}\<[23E]%
\>[26]{}\Varid{empty}\;\Varid{sp}{}\<[E]%
\\
\>[13]{}\mid {}\<[13E]%
\>[16]{}\Conid{Sum}{}\<[23]%
\>[23]{}\mathrel{\;:\;}{}\<[23E]%
\>[26]{}\Tyvarid{a}\;\Varid{sp}\mathbin{\times}\Tyvarid{b}\;\Varid{sp}\to (\Tyvarid{a},\,\Tyvarid{b})\;\Varid{sum}\;\Varid{sp}{}\<[E]%
\\
\>[13]{}\mid {}\<[13E]%
\>[16]{}\Conid{Unit}{}\<[23]%
\>[23]{}\mathrel{\;:\;}{}\<[23E]%
\>[26]{}\Varid{unit}\;\Varid{sp}{}\<[E]%
\\
\>[13]{}\mid {}\<[13E]%
\>[16]{}\Conid{Prod}{}\<[23]%
\>[23]{}\mathrel{\;:\;}{}\<[23E]%
\>[26]{}\Tyvarid{a}\;\Varid{sp}\mathbin{\times}\Tyvarid{b}\;\Varid{sp}\to (\Tyvarid{a}\mathbin{\times}\Tyvarid{b})\;\Varid{sp}{}\<[E]%
\ColumnHook
\end{hscode}\resethooks
With only those constructors, \ensuremath{\Tyvarid{a}\;\Varid{sp}} would only allow us to represent types \ensuremath{\Tyvarid{a}} built out of \ensuremath{\Varid{empty}}, \ensuremath{\Varid{sum}}, \ensuremath{\Varid{unit}} and \ensuremath{(\mathbin{\times})}. We extend it to user defined types (variants, records, {\em etc.}) by providing an isomorphism between the user type and a representation as a sum of products, built from the \ensuremath{\mathbf{type\;}\;(\Tyvarid{a},\,\Tyvarid{b})\;\Varid{\Conid{Fun}.iso}\mathrel{=}\{\mskip1.5mu \Varid{fwd}\mathrel{\;:\;}\Tyvarid{a}\to \Tyvarid{b}\mathbin{;\;}\Varid{bck}\mathrel{\;:\;}\Tyvarid{b}\to \Tyvarid{a}\mathbin{;\;}\mskip1.5mu\}}, whose values convert bijectively type \ensuremath{\Tyvarid{a}} to/from type \ensuremath{\Tyvarid{b}}.
\restorecolumns
\begin{hscode}\SaveRestoreHook
\column{B}{@{}>{\hspre}l<{\hspost}@{}}%
\column{13}{@{}>{\hspre}c<{\hspost}@{}}%
\column{13E}{@{}l@{}}%
\column{16}{@{}>{\hspre}l<{\hspost}@{}}%
\column{23}{@{}>{\hspre}l<{\hspost}@{}}%
\column{E}{@{}>{\hspre}l<{\hspost}@{}}%
\>[13]{}\mid {}\<[13E]%
\>[16]{}\Conid{Iso}{}\<[23]%
\>[23]{}\mathrel{\;:\;}\Tyvarid{a}\;\Varid{sp}\mathbin{\times}(\Tyvarid{a},\,\Tyvarid{b})\;\Varid{\Conid{Fun}.iso}\to \Tyvarid{b}\;\Varid{sp}{}\<[E]%
\ColumnHook
\end{hscode}\resethooks
Meta information may be attached to the representation, for instance we may provide the name of variant constructors and record fields with:
\restorecolumns
\begin{hscode}\SaveRestoreHook
\column{B}{@{}>{\hspre}l<{\hspost}@{}}%
\column{13}{@{}>{\hspre}c<{\hspost}@{}}%
\column{13E}{@{}l@{}}%
\column{16}{@{}>{\hspre}l<{\hspost}@{}}%
\column{23}{@{}>{\hspre}l<{\hspost}@{}}%
\column{E}{@{}>{\hspre}l<{\hspost}@{}}%
\>[13]{}\mid {}\<[13E]%
\>[16]{}\Conid{Con}{}\<[23]%
\>[23]{}\mathrel{\;:\;}\Varid{string}\mathbin{\times}\Tyvarid{a}\;\Varid{sp}\to \Tyvarid{a}\;\Varid{sp}{}\<[E]%
\\
\>[13]{}\mid {}\<[13E]%
\>[16]{}\Conid{Field}{}\<[23]%
\>[23]{}\mathrel{\;:\;}\Varid{string}\mathbin{\times}\Tyvarid{a}\;\Varid{sp}\to \Tyvarid{a}\;\Varid{sp}{}\<[E]%
\ColumnHook
\end{hscode}\resethooks
We provide a type witness for the types that cannot be represented as a sum of products, those are the base cases that require a specific behaviour: \ensuremath{\Varid{int}}, \ensuremath{\Varid{float}}, \ensuremath{\Varid{char}}, \ensuremath{\Varid{string}}, \ensuremath{\Varid{array}}, etc.
\restorecolumns
\begin{hscode}\SaveRestoreHook
\column{B}{@{}>{\hspre}l<{\hspost}@{}}%
\column{13}{@{}>{\hspre}c<{\hspost}@{}}%
\column{13E}{@{}l@{}}%
\column{16}{@{}>{\hspre}l<{\hspost}@{}}%
\column{23}{@{}>{\hspre}l<{\hspost}@{}}%
\column{E}{@{}>{\hspre}l<{\hspost}@{}}%
\>[13]{}\mid {}\<[13E]%
\>[16]{}\Conid{Base}{}\<[23]%
\>[23]{}\mathrel{\;:\;}\Tyvarid{a}\;\Varid{ty}\to \Tyvarid{a}\;\Varid{sp}{}\<[E]%
\ColumnHook
\end{hscode}\resethooks
Finally, we need a constructor to delay the computation of a view, similar in its usage to the \ensuremath{\Keyword{lazy}} keyword in that it prevents an infinite term to be computed.
\restorecolumns
\begin{hscode}\SaveRestoreHook
\column{B}{@{}>{\hspre}l<{\hspost}@{}}%
\column{13}{@{}>{\hspre}c<{\hspost}@{}}%
\column{13E}{@{}l@{}}%
\column{16}{@{}>{\hspre}l<{\hspost}@{}}%
\column{23}{@{}>{\hspre}l<{\hspost}@{}}%
\column{E}{@{}>{\hspre}l<{\hspost}@{}}%
\>[13]{}\mid {}\<[13E]%
\>[16]{}\Conid{Delay}{}\<[23]%
\>[23]{}\mathrel{\;:\;}\Tyvarid{a}\;\Varid{ty}\to \Tyvarid{a}\;\Varid{sp}{}\<[E]%
\ColumnHook
\end{hscode}\resethooks
\ensuremath{\Conid{Delay}} is used to explicitely mark possibles occurrences of recursion.
Intuitively, \ensuremath{\Conid{Delay}\;\Varid{t}} has some similarity with \ensuremath{\Keyword{lazy}\;(\Varid{view}\;\Varid{t})} in that it prevents from looping forever by delaying the recursive computation of the view. Without \ensuremath{\Conid{Delay}}, it would not be possible to define the function \ensuremath{\Varid{children}} for instance (see below).

\paragraph{Sum of Products View for Binary Trees}
The generic function \ensuremath{\Varid{sumprod}\mathrel{\;:\;}\Tyvarid{a}\;\Varid{ty}\to \Tyvarid{a}\;\Varid{sp}} computes the sum of products representation of any type, it is derived from the low level \ensuremath{\Varid{desc}} view. In the case of binary trees, the view is given by:
\begin{hscode}\SaveRestoreHook
\column{B}{@{}>{\hspre}l<{\hspost}@{}}%
\column{3}{@{}>{\hspre}l<{\hspost}@{}}%
\column{8}{@{}>{\hspre}l<{\hspost}@{}}%
\column{11}{@{}>{\hspre}c<{\hspost}@{}}%
\column{11E}{@{}l@{}}%
\column{14}{@{}>{\hspre}l<{\hspost}@{}}%
\column{15}{@{}>{\hspre}c<{\hspost}@{}}%
\column{15E}{@{}l@{}}%
\column{18}{@{}>{\hspre}l<{\hspost}@{}}%
\column{19}{@{}>{\hspre}l<{\hspost}@{}}%
\column{33}{@{}>{\hspre}l<{\hspost}@{}}%
\column{39}{@{}>{\hspre}c<{\hspost}@{}}%
\column{39E}{@{}l@{}}%
\column{40}{@{}>{\hspre}l<{\hspost}@{}}%
\column{42}{@{}>{\hspre}l<{\hspost}@{}}%
\column{49}{@{}>{\hspre}l<{\hspost}@{}}%
\column{52}{@{}>{\hspre}l<{\hspost}@{}}%
\column{53}{@{}>{\hspre}l<{\hspost}@{}}%
\column{58}{@{}>{\hspre}c<{\hspost}@{}}%
\column{58E}{@{}l@{}}%
\column{61}{@{}>{\hspre}l<{\hspost}@{}}%
\column{62}{@{}>{\hspre}l<{\hspost}@{}}%
\column{68}{@{}>{\hspre}l<{\hspost}@{}}%
\column{81}{@{}>{\hspre}l<{\hspost}@{}}%
\column{110}{@{}>{\hspre}l<{\hspost}@{}}%
\column{E}{@{}>{\hspre}l<{\hspost}@{}}%
\>[B]{}\Varid{sumprod}\;(\Conid{Btree}\;\Varid{a})\equiv {}\<[E]%
\\
\>[B]{}\hsindent{3}{}\<[3]%
\>[3]{}\Conid{Iso}\;{}\<[8]%
\>[8]{}(\Conid{Sum}\;{}\<[15]%
\>[15]{}({}\<[15E]%
\>[18]{}\Conid{Con}\;(\text{\tt \char34 Empty\char34},\,{}\<[33]%
\>[33]{}\Conid{Unit}){}\<[E]%
\\
\>[15]{},\,{}\<[15E]%
\>[18]{}\Conid{Con}\;(\text{\tt \char34 Node\char34},\,{}\<[33]%
\>[33]{}\Conid{Prod}\;{}\<[39]%
\>[39]{}({}\<[39E]%
\>[42]{}\Conid{Delay}\;{}\<[49]%
\>[49]{}(\Conid{Btree}\;\Varid{a}){}\<[E]%
\\
\>[39]{},\,{}\<[39E]%
\>[42]{}\Conid{Prod}\;{}\<[49]%
\>[49]{}({}\<[52]%
\>[52]{}\Conid{Delay}\;\Varid{a}{}\<[E]%
\\
\>[49]{},\,{}\<[52]%
\>[52]{}\Conid{Prod}\;{}\<[58]%
\>[58]{}({}\<[58E]%
\>[61]{}\Conid{Delay}\;{}\<[68]%
\>[68]{}(\Conid{Btree}\;\Varid{a}){}\<[E]%
\\
\>[58]{},\,{}\<[58E]%
\>[61]{}\Conid{Unit})))))){}\<[E]%
\\
\>[8]{},\,{}\<[11]%
\>[11]{}\{\mskip1.5mu {}\<[11E]%
\>[14]{}\Varid{fwd}{}\<[19]%
\>[19]{}\mathrel{=}(\Keyword{function}\;\Conid{Left}\;(){}\<[40]%
\>[40]{}\to \Conid{Empty}{}\<[53]%
\>[53]{}\mid \Conid{Right}\;{}\<[62]%
\>[62]{}(\Varid{l},\,(\Varid{x},\,(\Varid{r},\,()))){}\<[81]%
\>[81]{}\to \Conid{Node}\;(\Varid{l},\,\Varid{x},\,\Varid{r})){}\<[E]%
\\
\>[11]{}\mathbin{;\;}{}\<[11E]%
\>[14]{}\Varid{bck}{}\<[19]%
\>[19]{}\mathrel{=}(\Keyword{function}\;\Conid{Empty}{}\<[40]%
\>[40]{}\to \Conid{Left}\;(){}\<[53]%
\>[53]{}\mid \Conid{Node}\;{}\<[62]%
\>[62]{}(\Varid{l},\,\Varid{x},\,\Varid{r}){}\<[81]%
\>[81]{}\to \Conid{Right}\;(\Varid{l},\,(\Varid{x},\,(\Varid{r},\,())))){}\<[110]%
\>[110]{}\mskip1.5mu\}){}\<[E]%
\ColumnHook
\end{hscode}\resethooks

\paragraph{Generic Equality}
A common pattern when defining a generic function is to define two mutually recursive functions, one working on the type witness and the other one working on a generic view.
The first one may implement ad-hoc cases by pattern matching on the type witness and a generic case covering the other cases.

The function \ensuremath{\Varid{equal}} calls \ensuremath{\Varid{equal\char95 sp}} on the \ensuremath{\Varid{sumprod}} view:
\savecolumns
\begin{hscode}\SaveRestoreHook
\column{B}{@{}>{\hspre}l<{\hspost}@{}}%
\column{6}{@{}>{\hspre}l<{\hspost}@{}}%
\column{11}{@{}>{\hspre}l<{\hspost}@{}}%
\column{21}{@{}>{\hspre}l<{\hspost}@{}}%
\column{24}{@{}>{\hspre}l<{\hspost}@{}}%
\column{37}{@{}>{\hspre}l<{\hspost}@{}}%
\column{E}{@{}>{\hspre}l<{\hspost}@{}}%
\>[B]{}\Keyword{let}\;{}\<[6]%
\>[6]{}\Keyword{rec}\;{}\<[11]%
\>[11]{}\Varid{equal}{}\<[21]%
\>[21]{}\mathrel{\;:\;}{}\<[24]%
\>[24]{}\mathbf{type\;}\;\Varid{a}\mathbin{°.°}\Varid{a}\;\Varid{ty}\to \Varid{a}\to \Varid{a}\to \Varid{bool}{}\<[E]%
\\
\>[21]{}\mathrel{=}{}\<[24]%
\>[24]{}\Keyword{fun}\;\Varid{t}\to \Varid{equal\char95 sp}\;(\Varid{sumprod}\;\Varid{t}){}\<[E]%
\\
\>[6]{}\Keyword{and}\;{}\<[11]%
\>[11]{}\Varid{equal\char95 sp}{}\<[21]%
\>[21]{}\mathrel{\;:\;}{}\<[24]%
\>[24]{}\mathbf{type\;}\;\Varid{a}\mathbin{°.°}\Varid{a}\;\Varid{sp}\to \Varid{a}\to \Varid{a}\to \Varid{bool}{}\<[E]%
\\
\>[21]{}\mathrel{=}\Keyword{fun}\;\Varid{s}\;\Varid{x}\;\Varid{y}\to {}\<[37]%
\>[37]{}\Keyword{match}\;\Varid{s}\;\Keyword{with}{}\<[E]%
\ColumnHook
\end{hscode}\resethooks
\ensuremath{\Varid{equal\char95 sp}} is by induction on the type structure. Equality for the unit type holds trivially.
\restorecolumns
\begin{hscode}\SaveRestoreHook
\column{B}{@{}>{\hspre}l<{\hspost}@{}}%
\column{37}{@{}>{\hspre}l<{\hspost}@{}}%
\column{E}{@{}>{\hspre}l<{\hspost}@{}}%
\>[37]{}\mid \Conid{Unit}\to \Keyword{true}{}\<[E]%
\ColumnHook
\end{hscode}\resethooks
Equality for products is component wise.
\restorecolumns
\begin{hscode}\SaveRestoreHook
\column{B}{@{}>{\hspre}l<{\hspost}@{}}%
\column{37}{@{}>{\hspre}l<{\hspost}@{}}%
\column{57}{@{}>{\hspre}l<{\hspost}@{}}%
\column{E}{@{}>{\hspre}l<{\hspost}@{}}%
\>[37]{}\mid \Conid{Prod}\;(\Varid{a},\,\Varid{b})\to ({}\<[57]%
\>[57]{}\Keyword{match}\;\Varid{x},\,\Varid{y}\;\Keyword{with}\;(\Varid{xa},\,\Varid{xb}),\,(\Varid{ya},\,\Varid{yb}){}\<[E]%
\\
\>[57]{}\to \Varid{equal\char95 sp}\;\Varid{a}\;\Varid{xa}\;\Varid{ya}\mathrel{\wedge}\Varid{equal\char95 sp}\;\Varid{b}\;\Varid{xb}\;\Varid{yb}){}\<[E]%
\ColumnHook
\end{hscode}\resethooks
\noindent
Values of sum types are equal when they have the same constructor and their arguments are equal.
\restorecolumns
\begin{hscode}\SaveRestoreHook
\column{B}{@{}>{\hspre}l<{\hspost}@{}}%
\column{37}{@{}>{\hspre}l<{\hspost}@{}}%
\column{56}{@{}>{\hspre}l<{\hspost}@{}}%
\column{65}{@{}>{\hspre}l<{\hspost}@{}}%
\column{69}{@{}>{\hspre}l<{\hspost}@{}}%
\column{70}{@{}>{\hspre}l<{\hspost}@{}}%
\column{77}{@{}>{\hspre}l<{\hspost}@{}}%
\column{81}{@{}>{\hspre}c<{\hspost}@{}}%
\column{81E}{@{}l@{}}%
\column{85}{@{}>{\hspre}l<{\hspost}@{}}%
\column{E}{@{}>{\hspre}l<{\hspost}@{}}%
\>[37]{}\mid \Conid{Sum}\;(\Varid{a},\,\Varid{b})\to ({}\<[56]%
\>[56]{}\Keyword{match}\;\Varid{x},\,\Varid{y}\;\Keyword{with}{}\<[E]%
\\
\>[56]{}\mid \Conid{Left}\;{}\<[65]%
\>[65]{}\Varid{xa},\,{}\<[70]%
\>[70]{}\Conid{Left}\;{}\<[77]%
\>[77]{}\Varid{ya}{}\<[81]%
\>[81]{}\to {}\<[81E]%
\>[85]{}\Varid{equal\char95 sp}\;\Varid{a}\;\Varid{xa}\;\Varid{ya}{}\<[E]%
\\
\>[56]{}\mid \Conid{Right}\;{}\<[65]%
\>[65]{}\Varid{xb},\,{}\<[70]%
\>[70]{}\Conid{Right}\;{}\<[77]%
\>[77]{}\Varid{yb}{}\<[81]%
\>[81]{}\to {}\<[81E]%
\>[85]{}\Varid{equal\char95 sp}\;\Varid{b}\;\Varid{xb}\;\Varid{yb}{}\<[E]%
\\
\>[56]{}\mid \anonymous ,\,{}\<[69]%
\>[69]{}\anonymous {}\<[81]%
\>[81]{}\to {}\<[81E]%
\>[85]{}\Keyword{false}\;){}\<[E]%
\ColumnHook
\end{hscode}\resethooks
Meta information is ignored.\vspace{-24.8pt}
\restorecolumns
\begin{hscode}\SaveRestoreHook
\column{B}{@{}>{\hspre}l<{\hspost}@{}}%
\column{37}{@{}>{\hspre}l<{\hspost}@{}}%
\column{46}{@{}>{\hspre}l<{\hspost}@{}}%
\column{E}{@{}>{\hspre}l<{\hspost}@{}}%
\>[37]{}\mid \Conid{Con}\;{}\<[46]%
\>[46]{}(\anonymous ,\,\Varid{a})\to \Varid{equal\char95 sp}\;\Varid{a}\;\Varid{x}\;\Varid{y}{}\<[E]%
\\
\>[37]{}\mid \Conid{Field}\;{}\<[46]%
\>[46]{}(\anonymous ,\,\Varid{a})\to \Varid{equal\char95 sp}\;\Varid{a}\;\Varid{x}\;\Varid{y}{}\<[E]%
\ColumnHook
\end{hscode}\resethooks
Values of user types are equal if their sum of products representations are equal.
\restorecolumns
\begin{hscode}\SaveRestoreHook
\column{B}{@{}>{\hspre}l<{\hspost}@{}}%
\column{37}{@{}>{\hspre}l<{\hspost}@{}}%
\column{E}{@{}>{\hspre}l<{\hspost}@{}}%
\>[37]{}\mid \Conid{Iso}\;(\Varid{s},\,\{\mskip1.5mu \Varid{bck}\mathbin{;\;}\anonymous \mskip1.5mu\})\to \Varid{equal\char95 sp}\;\Varid{s}\;(\Varid{bck}\;\Varid{x})\;(\Varid{bck}\;\Varid{y}){}\<[E]%
\ColumnHook
\end{hscode}\resethooks
The constructor \ensuremath{\Conid{Delay}} is used to avoid producing infinite representations. Note the mutual recursion with \ensuremath{\Varid{equal}} since \ensuremath{\Varid{t}} is a type witness rather than a sum of product representation.
\restorecolumns
\begin{hscode}\SaveRestoreHook
\column{B}{@{}>{\hspre}l<{\hspost}@{}}%
\column{37}{@{}>{\hspre}l<{\hspost}@{}}%
\column{E}{@{}>{\hspre}l<{\hspost}@{}}%
\>[37]{}\mid \Conid{Delay}\;\Varid{t}\to \Varid{equal}\;\Varid{t}\;\Varid{x}\;\Varid{y}{}\<[E]%
\ColumnHook
\end{hscode}\resethooks
Equality over the empty type is the empty function. We use \ensuremath{\Varid{empty\char95 elim}\mathrel{\;:\;}\forall\!\;\Tyvarid{a}\mathbin{°.°}\Conid{Empty}\to \Tyvarid{a}}.
\restorecolumns
\begin{hscode}\SaveRestoreHook
\column{B}{@{}>{\hspre}l<{\hspost}@{}}%
\column{37}{@{}>{\hspre}l<{\hspost}@{}}%
\column{E}{@{}>{\hspre}l<{\hspost}@{}}%
\>[37]{}\mid \Conid{Empty}\to \Varid{empty\char95 elim}\;\Varid{x}{}\<[E]%
\ColumnHook
\end{hscode}\resethooks
For the basic types (\ensuremath{\Varid{char}}, \ensuremath{\Varid{float}}, etc) we resort to the built-in equality:
\restorecolumns
\begin{hscode}\SaveRestoreHook
\column{B}{@{}>{\hspre}l<{\hspost}@{}}%
\column{37}{@{}>{\hspre}l<{\hspost}@{}}%
\column{E}{@{}>{\hspre}l<{\hspost}@{}}%
\>[37]{}\mid \Conid{Base}\;\Varid{t}\to \Varid{x}\mathrel{=}\Varid{y}{}\<[E]%
\ColumnHook
\end{hscode}\resethooks
\paragraph{Children}
The function \ensuremath{\Varid{children}}, that computes the list of immediate subnodes of a value, makes crucial use of the \ensuremath{\Conid{Delay}} constructor to check whether the delayed type is the same as the type of the term.

\savecolumns
\begin{hscode}\SaveRestoreHook
\column{B}{@{}>{\hspre}l<{\hspost}@{}}%
\column{6}{@{}>{\hspre}l<{\hspost}@{}}%
\column{11}{@{}>{\hspre}l<{\hspost}@{}}%
\column{24}{@{}>{\hspre}c<{\hspost}@{}}%
\column{24E}{@{}l@{}}%
\column{27}{@{}>{\hspre}l<{\hspost}@{}}%
\column{41}{@{}>{\hspre}l<{\hspost}@{}}%
\column{E}{@{}>{\hspre}l<{\hspost}@{}}%
\>[B]{}\Keyword{let}\;{}\<[6]%
\>[6]{}\Keyword{rec}\;{}\<[11]%
\>[11]{}\Varid{children}\;\Varid{t}{}\<[24]%
\>[24]{}\mathrel{=}{}\<[24E]%
\>[27]{}\Varid{children\char95 sp}\;\Varid{t}\;(\Varid{sumprod}\;\Varid{t}){}\<[E]%
\\
\>[6]{}\Keyword{and}\;{}\<[11]%
\>[11]{}\Varid{children\char95 sp}{}\<[24]%
\>[24]{}\mathrel{\;:\;}{}\<[24E]%
\>[27]{}\mathbf{type\;}\;\Varid{b}\mathbin{°.°}\Tyvarid{a}\;\Varid{ty}\to \Varid{b}\;\Varid{sp}\to \Varid{b}\to \Tyvarid{a}\;\Varid{list}{}\<[E]%
\\
\>[24]{}\mathrel{=}{}\<[24E]%
\>[27]{}\Keyword{fun}\;\Varid{t}\;\Varid{s}\;\Varid{x}\to {}\<[41]%
\>[41]{}\Keyword{match}\;\Varid{s}\;\Keyword{with}{}\<[E]%
\\
\>[41]{}\mid \Conid{Delay}\;\Varid{t'}\to \Varid{child}\;\Varid{t}\;\Varid{t'}\;\Varid{x}{}\<[E]%
\ColumnHook
\end{hscode}\resethooks
The children lists of both components of a product are concatenated.
\restorecolumns
\begin{hscode}\SaveRestoreHook
\column{B}{@{}>{\hspre}l<{\hspost}@{}}%
\column{41}{@{}>{\hspre}l<{\hspost}@{}}%
\column{59}{@{}>{\hspre}l<{\hspost}@{}}%
\column{86}{@{}>{\hspre}c<{\hspost}@{}}%
\column{86E}{@{}l@{}}%
\column{90}{@{}>{\hspre}l<{\hspost}@{}}%
\column{E}{@{}>{\hspre}l<{\hspost}@{}}%
\>[41]{}\mid \Conid{Prod}\;(\Varid{a},\,\Varid{b}){}\<[59]%
\>[59]{}\to (\Keyword{match}\;\Varid{x}\;\Keyword{with}\;(\Varid{xa},\,\Varid{xb}){}\<[86]%
\>[86]{}\to {}\<[86E]%
\>[90]{}\Varid{children\char95 sp}\;\Varid{t}\;\Varid{a}\;\Varid{xa}{}\<[E]%
\\
\>[86]{}\mathbin{@}{}\<[86E]%
\>[90]{}\Varid{children\char95 sp}\;\Varid{t}\;\Varid{b}\;\Varid{xb}){}\<[E]%
\ColumnHook
\end{hscode}\resethooks
All the remaining cases simply recurse following the type structure.
\restorecolumns
\begin{hscode}\SaveRestoreHook
\column{B}{@{}>{\hspre}l<{\hspost}@{}}%
\column{41}{@{}>{\hspre}l<{\hspost}@{}}%
\column{59}{@{}>{\hspre}l<{\hspost}@{}}%
\column{73}{@{}>{\hspre}l<{\hspost}@{}}%
\column{77}{@{}>{\hspre}l<{\hspost}@{}}%
\column{89}{@{}>{\hspre}l<{\hspost}@{}}%
\column{E}{@{}>{\hspre}l<{\hspost}@{}}%
\>[41]{}\mid \Conid{Sum}\;(\Varid{a},\,\Varid{b}){}\<[59]%
\>[59]{}\to (\Keyword{match}\;\Varid{x}\;\Keyword{with}{}\<[77]%
\>[77]{}\mid \Conid{Left}\;\Varid{xa}{}\<[89]%
\>[89]{}\to \Varid{children\char95 sp}\;\Varid{t}\;\Varid{a}\;\Varid{xa}{}\<[E]%
\\
\>[77]{}\mid \Conid{Right}\;\Varid{xb}{}\<[89]%
\>[89]{}\to \Varid{children\char95 sp}\;\Varid{t}\;\Varid{b}\;\Varid{xb}){}\<[E]%
\\
\>[41]{}\mid \Conid{Con}\;(\anonymous ,\,\Varid{s'})\mid \Conid{Field}\;(\anonymous ,\,\Varid{s'}){}\<[73]%
\>[73]{}\to \Varid{children\char95 sp}\;\Varid{t}\;\Varid{s'}\;\Varid{x}{}\<[E]%
\\
\>[41]{}\mid \Conid{Iso}\;(\Varid{s'},\,\Varid{fb}){}\<[59]%
\>[59]{}\to \Varid{children\char95 sp}\;\Varid{t}\;\Varid{s'}\;(\Varid{fb}.\Varid{bck}\;\Varid{x}){}\<[E]%
\ColumnHook
\end{hscode}\resethooks
\ensuremath{\Conid{Base}} and \ensuremath{\Conid{Empty}} have no children.
\restorecolumns
\begin{hscode}\SaveRestoreHook
\column{B}{@{}>{\hspre}l<{\hspost}@{}}%
\column{41}{@{}>{\hspre}l<{\hspost}@{}}%
\column{E}{@{}>{\hspre}l<{\hspost}@{}}%
\>[41]{}\mid \Conid{Base}\;\anonymous \mid \Conid{Empty}\to [\mskip1.5mu \mskip1.5mu]{}\<[E]%
\ColumnHook
\end{hscode}\resethooks
\noindent \ensuremath{\Varid{child}} builds a singleton list only if the two type witnesses are equal.
\label{prog:child}
\begin{hscode}\SaveRestoreHook
\column{B}{@{}>{\hspre}l<{\hspost}@{}}%
\column{E}{@{}>{\hspre}l<{\hspost}@{}}%
\>[B]{}\Keyword{val}\;\Varid{child}\mathrel{\;:\;}\Tyvarid{a}\;\Varid{ty}\to \Tyvarid{b}\;\Varid{ty}\to \Tyvarid{b}\to \Tyvarid{a}\;\Varid{list}{}\<[E]%
\ColumnHook
\end{hscode}\resethooks
To implement \ensuremath{\Varid{child}} we use a function to coerce a value from a type to another if they are equal,
\ensuremath{\Varid{coerce}} is a primitive of the library discussed in section \ref{sec:coercion}.
\begin{hscode}\SaveRestoreHook
\column{B}{@{}>{\hspre}l<{\hspost}@{}}%
\column{E}{@{}>{\hspre}l<{\hspost}@{}}%
\>[B]{}\Keyword{val}\;\Varid{coerce}\mathrel{\;:\;}\Tyvarid{a}\;\Varid{ty}\to \Tyvarid{b}\;\Varid{ty}\to \Tyvarid{a}\to \Tyvarid{b}\;\Varid{option}{}\<[E]%
\\
\>[B]{}\Keyword{let}\;\Varid{child}\;\Varid{a}\;\Varid{b}\;\Varid{x}\mathrel{=}\Varid{list\char95 of\char95 opt}\;(\Varid{coerce}\;\Varid{b}\;\Varid{a}\;\Varid{x}){}\<[E]%
\ColumnHook
\end{hscode}\resethooks
\ensuremath{\Varid{list\char95 of\char95 opt}} does what its name suggests:
\begin{hscode}\SaveRestoreHook
\column{B}{@{}>{\hspre}l<{\hspost}@{}}%
\column{E}{@{}>{\hspre}l<{\hspost}@{}}%
\>[B]{}\Keyword{let}\;\Varid{list\char95 of\char95 opt}\mathrel{=}\Keyword{function}\;\Conid{None}\to [\mskip1.5mu \mskip1.5mu]\mid \Conid{Some}\;\Varid{x}\to [\mskip1.5mu \Varid{x}\mskip1.5mu]{}\<[E]%
\ColumnHook
\end{hscode}\resethooks
\subsubsection{Spine}
\label{sec:spine}
The spine view underlies the Scrap Your Boilerplate library for Haskell \cite{hinze:06:syb-revolutions}.
It allows to write generic functions very concisely compared to other views like the sum of product view. However it has some limitations, for instance it is not possible to define generators (such as a generic parser for instance). It is only really useful to define consumers (such as a generic print function).

The spine view is unusual in that it gives a view on a typed value rather than on the type alone:

\begin{hscode}\SaveRestoreHook
\column{B}{@{}>{\hspre}l<{\hspost}@{}}%
\column{E}{@{}>{\hspre}l<{\hspost}@{}}%
\>[B]{}\Keyword{val}\;\Varid{spine}\mathrel{\;:\;}\Tyvarid{a}\;\Varid{ty}\to \Tyvarid{a}\to \Tyvarid{a}\;\Varid{spine}{}\<[E]%
\ColumnHook
\end{hscode}\resethooks
The spine representation shows the applicative structure of a value as a constructor applied to its arguments:
\begin{hscode}\SaveRestoreHook
\column{B}{@{}>{\hspre}l<{\hspost}@{}}%
\column{16}{@{}>{\hspre}c<{\hspost}@{}}%
\column{16E}{@{}l@{}}%
\column{19}{@{}>{\hspre}l<{\hspost}@{}}%
\column{24}{@{}>{\hspre}c<{\hspost}@{}}%
\column{24E}{@{}l@{}}%
\column{27}{@{}>{\hspre}l<{\hspost}@{}}%
\column{E}{@{}>{\hspre}l<{\hspost}@{}}%
\>[B]{}\mathbf{type\;}\;\Tyvarid{a}\;\Varid{spine}{}\<[16]%
\>[16]{}\mathrel{=}{}\<[16E]%
\>[19]{}\Conid{Con}{}\<[24]%
\>[24]{}\mathrel{\;:\;}{}\<[24E]%
\>[27]{}\Tyvarid{a}\to \Tyvarid{a}\;\Varid{spine}{}\<[E]%
\\
\>[16]{}\mid {}\<[16E]%
\>[19]{}\Conid{App}{}\<[24]%
\>[24]{}\mathrel{\;:\;}{}\<[24E]%
\>[27]{}(\Tyvarid{a}\to \Tyvarid{b})\;\Varid{spine}\mathbin{\times}\Tyvarid{a}\;\Varid{ty}\mathbin{\times}\Tyvarid{a}\to \Tyvarid{b}\;\Varid{spine}{}\<[E]%
\ColumnHook
\end{hscode}\resethooks
\noindent
For instance, the value \ensuremath{\Conid{Node}\;(\Conid{Empty},\,\mathrm{1},\,\Conid{Node}\;(\Conid{Empty},\,\mathrm{2},\,\Conid{Empty}))} is represented as
\begin{hscode}\SaveRestoreHook
\column{B}{@{}>{\hspre}l<{\hspost}@{}}%
\column{E}{@{}>{\hspre}l<{\hspost}@{}}%
\>[B]{}\Conid{App}\;(\Conid{App}\;(\Conid{App}\;(\Conid{Con}\;\Varid{node},\,\Conid{Btree}\;\Conid{Int},\,\Conid{Empty}),\,\Conid{Int},\,\mathrm{1}),\,\Conid{Btree}\;\Conid{Int},\,\Conid{Node}\;(\Conid{Empty},\,\mathrm{2},\,\Conid{Empty})){}\<[E]%
\ColumnHook
\end{hscode}\resethooks
Where \ensuremath{\Varid{node}} is a curried constructor function.
\begin{hscode}\SaveRestoreHook
\column{B}{@{}>{\hspre}l<{\hspost}@{}}%
\column{E}{@{}>{\hspre}l<{\hspost}@{}}%
\>[B]{}\Keyword{let}\;\Varid{node}\;\Varid{l}\;\Varid{x}\;\Varid{r}\mathrel{=}\Conid{Node}\;(\Varid{l},\,\Varid{x},\,\Varid{r}){}\<[E]%
\ColumnHook
\end{hscode}\resethooks
\paragraph{Spine View for Binary Trees}
Let us look at the spine view of binary trees. We define a function \ensuremath{\Varid{spine\char95 btree}}:
\begin{hscode}\SaveRestoreHook
\column{B}{@{}>{\hspre}l<{\hspost}@{}}%
\column{E}{@{}>{\hspre}l<{\hspost}@{}}%
\>[B]{}\Keyword{val}\;\Varid{spine\char95 btree}\mathrel{\;:\;}\Tyvarid{a}\;\Varid{ty}\to \Tyvarid{a}\;\Varid{btree}\to \Tyvarid{a}\;\Varid{btree}\;\Varid{spine}{}\<[E]%
\ColumnHook
\end{hscode}\resethooks
such that \ensuremath{\Varid{spine}\;(\Conid{Btree}\;\Varid{a})\mathrel{=}\Varid{spine\char95 btree}\;\Varid{a}}.
Note that \ensuremath{\Varid{spine}}---like all the high level views---is in fact generically defined in terms of the low level view.
\begin{hscode}\SaveRestoreHook
\column{B}{@{}>{\hspre}l<{\hspost}@{}}%
\column{3}{@{}>{\hspre}c<{\hspost}@{}}%
\column{3E}{@{}l@{}}%
\column{6}{@{}>{\hspre}l<{\hspost}@{}}%
\column{20}{@{}>{\hspre}l<{\hspost}@{}}%
\column{29}{@{}>{\hspre}l<{\hspost}@{}}%
\column{37}{@{}>{\hspre}c<{\hspost}@{}}%
\column{37E}{@{}l@{}}%
\column{40}{@{}>{\hspre}l<{\hspost}@{}}%
\column{45}{@{}>{\hspre}c<{\hspost}@{}}%
\column{45E}{@{}l@{}}%
\column{48}{@{}>{\hspre}l<{\hspost}@{}}%
\column{53}{@{}>{\hspre}l<{\hspost}@{}}%
\column{81}{@{}>{\hspre}l<{\hspost}@{}}%
\column{89}{@{}>{\hspre}l<{\hspost}@{}}%
\column{E}{@{}>{\hspre}l<{\hspost}@{}}%
\>[B]{}\Keyword{let}\;\Varid{spine\char95 btree}\;\Varid{a}\;{}\<[20]%
\>[20]{}\Varid{x}\mathrel{=}\Keyword{match}\;\Varid{x}\;\Keyword{with}{}\<[E]%
\\
\>[B]{}\hsindent{3}{}\<[3]%
\>[3]{}\mid {}\<[3E]%
\>[6]{}\Conid{Empty}{}\<[29]%
\>[29]{}\to \Conid{Con}\;\Conid{Empty}{}\<[E]%
\\
\>[B]{}\hsindent{3}{}\<[3]%
\>[3]{}\mid {}\<[3E]%
\>[6]{}\Conid{Node}\;(\Varid{left},\,\Varid{x},\,\Varid{right}){}\<[29]%
\>[29]{}\to \Conid{App}\;{}\<[37]%
\>[37]{}({}\<[37E]%
\>[40]{}\Conid{App}\;{}\<[45]%
\>[45]{}({}\<[45E]%
\>[48]{}\Conid{App}\;{}\<[53]%
\>[53]{}(\Conid{Con}\;\Varid{node},\,\Conid{Btree}\;\Varid{a},\,\Varid{left}),\,{}\<[81]%
\>[81]{}\Varid{a},\,\Varid{x}),\,{}\<[89]%
\>[89]{}\Conid{Btree}\;\Varid{a},\,\Varid{right}){}\<[E]%
\ColumnHook
\end{hscode}\resethooks
\paragraph{Equality}
Generic equality can be implemented using the spine view, at the price of a small modification of the \ensuremath{\Varid{spine}} type. With the current definition of \ensuremath{\Varid{spine}}, we would write:
\savecolumns
\begin{hscode}\SaveRestoreHook
\column{B}{@{}>{\hspre}l<{\hspost}@{}}%
\column{6}{@{}>{\hspre}l<{\hspost}@{}}%
\column{11}{@{}>{\hspre}l<{\hspost}@{}}%
\column{24}{@{}>{\hspre}c<{\hspost}@{}}%
\column{24E}{@{}l@{}}%
\column{27}{@{}>{\hspre}l<{\hspost}@{}}%
\column{30}{@{}>{\hspre}l<{\hspost}@{}}%
\column{46}{@{}>{\hspre}l<{\hspost}@{}}%
\column{61}{@{}>{\hspre}c<{\hspost}@{}}%
\column{61E}{@{}l@{}}%
\column{65}{@{}>{\hspre}l<{\hspost}@{}}%
\column{75}{@{}>{\hspre}c<{\hspost}@{}}%
\column{75E}{@{}l@{}}%
\column{78}{@{}>{\hspre}l<{\hspost}@{}}%
\column{87}{@{}>{\hspre}l<{\hspost}@{}}%
\column{E}{@{}>{\hspre}l<{\hspost}@{}}%
\>[B]{}\Keyword{let}\;{}\<[6]%
\>[6]{}\Keyword{rec}\;{}\<[11]%
\>[11]{}\Varid{equal}{}\<[24]%
\>[24]{}\mathrel{\;:\;}{}\<[24E]%
\>[27]{}\mathbf{type\;}\;\Varid{a}\mathbin{°.°}\Varid{a}\;\Varid{ty}\to \Varid{a}\to \Varid{a}\to \Varid{bool}{}\<[75]%
\>[75]{}\mathrel{=}{}\<[75E]%
\>[78]{}\Keyword{fun}\;\Varid{t}\to \Varid{equal\char95 het}\;\Varid{t}\;\Varid{t}{}\<[E]%
\\
\>[6]{}\Keyword{and}\;{}\<[11]%
\>[11]{}\Varid{equal\char95 het}{}\<[24]%
\>[24]{}\mathrel{\;:\;}{}\<[24E]%
\>[27]{}\mathbf{type\;}\;\Varid{a}\;\Varid{b}\mathbin{°.°}\Varid{a}\;\Varid{ty}\to \Varid{b}\;\Varid{ty}\to \Varid{a}\to \Varid{b}\to \Varid{bool}{}\<[75]%
\>[75]{}\mathrel{=}{}\<[75E]%
\>[78]{}\Keyword{fun}\;\Varid{a}\;\Varid{b}\;\Varid{x}\;\Varid{y}{}\<[E]%
\\
\>[78]{}\to \Varid{equal\char95 spine}\;(\Varid{view}\;\Varid{a}\;\Varid{x},\,\Varid{view}\;\Varid{b}\;\Varid{y}){}\<[E]%
\\
\>[6]{}\Keyword{and}\;{}\<[11]%
\>[11]{}\Varid{equal\char95 spine}{}\<[24]%
\>[24]{}\mathrel{\;:\;}{}\<[24E]%
\>[27]{}\mathbf{type\;}\;\Varid{a}\;\Varid{b}\mathbin{°.°}\Varid{a}\;\Varid{spine}\mathbin{\times}\Varid{b}\;\Varid{spine}\to \Varid{bool}{}\<[E]%
\\
\>[24]{}\mathrel{=}{}\<[24E]%
\>[27]{}\Keyword{function}{}\<[E]%
\\
\>[27]{}\mid {}\<[30]%
\>[30]{}\Conid{App}\;(\Varid{f},\,\Varid{a},\,\Varid{x}),\,{}\<[46]%
\>[46]{}\Conid{App}\;(\Varid{g},\,\Varid{b},\,\Varid{y}){}\<[61]%
\>[61]{}\to {}\<[61E]%
\>[65]{}\Varid{equal\char95 het}\;\Varid{a}\;\Varid{b}\;\Varid{x}\;\Varid{y}\mathrel{\wedge}{}\<[87]%
\>[87]{}\Varid{equal\char95 spine}\;(\Varid{f},\,\Varid{g}){}\<[E]%
\\
\>[27]{}\mid {}\<[30]%
\>[30]{}\Conid{Con}\;\Varid{x},\,{}\<[46]%
\>[46]{}\Conid{Con}\;\Varid{y}{}\<[61]%
\>[61]{}\to {}\<[61E]%
\>[65]{}\mathbin{...}\mbox{\commentbegin  we have no way to proceed  \commentend}{}\<[E]%
\\
\>[27]{}\mid {}\<[30]%
\>[30]{}\anonymous ,\,{}\<[46]%
\>[46]{}\anonymous {}\<[61]%
\>[61]{}\to {}\<[61E]%
\>[65]{}\Keyword{false}{}\<[E]%
\ColumnHook
\end{hscode}\resethooks
First, notice that we need to generalise the type of equality to arguments of different types (heterogeneous equality). This is because the spine view for each argument gives rise to independent existentially quantified variables.
As a result, when comparing two constructors for equality, we are left stuck with two values of different types, since \ensuremath{\Conid{Con}\mathrel{\;:\;}\Tyvarid{a}\to \Tyvarid{a}\;\Varid{spine}}, that even built-in equality could not compare because of their different types.
To fix our problem, we need to extend the spine datatype in order to carry more information about constructors, such as their name, arity, module, file location, and so on. (We do not give here the precise nature of this rather bureaucratic meta information.)
If we make sure that the meta information associated to a constructor uniquely identifies it, then we can complete our implementation of \ensuremath{\Varid{equal}} by changing the erroneous line with:
\restorecolumns
\begin{hscode}\SaveRestoreHook
\column{B}{@{}>{\hspre}l<{\hspost}@{}}%
\column{27}{@{}>{\hspre}l<{\hspost}@{}}%
\column{E}{@{}>{\hspre}l<{\hspost}@{}}%
\>[27]{}\mid \Conid{Con}\;(\Varid{x},\,\Varid{meta\char95 x}),\,\Conid{Con}\;(\Varid{y},\,\Varid{meta\char95 y})\to \Varid{meta\char95 x}\mathrel{=}\Varid{meta\char95 y}{}\<[E]%
\ColumnHook
\end{hscode}\resethooks
Using built-in equality to check that the meta information is indeed the same.

\paragraph{Children}
Let us implement the \ensuremath{\Varid{children}} function from the introduction using the spine view.
\begin{hscode}\SaveRestoreHook
\column{B}{@{}>{\hspre}l<{\hspost}@{}}%
\column{E}{@{}>{\hspre}l<{\hspost}@{}}%
\>[B]{}\Keyword{let}\;\Varid{children}\;\Varid{a}\;\Varid{x}\mathrel{=}\Varid{children\char95 spine}\;\Varid{a}\;(\Varid{spine}\;\Varid{a}\;\Varid{x}){}\<[E]%
\ColumnHook
\end{hscode}\resethooks
\ensuremath{\Varid{children\char95 spine}} takes the type witness of the tree---this is also the type of the children---and a spine whose type is different.
This is because when we recursively go through the spine, the type of the spine changes.
\begin{hscode}\SaveRestoreHook
\column{B}{@{}>{\hspre}l<{\hspost}@{}}%
\column{E}{@{}>{\hspre}l<{\hspost}@{}}%
\>[B]{}\Keyword{val}\;\Varid{children\char95 spine}\mathrel{\;:\;}\Tyvarid{a}\;\Varid{ty}\to \Tyvarid{b}\;\Varid{spine}\to \Tyvarid{a}\;\Varid{list}{}\<[E]%
\ColumnHook
\end{hscode}\resethooks
If the spine is a constructor, it contains no child and we return the empty list.
If the spine is an application \ensuremath{\Varid{f}\;\Varid{x}}, we collect the children of both \ensuremath{\Varid{f}} and \ensuremath{\Varid{x}}.
A type annotation is necessary because the recursive calls change the type of \ensuremath{\Varid{b}\;\Varid{spine}}.
\begin{hscode}\SaveRestoreHook
\column{B}{@{}>{\hspre}l<{\hspost}@{}}%
\column{25}{@{}>{\hspre}l<{\hspost}@{}}%
\column{28}{@{}>{\hspre}l<{\hspost}@{}}%
\column{37}{@{}>{\hspre}l<{\hspost}@{}}%
\column{54}{@{}>{\hspre}c<{\hspost}@{}}%
\column{54E}{@{}l@{}}%
\column{58}{@{}>{\hspre}l<{\hspost}@{}}%
\column{E}{@{}>{\hspre}l<{\hspost}@{}}%
\>[B]{}\Keyword{let}\;\Keyword{rec}\;\Varid{children\char95 spine}{}\<[25]%
\>[25]{}\mathrel{\;:\;}{}\<[28]%
\>[28]{}\mathbf{type\;}\;\Varid{b}\mathbin{°.°}\Tyvarid{a}\;\Varid{ty}\to \Varid{b}\;\Varid{spine}\to \Tyvarid{a}\;\Varid{list}{}\<[E]%
\\
\>[25]{}\mathrel{=}\Keyword{fun}\;\Varid{t}\to {}\<[37]%
\>[37]{}\Keyword{function}{}\<[E]%
\\
\>[37]{}\mid \Conid{Con}\;\anonymous {}\<[54]%
\>[54]{}\to {}\<[54E]%
\>[58]{}[\mskip1.5mu \mskip1.5mu]{}\<[E]%
\\
\>[37]{}\mid \Conid{App}\;(\Varid{f},\,\Varid{a},\,\Varid{x}){}\<[54]%
\>[54]{}\to {}\<[54E]%
\>[58]{}\Varid{children\char95 spine}\;\Varid{t}\;\Varid{f}\mathbin{@}\Varid{child}\;\Varid{t}\;\Varid{a}\;\Varid{x}{}\<[E]%
\ColumnHook
\end{hscode}\resethooks
\noindent
See how much simpler that definition of \ensuremath{\Varid{children}} is in comparison to the one using the sum of products view.

\subsubsection{Low Level View}
\label{sec:view/desc}
Whereas the high level views give a uniform structural representation of types, the low level view \ensuremath{\Varid{desc}} captures the particularities.

\ocaml{} types are grouped in categories. Each of them is identified by a constructor of the \ensuremath{\Varid{desc}} view:
\begin{hscode}\SaveRestoreHook
\column{B}{@{}>{\hspre}l<{\hspost}@{}}%
\column{3}{@{}>{\hspre}l<{\hspost}@{}}%
\column{17}{@{}>{\hspre}c<{\hspost}@{}}%
\column{17E}{@{}l@{}}%
\column{21}{@{}>{\hspre}l<{\hspost}@{}}%
\column{52}{@{}>{\hspre}l<{\hspost}@{}}%
\column{56}{@{}>{\hspre}l<{\hspost}@{}}%
\column{85}{@{}>{\hspre}c<{\hspost}@{}}%
\column{85E}{@{}l@{}}%
\column{89}{@{}>{\hspre}l<{\hspost}@{}}%
\column{E}{@{}>{\hspre}l<{\hspost}@{}}%
\>[B]{}\mathbf{type\;}\;\Tyvarid{a}\;\Varid{desc}\mathrel{=}{}\<[E]%
\\
\>[B]{}\hsindent{3}{}\<[3]%
\>[3]{}\mid \Conid{Array}{}\<[17]%
\>[17]{}\mathrel{\;:\;}{}\<[17E]%
\>[21]{}\Tyvarid{b}\;\Varid{ty}\mathbin{\times}(\Keyword{module}\;\Conid{Array\char95 intf}\;\Keyword{with}\;\mathbf{type\;}\;\Varid{t}\mathrel{=}\Tyvarid{a}\;\Keyword{and}\;\mathbf{type\;}\;\Varid{elt}\mathrel{=}\Tyvarid{b}){}\<[85]%
\>[85]{}\to {}\<[85E]%
\>[89]{}\Tyvarid{a}\;\Varid{desc}{}\<[E]%
\\
\>[B]{}\hsindent{3}{}\<[3]%
\>[3]{}\mid \Conid{Product}{}\<[17]%
\>[17]{}\mathrel{\;:\;}{}\<[17E]%
\>[21]{}\Tyvarid{b}\;\Varid{product}\mathbin{\times}(\Tyvarid{b},\,\Tyvarid{a})\;\Varid{iso}{}\<[52]%
\>[52]{}\to {}\<[56]%
\>[56]{}\Tyvarid{a}\;\Varid{desc}{}\<[E]%
\\
\>[B]{}\hsindent{3}{}\<[3]%
\>[3]{}\mid \Conid{Record}{}\<[17]%
\>[17]{}\mathrel{\;:\;}{}\<[17E]%
\>[21]{}(\Tyvarid{b},\,\Tyvarid{a})\;\Varid{record}{}\<[52]%
\>[52]{}\to {}\<[56]%
\>[56]{}\Tyvarid{a}\;\Varid{desc}{}\<[E]%
\\
\>[B]{}\hsindent{3}{}\<[3]%
\>[3]{}\mid \Conid{Variant}{}\<[17]%
\>[17]{}\mathrel{\;:\;}{}\<[17E]%
\>[21]{}\Tyvarid{a}\;\Varid{variant}{}\<[52]%
\>[52]{}\to \Tyvarid{a}\;\Varid{desc}{}\<[E]%
\\
\>[B]{}\hsindent{3}{}\<[3]%
\>[3]{}\mid \Conid{Extensible}{}\<[17]%
\>[17]{}\mathrel{\;:\;}{}\<[17E]%
\>[21]{}\Tyvarid{a}\;\Varid{ext}{}\<[52]%
\>[52]{}\to \Tyvarid{a}\;\Varid{desc}{}\<[E]%
\\
\>[B]{}\hsindent{3}{}\<[3]%
\>[3]{}\mid \Conid{Custom}{}\<[17]%
\>[17]{}\mathrel{\;:\;}{}\<[17E]%
\>[21]{}\Tyvarid{a}\;\Varid{custom}{}\<[52]%
\>[52]{}\to \Tyvarid{a}\;\Varid{desc}{}\<[E]%
\\
\>[B]{}\hsindent{3}{}\<[3]%
\>[3]{}\mid \Conid{Class}{}\<[17]%
\>[17]{}\mathrel{\;:\;}{}\<[17E]%
\>[21]{}\Tyvarid{a}\;\Varid{class\char95 t}{}\<[52]%
\>[52]{}\to \Tyvarid{a}\;\Varid{desc}{}\<[E]%
\\
\>[B]{}\hsindent{3}{}\<[3]%
\>[3]{}\mid \Conid{Synonym}{}\<[17]%
\>[17]{}\mathrel{\;:\;}{}\<[17E]%
\>[21]{}\Tyvarid{a}\;\Varid{synonym}{}\<[52]%
\>[52]{}\to \Tyvarid{a}\;\Varid{desc}{}\<[E]%
\\
\>[B]{}\hsindent{3}{}\<[3]%
\>[3]{}\mid \Conid{Abstract}{}\<[17]%
\>[17]{}\mathrel{\;:\;}{}\<[17E]%
\>[21]{}\Tyvarid{a}\;\Varid{abstract}{}\<[52]%
\>[52]{}\to \Tyvarid{a}\;\Varid{desc}{}\<[E]%
\\
\>[B]{}\hsindent{3}{}\<[3]%
\>[3]{}\mid \Conid{NoDesc}{}\<[17]%
\>[17]{}\mathrel{\;:\;}{}\<[17E]%
\>[21]{}\Tyvarid{a}\;\Varid{desc}{}\<[E]%
\ColumnHook
\end{hscode}\resethooks

\paragraph{Array}
\ocaml{} has a few array-like types: \ensuremath{\Varid{array}}, \ensuremath{\Varid{string}} and \ensuremath{\Varid{bytes}}. They can be handled generically using a common interface.
\begin{hscode}\SaveRestoreHook
\column{B}{@{}>{\hspre}l<{\hspost}@{}}%
\column{5}{@{}>{\hspre}l<{\hspost}@{}}%
\column{11}{@{}>{\hspre}l<{\hspost}@{}}%
\column{23}{@{}>{\hspre}c<{\hspost}@{}}%
\column{23E}{@{}l@{}}%
\column{26}{@{}>{\hspre}l<{\hspost}@{}}%
\column{E}{@{}>{\hspre}l<{\hspost}@{}}%
\>[B]{}\Keyword{module}\;\mathbf{type\;}\;\Conid{Array\char95 intf}\mathrel{=}\Keyword{sig}{}\<[E]%
\\
\>[B]{}\hsindent{5}{}\<[5]%
\>[5]{}\mathbf{type\;}\;{}\<[11]%
\>[11]{}\Varid{t}{}\<[E]%
\\
\>[B]{}\hsindent{5}{}\<[5]%
\>[5]{}\mathbf{type\;}\;{}\<[11]%
\>[11]{}\Varid{elt}{}\<[E]%
\\
\>[B]{}\hsindent{5}{}\<[5]%
\>[5]{}\Keyword{val}\;{}\<[11]%
\>[11]{}\Varid{length}{}\<[23]%
\>[23]{}\mathrel{\;:\;}{}\<[23E]%
\>[26]{}\Varid{t}\to \Varid{int}{}\<[E]%
\\
\>[B]{}\hsindent{5}{}\<[5]%
\>[5]{}\Keyword{val}\;{}\<[11]%
\>[11]{}\Varid{get}{}\<[23]%
\>[23]{}\mathrel{\;:\;}{}\<[23E]%
\>[26]{}\Varid{t}\to \Varid{int}\to \Varid{elt}{}\<[E]%
\\
\>[B]{}\hsindent{5}{}\<[5]%
\>[5]{}\Keyword{val}\;{}\<[11]%
\>[11]{}\Varid{set}{}\<[23]%
\>[23]{}\mathrel{\;:\;}{}\<[23E]%
\>[26]{}\Varid{t}\to \Varid{int}\to \Varid{elt}\to \Varid{unit}{}\<[E]%
\\
\>[B]{}\hsindent{5}{}\<[5]%
\>[5]{}\Keyword{val}\;{}\<[11]%
\>[11]{}\Varid{init}{}\<[23]%
\>[23]{}\mathrel{\;:\;}{}\<[23E]%
\>[26]{}\Varid{int}\to (\Varid{int}\to \Varid{elt})\to \Varid{t}{}\<[E]%
\\
\>[B]{}\hsindent{5}{}\<[5]%
\>[5]{}\Keyword{val}\;{}\<[11]%
\>[11]{}\Varid{max\char95 length}{}\<[23]%
\>[23]{}\mathrel{\;:\;}{}\<[23E]%
\>[26]{}\Varid{int}{}\<[E]%
\\
\>[B]{}\Keyword{end}{}\<[E]%
\ColumnHook
\end{hscode}\resethooks
The rest of the array operations can be derived from this minimal interface.

The \ensuremath{\Varid{desc}} view for arrays consists of a witness for the type of the array elements, and a first class module of type \ensuremath{\Conid{Array\char95 intf}}.

\paragraph{Product}
\label{prog:product}
Tuple types are a family of built-in types. The \ensuremath{\Varid{desc}} view for \ensuremath{\Varid{n}}-ary products consists of an isomorphism between the product and \ensuremath{\Varid{n}} nested binary products,
\emph{i.e.} \ensuremath{\Tyvarid{a}\mathbin{\times}\Tyvarid{b}\mathbin{\times}\Tyvarid{c}} $\cong$ \ensuremath{\Tyvarid{a}\mathbin{\times}(\Tyvarid{b}\mathbin{\times}(\Tyvarid{c}\mathbin{\times}\Varid{unit}))}.

Right-nested binary products are fully captured by the following indexed-type:
\begin{hscode}\SaveRestoreHook
\column{B}{@{}>{\hspre}l<{\hspost}@{}}%
\column{18}{@{}>{\hspre}c<{\hspost}@{}}%
\column{18E}{@{}l@{}}%
\column{21}{@{}>{\hspre}l<{\hspost}@{}}%
\column{27}{@{}>{\hspre}l<{\hspost}@{}}%
\column{52}{@{}>{\hspre}l<{\hspost}@{}}%
\column{57}{@{}>{\hspre}l<{\hspost}@{}}%
\column{64}{@{}>{\hspre}l<{\hspost}@{}}%
\column{E}{@{}>{\hspre}l<{\hspost}@{}}%
\>[B]{}\mathbf{type\;}\;\Tyvarid{a}\;\Varid{product}{}\<[18]%
\>[18]{}\mathrel{=}{}\<[18E]%
\>[21]{}\Conid{Nil}{}\<[27]%
\>[27]{}\mathrel{\;:\;}{}\<[57]%
\>[57]{}\Varid{unit}\;{}\<[64]%
\>[64]{}\Varid{product}{}\<[E]%
\\
\>[18]{}\mid {}\<[18E]%
\>[21]{}\Conid{Cons}{}\<[27]%
\>[27]{}\mathrel{\;:\;}\Tyvarid{a}\;\Varid{ty}\mathbin{\times}\Tyvarid{b}\;\Varid{product}\to {}\<[52]%
\>[52]{}(\Tyvarid{a}{}\<[57]%
\>[57]{}\mathbin{\times}\Tyvarid{b})\;{}\<[64]%
\>[64]{}\Varid{product}{}\<[E]%
\ColumnHook
\end{hscode}\resethooks
The isomorphism is given by two functions which are required to be each other's inverse.
\begin{hscode}\SaveRestoreHook
\column{B}{@{}>{\hspre}l<{\hspost}@{}}%
\column{E}{@{}>{\hspre}l<{\hspost}@{}}%
\>[B]{}\mathbf{type\;}\;(\Tyvarid{a},\,\Tyvarid{b})\;\Varid{iso}\mathrel{=}\{\mskip1.5mu \Varid{fwd}\mathrel{\;:\;}\Tyvarid{a}\to \Tyvarid{b}\mathbin{;\;}\quad \Varid{bck}\mathrel{\;:\;}\Tyvarid{b}\to \Tyvarid{a}\mskip1.5mu\}{}\<[E]%
\ColumnHook
\end{hscode}\resethooks
\paragraph{Record}
A record type is described by a set of fields, parametrised by a product type.
An isomorphism is provided to convert between records and products.
\savecolumns
\begin{hscode}\SaveRestoreHook
\column{B}{@{}>{\hspre}l<{\hspost}@{}}%
\column{16}{@{}>{\hspre}l<{\hspost}@{}}%
\column{24}{@{}>{\hspre}c<{\hspost}@{}}%
\column{24E}{@{}l@{}}%
\column{27}{@{}>{\hspre}l<{\hspost}@{}}%
\column{42}{@{}>{\hspre}c<{\hspost}@{}}%
\column{42E}{@{}l@{}}%
\column{45}{@{}>{\hspre}l<{\hspost}@{}}%
\column{E}{@{}>{\hspre}l<{\hspost}@{}}%
\>[B]{}\mathbf{type\;}\;(\Tyvarid{b},\,\Tyvarid{a})\;{}\<[16]%
\>[16]{}\Varid{record}{}\<[24]%
\>[24]{}\mathrel{=}{}\<[24E]%
\>[27]{}\{\mskip1.5mu \Varid{name}{}\<[42]%
\>[42]{}\mathrel{\;:\;}{}\<[42E]%
\>[45]{}\Varid{string}{}\<[E]%
\\
\>[27]{}\mathbin{;\;}\Varid{module\char95 path}{}\<[42]%
\>[42]{}\mathrel{\;:\;}{}\<[42E]%
\>[45]{}\Varid{string}\;\Varid{list}{}\<[E]%
\\
\>[27]{}\mathbin{;\;}\Varid{fields}{}\<[42]%
\>[42]{}\mathrel{\;:\;}{}\<[42E]%
\>[45]{}(\Tyvarid{b},\,\Tyvarid{a})\;\Varid{fields}{}\<[E]%
\\
\>[27]{}\mathbin{;\;}\Varid{iso}{}\<[42]%
\>[42]{}\mathrel{\;:\;}{}\<[42E]%
\>[45]{}(\Tyvarid{b},\,\Tyvarid{a})\;\Varid{iso}\mskip1.5mu\}{}\<[E]%
\ColumnHook
\end{hscode}\resethooks
The type \ensuremath{\Varid{fields}} is indexed by the product of the types of the fields.
\restorecolumns
\begin{hscode}\SaveRestoreHook
\column{B}{@{}>{\hspre}l<{\hspost}@{}}%
\column{16}{@{}>{\hspre}l<{\hspost}@{}}%
\column{24}{@{}>{\hspre}c<{\hspost}@{}}%
\column{24E}{@{}l@{}}%
\column{27}{@{}>{\hspre}l<{\hspost}@{}}%
\column{33}{@{}>{\hspre}c<{\hspost}@{}}%
\column{33E}{@{}l@{}}%
\column{36}{@{}>{\hspre}l<{\hspost}@{}}%
\column{73}{@{}>{\hspre}l<{\hspost}@{}}%
\column{88}{@{}>{\hspre}l<{\hspost}@{}}%
\column{E}{@{}>{\hspre}l<{\hspost}@{}}%
\>[B]{}\mathbf{type\;}\;(\Tyvarid{b},\,\Tyvarid{a})\;{}\<[16]%
\>[16]{}\Varid{fields}{}\<[24]%
\>[24]{}\mathrel{=}{}\<[24E]%
\>[27]{}\Conid{Nil}{}\<[33]%
\>[33]{}\mathrel{\;:\;}{}\<[33E]%
\>[73]{}(\Varid{unit},\,\Tyvarid{a})\;{}\<[88]%
\>[88]{}\Varid{fields}{}\<[E]%
\\
\>[24]{}\mid {}\<[24E]%
\>[27]{}\Conid{Cons}{}\<[33]%
\>[33]{}\mathrel{\;:\;}{}\<[33E]%
\>[36]{}(\Tyvarid{b},\,\Tyvarid{a})\;\Varid{field}\mathbin{\times}(\Tyvarid{c},\,\Tyvarid{a})\;\Varid{fields}\to {}\<[73]%
\>[73]{}(\Tyvarid{b}\mathbin{\times}\Tyvarid{c},\,\Tyvarid{a})\;{}\<[88]%
\>[88]{}\Varid{fields}{}\<[E]%
\ColumnHook
\end{hscode}\resethooks
Each field is described by its name, type and a procedure to update its value if it is mutable.
It is indexed by the type of the field and the type of the record to which it belongs.
\restorecolumns
\begin{hscode}\SaveRestoreHook
\column{B}{@{}>{\hspre}l<{\hspost}@{}}%
\column{16}{@{}>{\hspre}l<{\hspost}@{}}%
\column{24}{@{}>{\hspre}c<{\hspost}@{}}%
\column{24E}{@{}l@{}}%
\column{27}{@{}>{\hspre}l<{\hspost}@{}}%
\column{35}{@{}>{\hspre}c<{\hspost}@{}}%
\column{35E}{@{}l@{}}%
\column{38}{@{}>{\hspre}l<{\hspost}@{}}%
\column{E}{@{}>{\hspre}l<{\hspost}@{}}%
\>[B]{}\mathbf{type\;}\;(\Tyvarid{a},\,\Tyvarid{r})\;{}\<[16]%
\>[16]{}\Varid{field}{}\<[24]%
\>[24]{}\mathrel{=}{}\<[24E]%
\>[27]{}\{\mskip1.5mu \Varid{name}{}\<[35]%
\>[35]{}\mathrel{\;:\;}{}\<[35E]%
\>[38]{}\Varid{string}{}\<[E]%
\\
\>[27]{}\mathbin{;\;}\Varid{ty}{}\<[35]%
\>[35]{}\mathrel{\;:\;}{}\<[35E]%
\>[38]{}\Tyvarid{a}\;\Varid{ty}{}\<[E]%
\\
\>[27]{}\mathbin{;\;}\Varid{set}{}\<[35]%
\>[35]{}\mathrel{\;:\;}{}\<[35E]%
\>[38]{}(\Tyvarid{r}\to \Tyvarid{a}\to \Varid{unit})\;\Varid{option}\mskip1.5mu\}{}\<[E]%
\ColumnHook
\end{hscode}\resethooks
\paragraph{Variant}
A variant is described as a set of constructors.
\begin{hscode}\SaveRestoreHook
\column{B}{@{}>{\hspre}l<{\hspost}@{}}%
\column{18}{@{}>{\hspre}c<{\hspost}@{}}%
\column{18E}{@{}l@{}}%
\column{21}{@{}>{\hspre}l<{\hspost}@{}}%
\column{36}{@{}>{\hspre}c<{\hspost}@{}}%
\column{36E}{@{}l@{}}%
\column{39}{@{}>{\hspre}l<{\hspost}@{}}%
\column{E}{@{}>{\hspre}l<{\hspost}@{}}%
\>[B]{}\mathbf{type\;}\;\Tyvarid{a}\;\Varid{variant}{}\<[18]%
\>[18]{}\mathrel{=}{}\<[18E]%
\>[21]{}\{\mskip1.5mu \Varid{name}{}\<[36]%
\>[36]{}\mathrel{\;:\;}{}\<[36E]%
\>[39]{}\Varid{string}{}\<[E]%
\\
\>[21]{}\mathbin{;\;}\Varid{module\char95 path}{}\<[36]%
\>[36]{}\mathrel{\;:\;}{}\<[36E]%
\>[39]{}\Varid{string}\;\Varid{list}{}\<[E]%
\\
\>[21]{}\mathbin{;\;}\Varid{cons}{}\<[36]%
\>[36]{}\mathrel{\;:\;}{}\<[36E]%
\>[39]{}\Tyvarid{a}\;\Varid{cons}\mskip1.5mu\}{}\<[E]%
\ColumnHook
\end{hscode}\resethooks
The set of constructors is an abstract type \ensuremath{\Varid{cons}}.
The primitive way to build a description of the set of constructors is through the function \ensuremath{\Varid{cons}} that turns a list of single constructor descriptions into the abstract set \ensuremath{\Varid{cons}}.
\savecolumns
\begin{hscode}\SaveRestoreHook
\column{B}{@{}>{\hspre}l<{\hspost}@{}}%
\column{15}{@{}>{\hspre}c<{\hspost}@{}}%
\column{15E}{@{}l@{}}%
\column{18}{@{}>{\hspre}l<{\hspost}@{}}%
\column{E}{@{}>{\hspre}l<{\hspost}@{}}%
\>[B]{}\Keyword{val}\;\Varid{cons}{}\<[15]%
\>[15]{}\mathrel{\;:\;}{}\<[15E]%
\>[18]{}\Tyvarid{a}\;\Varid{con}\;\Varid{list}\to \Tyvarid{a}\;\Varid{cons}{}\<[E]%
\ColumnHook
\end{hscode}\resethooks
Each constructor is described by its name, the types of its arguments given as a nested product, a function to embed a value of the nested product to the variant type, and a partial projection function that only succeeds when its argument is built with precisely this constructor.

{\setlength{\commentwidth}{10cm}
\begin{hscode}\SaveRestoreHook
\column{B}{@{}>{\hspre}l<{\hspost}@{}}%
\column{4}{@{}>{\hspre}l<{\hspost}@{}}%
\column{13}{@{}>{\hspre}c<{\hspost}@{}}%
\column{13E}{@{}l@{}}%
\column{16}{@{}>{\hspre}l<{\hspost}@{}}%
\column{39}{@{}>{\hspre}l<{\hspost}@{}}%
\column{E}{@{}>{\hspre}l<{\hspost}@{}}%
\>[B]{}\mathbf{type\;}\;(\Tyvarid{b},\,\Tyvarid{a})\;\Varid{con\char95 desc}\mathrel{=}{}\<[E]%
\\
\>[B]{}\hsindent{4}{}\<[4]%
\>[4]{}\{\mskip1.5mu \Varid{name}{}\<[13]%
\>[13]{}\mathrel{\;:\;}{}\<[13E]%
\>[16]{}\Varid{string}{}\<[39]%
\>[39]{}\mbox{\commentbegin  name of the constructor  \commentend}{}\<[E]%
\\
\>[B]{}\hsindent{4}{}\<[4]%
\>[4]{}\mathbin{;\;}\Varid{args}{}\<[13]%
\>[13]{}\mathrel{\;:\;}{}\<[13E]%
\>[16]{}(\Tyvarid{b},\,\Tyvarid{a})\;\Varid{fields}{}\<[39]%
\>[39]{}\mbox{\commentbegin  arguments of the constructor  \commentend}{}\<[E]%
\\
\>[B]{}\hsindent{4}{}\<[4]%
\>[4]{}\mathbin{;\;}\Varid{embed}{}\<[13]%
\>[13]{}\mathrel{\;:\;}{}\<[13E]%
\>[16]{}\Tyvarid{b}\to \Tyvarid{a}{}\<[39]%
\>[39]{}\mbox{\commentbegin  applies the constructor to the arguments.  \commentend}{}\<[E]%
\\
\>[B]{}\hsindent{4}{}\<[4]%
\>[4]{}\mathbin{;\;}\Varid{proj}{}\<[13]%
\>[13]{}\mathrel{\;:\;}{}\<[13E]%
\>[16]{}\Tyvarid{a}\to \Tyvarid{b}\;\Varid{option}\mskip1.5mu\}{}\<[39]%
\>[39]{}\mbox{\commentbegin  tries to deconstruct that constructor  \commentend}{}\<[E]%
\\
\>[B]{}\mathbf{type\;}\;\Tyvarid{v}\;\Varid{con}\mathrel{=}\Conid{Con}\mathrel{\;:\;}(\Tyvarid{t},\,\Tyvarid{v})\;\Varid{desc}\to \Tyvarid{v}\;\Varid{con}{}\<[E]%
\ColumnHook
\end{hscode}\resethooks
}
Through the public functions, non-constant constructors ({\em i.e.} whose arity is $> 0$) and constant constructors are accessed separately and ordered by their internal tag. This information is crucial for checking the compatibility of runtime values with a given variant and is needed in the implementation of type-safe deserialisation, see section~\ref{sec:unmarshal}.

The interface consists of functions to get the number of constant constructors, access a constant constructor of a given tag and construct the list of all constant constructors ordered by their tags.

\restorecolumns
\begin{hscode}\SaveRestoreHook
\column{B}{@{}>{\hspre}l<{\hspost}@{}}%
\column{15}{@{}>{\hspre}c<{\hspost}@{}}%
\column{15E}{@{}l@{}}%
\column{18}{@{}>{\hspre}l<{\hspost}@{}}%
\column{E}{@{}>{\hspre}l<{\hspost}@{}}%
\>[B]{}\Keyword{val}\;\Varid{cst\char95 len}{}\<[15]%
\>[15]{}\mathrel{\;:\;}{}\<[15E]%
\>[18]{}\Tyvarid{a}\;\Varid{cons}\to \Varid{int}{}\<[E]%
\\
\>[B]{}\Keyword{val}\;\Varid{cst\char95 get}{}\<[15]%
\>[15]{}\mathrel{\;:\;}{}\<[15E]%
\>[18]{}\Tyvarid{a}\;\Varid{cons}\to \Varid{int}\to \Tyvarid{a}\;\Varid{con}{}\<[E]%
\\
\>[B]{}\Keyword{val}\;\Varid{cst}{}\<[15]%
\>[15]{}\mathrel{\;:\;}{}\<[15E]%
\>[18]{}\Tyvarid{a}\;\Varid{cons}\to \Tyvarid{a}\;\Varid{con}\;\Varid{list}{}\<[E]%
\ColumnHook
\end{hscode}\resethooks
The same set of functions is provided for non-constant constructors.
\restorecolumns
\begin{hscode}\SaveRestoreHook
\column{B}{@{}>{\hspre}l<{\hspost}@{}}%
\column{15}{@{}>{\hspre}c<{\hspost}@{}}%
\column{15E}{@{}l@{}}%
\column{18}{@{}>{\hspre}l<{\hspost}@{}}%
\column{E}{@{}>{\hspre}l<{\hspost}@{}}%
\>[B]{}\Keyword{val}\;\Varid{ncst\char95 len}{}\<[15]%
\>[15]{}\mathrel{\;:\;}{}\<[15E]%
\>[18]{}\Tyvarid{a}\;\Varid{cons}\to \Varid{int}{}\<[E]%
\\
\>[B]{}\Keyword{val}\;\Varid{ncst\char95 get}{}\<[15]%
\>[15]{}\mathrel{\;:\;}{}\<[15E]%
\>[18]{}\Tyvarid{a}\;\Varid{cons}\to \Varid{int}\to \Tyvarid{a}\;\Varid{con}{}\<[E]%
\\
\>[B]{}\Keyword{val}\;\Varid{ncst}{}\<[15]%
\>[15]{}\mathrel{\;:\;}{}\<[15E]%
\>[18]{}\Tyvarid{a}\;\Varid{cons}\to \Tyvarid{a}\;\Varid{con}\;\Varid{list}{}\<[E]%
\ColumnHook
\end{hscode}\resethooks
The list of all constructors (constant and non-constant) may be computed with \ensuremath{\Varid{con\char95 list}}.
\restorecolumns
\begin{hscode}\SaveRestoreHook
\column{B}{@{}>{\hspre}l<{\hspost}@{}}%
\column{15}{@{}>{\hspre}c<{\hspost}@{}}%
\column{15E}{@{}l@{}}%
\column{18}{@{}>{\hspre}l<{\hspost}@{}}%
\column{E}{@{}>{\hspre}l<{\hspost}@{}}%
\>[B]{}\Keyword{val}\;\Varid{con\char95 list}{}\<[15]%
\>[15]{}\mathrel{\;:\;}{}\<[15E]%
\>[18]{}\Tyvarid{a}\;\Varid{cons}\to \Tyvarid{a}\;\Varid{con}\;\Varid{list}{}\<[E]%
\ColumnHook
\end{hscode}\resethooks
Finally, the function \ensuremath{\Varid{conap}\mathrel{\;:\;}\Tyvarid{a}\;\Varid{cons}\to \Tyvarid{a}\to \Tyvarid{a}\;\Varid{conap}}
deconstructs---in constant time---a value into a pair of a constructor and its arguments.
\begin{hscode}\SaveRestoreHook
\column{B}{@{}>{\hspre}l<{\hspost}@{}}%
\column{E}{@{}>{\hspre}l<{\hspost}@{}}%
\>[B]{}\mathbf{type\;}\;\Tyvarid{a}\;\Varid{conap}\mathrel{=}\Conid{Conap}\mathrel{\;:\;}(\Tyvarid{b},\,\Tyvarid{a})\;\Varid{con\char95 desc}\mathbin{\times}\Tyvarid{b}\to \Tyvarid{a}\;\Varid{conap}{}\<[E]%
\ColumnHook
\end{hscode}\resethooks
The function \ensuremath{\Varid{conap}} enjoys the following property:
\ensuremath{\Varid{conap}\;\Varid{cs}\;\Varid{x}\mathrel{=}\Conid{Conap}\;(\Varid{c},\,\Varid{y})} $\implies$ \ensuremath{\Varid{c}.\Varid{embed}\;\Varid{y}\mathrel{=}\Varid{x}}.

Note that GADTs may be described as variants. The set of constructors may vary depending on the concrete type index of the GADT.

\paragraph{Extensible}
Extensible variants allow new constructors to be added to a type after it has been defined. The generic view for extensible variants must also be extensible so that the description of the new constructors may be added to the description of the extensible variant.

\begin{hscode}\SaveRestoreHook
\column{B}{@{}>{\hspre}l<{\hspost}@{}}%
\column{21}{@{}>{\hspre}c<{\hspost}@{}}%
\column{21E}{@{}l@{}}%
\column{24}{@{}>{\hspre}l<{\hspost}@{}}%
\column{39}{@{}>{\hspre}c<{\hspost}@{}}%
\column{39E}{@{}l@{}}%
\column{42}{@{}>{\hspre}l<{\hspost}@{}}%
\column{E}{@{}>{\hspre}l<{\hspost}@{}}%
\>[B]{}\mathbf{type\;}\;\Tyvarid{a}\;\Varid{extensible}{}\<[21]%
\>[21]{}\mathrel{=}{}\<[21E]%
\>[24]{}\{\mskip1.5mu \Varid{name}{}\<[39]%
\>[39]{}\mathrel{\;:\;}{}\<[39E]%
\>[42]{}\Varid{string}{}\<[E]%
\\
\>[24]{}\mathbin{;\;}\Varid{module\char95 path}{}\<[39]%
\>[39]{}\mathrel{\;:\;}{}\<[39E]%
\>[42]{}\Varid{string}\;\Varid{list}{}\<[E]%
\\
\>[24]{}\mathbin{;\;}\Varid{ty}{}\<[39]%
\>[39]{}\mathrel{\;:\;}{}\<[39E]%
\>[42]{}\Tyvarid{a}\;\Varid{ty}{}\<[E]%
\\
\>[24]{}\mathbin{;\;}\Varid{cons}{}\<[39]%
\>[39]{}\mathrel{\;:\;}{}\<[39E]%
\>[42]{}\Varid{ext\char95 cons}\mskip1.5mu\}{}\<[E]%
\ColumnHook
\end{hscode}\resethooks
The extensible set of constructors for the type \ensuremath{\Tyvarid{a}} is given by the field \ensuremath{\Varid{cons}\mathrel{\;:\;}\Varid{ext\char95 cons}} where
\ensuremath{\Varid{ext\char95 cons}} is an abstract type. An initially empty set is created with:
\begin{hscode}\SaveRestoreHook
\column{B}{@{}>{\hspre}l<{\hspost}@{}}%
\column{E}{@{}>{\hspre}l<{\hspost}@{}}%
\>[B]{}\Keyword{val}\;\Varid{create}\mathrel{\;:\;}\Varid{unit}\to \Varid{ext\char95 cons}{}\<[E]%
\ColumnHook
\end{hscode}\resethooks
A public interface allows us to modify and query the set of constructors:
{\setlength{\commentwidth}{8.5cm}
\begin{hscode}\SaveRestoreHook
\column{B}{@{}>{\hspre}l<{\hspost}@{}}%
\column{6}{@{}>{\hspre}l<{\hspost}@{}}%
\column{16}{@{}>{\hspre}l<{\hspost}@{}}%
\column{51}{@{}>{\hspre}l<{\hspost}@{}}%
\column{E}{@{}>{\hspre}l<{\hspost}@{}}%
\>[B]{}\mathbf{type\;}\;\Varid{con\char95 fn}\mathrel{=}\{\mskip1.5mu \Varid{con}\mathrel{\;:\;}\forall\!\;\Tyvarid{a}\mathbin{°.°}\Tyvarid{a}\;\Varid{ty}\to \Tyvarid{a}\;\Varid{con}\mathbin{;\;}\mskip1.5mu\}{}\<[E]%
\\
\>[B]{}\Keyword{val}\;{}\<[6]%
\>[6]{}\Varid{add\char95 con}{}\<[16]%
\>[16]{}\mathrel{\;:\;}\Tyvarid{a}\;\Varid{extensible}\to \Varid{con\char95 fn}\to \Varid{unit}{}\<[51]%
\>[51]{}\mbox{\commentbegin  Add the description of a new constructor.  \commentend}{}\<[E]%
\\
\>[B]{}\Keyword{val}\;{}\<[6]%
\>[6]{}\Varid{con\char95 list}{}\<[16]%
\>[16]{}\mathrel{\;:\;}\Tyvarid{a}\;\Varid{extensible}\to \Tyvarid{a}\;\Varid{con}\;\Varid{list}{}\<[51]%
\>[51]{}\mbox{\commentbegin  Return the list of existing constructors.  \commentend}{}\<[E]%
\\
\>[B]{}\Keyword{val}\;{}\<[6]%
\>[6]{}\Varid{con}{}\<[16]%
\>[16]{}\mathrel{\;:\;}\Tyvarid{a}\;\Varid{extensible}\to \Tyvarid{a}\to \Tyvarid{a}\;\Varid{con}{}\<[51]%
\>[51]{}\mbox{\commentbegin  Find the constructor of a given value.  \commentend}{}\<[E]%
\\
\>[B]{}\Keyword{val}\;{}\<[6]%
\>[6]{}\Varid{conap}{}\<[16]%
\>[16]{}\mathrel{\;:\;}\Tyvarid{a}\;\Varid{extensible}\to \Tyvarid{a}\to \Tyvarid{a}\;\Varid{conap}{}\<[51]%
\>[51]{}\mbox{\commentbegin  Deconstruct a value as a constructor application.  \commentend}{}\<[E]%
\ColumnHook
\end{hscode}\resethooks
}
\noindent Finally, the function \ensuremath{\Varid{reinstate}} is of particular interest for deserialisation:
\begin{hscode}\SaveRestoreHook
\column{B}{@{}>{\hspre}l<{\hspost}@{}}%
\column{E}{@{}>{\hspre}l<{\hspost}@{}}%
\>[B]{}\Keyword{val}\;\Varid{reinstate}\mathrel{\;:\;}\Tyvarid{a}\;\Varid{extensible}\to \Tyvarid{a}\to \Tyvarid{a}{}\<[E]%
\ColumnHook
\end{hscode}\resethooks
Serialising then deserializing an extensible value (including exceptions) does not preserve structural equality: \ensuremath{\Varid{umarshall}\;(\Varid{marshall}\;\Varid{x})\mathbin{<>}\Varid{x}}.
In particular, pattern matching the deserialised value does not work as expected. The function \ensuremath{\Varid{reinstate}} fixes that: when given a deserialised value, it returns a value structurally equal to the one that was serialised: \ensuremath{\Varid{reinstate}\;(\Varid{unmarshall}\;(\Varid{marshall}\;\Varid{x}))\mathrel{=}\Varid{x}}

In details, when an extensible value has been deserialised, its memory representation will be different from that of the value before it was serialised. This is because constructors of extensible variants are implemented as object blocks (same as OCaml objects), and they get assigned a unique identifier when created. The expression \ensuremath{\Varid{reinstate}\;\Varid{ext}\;\Varid{x}} replaces the constructor of \ensuremath{\Varid{x}} with the original constructor object that is stored in the \ensuremath{\Tyvarid{a}\;\Varid{extensible}} data structure \ensuremath{\Varid{ext}}.

\paragraph{Custom}
Custom data are defined outside of \ocaml{}, typically in the language C. They are considered as abstract from an \ocaml{} perspective.
The only information available is their identifier, given by the homonymous field of the C-struct \ensuremath{\Varid{custom\char95 operations}} defined in \text{\tt \char60{}caml\char47{}custom\char46{}h\char62{}}.

\begin{hscode}\SaveRestoreHook
\column{B}{@{}>{\hspre}l<{\hspost}@{}}%
\column{17}{@{}>{\hspre}c<{\hspost}@{}}%
\column{17E}{@{}l@{}}%
\column{20}{@{}>{\hspre}l<{\hspost}@{}}%
\column{35}{@{}>{\hspre}c<{\hspost}@{}}%
\column{35E}{@{}l@{}}%
\column{38}{@{}>{\hspre}l<{\hspost}@{}}%
\column{E}{@{}>{\hspre}l<{\hspost}@{}}%
\>[B]{}\mathbf{type\;}\;\Tyvarid{a}\;\Varid{custom}{}\<[17]%
\>[17]{}\mathrel{=}{}\<[17E]%
\>[20]{}\{\mskip1.5mu \Varid{name}{}\<[35]%
\>[35]{}\mathrel{\;:\;}{}\<[35E]%
\>[38]{}\Varid{string}{}\<[E]%
\\
\>[20]{}\mathbin{;\;}\Varid{module\char95 path}{}\<[35]%
\>[35]{}\mathrel{\;:\;}{}\<[35E]%
\>[38]{}\Varid{string}\;\Varid{list}{}\<[E]%
\\
\>[20]{}\mathbin{;\;}\Varid{identifier}{}\<[35]%
\>[35]{}\mathrel{\;:\;}{}\<[35E]%
\>[38]{}\Varid{string}\mskip1.5mu\}{}\<[E]%
\ColumnHook
\end{hscode}\resethooks
Generic support for custom types comes from the use of a public representation, just like with abstract types, see section~\ref{sec:abstract}.

\paragraph{Class}
A class is given as a list of methods.
\begin{hscode}\SaveRestoreHook
\column{B}{@{}>{\hspre}l<{\hspost}@{}}%
\column{18}{@{}>{\hspre}c<{\hspost}@{}}%
\column{18E}{@{}l@{}}%
\column{21}{@{}>{\hspre}l<{\hspost}@{}}%
\column{36}{@{}>{\hspre}c<{\hspost}@{}}%
\column{36E}{@{}l@{}}%
\column{39}{@{}>{\hspre}l<{\hspost}@{}}%
\column{E}{@{}>{\hspre}l<{\hspost}@{}}%
\>[B]{}\mathbf{type\;}\;\Tyvarid{a}\;\Varid{class\char95 t}{}\<[18]%
\>[18]{}\mathrel{=}{}\<[18E]%
\>[21]{}\{\mskip1.5mu \Varid{name}{}\<[36]%
\>[36]{}\mathrel{\;:\;}{}\<[36E]%
\>[39]{}\Varid{string}{}\<[E]%
\\
\>[21]{}\mathbin{;\;}\Varid{module\char95 path}{}\<[36]%
\>[36]{}\mathrel{\;:\;}{}\<[36E]%
\>[39]{}\Varid{string}\;\Varid{list}{}\<[E]%
\\
\>[21]{}\mathbin{;\;}\Varid{methods}{}\<[36]%
\>[36]{}\mathrel{\;:\;}{}\<[36E]%
\>[39]{}\Tyvarid{a}\;\Varid{method\char95 t}\;\Varid{list}\mskip1.5mu\}{}\<[E]%
\ColumnHook
\end{hscode}\resethooks
A method has a name, a type, and a function that executes the corresponding method when called on an object of the class.
\begin{hscode}\SaveRestoreHook
\column{B}{@{}>{\hspre}l<{\hspost}@{}}%
\column{16}{@{}>{\hspre}l<{\hspost}@{}}%
\column{29}{@{}>{\hspre}c<{\hspost}@{}}%
\column{29E}{@{}l@{}}%
\column{32}{@{}>{\hspre}l<{\hspost}@{}}%
\column{40}{@{}>{\hspre}c<{\hspost}@{}}%
\column{40E}{@{}l@{}}%
\column{43}{@{}>{\hspre}l<{\hspost}@{}}%
\column{69}{@{}>{\hspre}c<{\hspost}@{}}%
\column{69E}{@{}l@{}}%
\column{72}{@{}>{\hspre}l<{\hspost}@{}}%
\column{100}{@{}>{\hspre}c<{\hspost}@{}}%
\column{100E}{@{}l@{}}%
\column{103}{@{}>{\hspre}l<{\hspost}@{}}%
\column{E}{@{}>{\hspre}l<{\hspost}@{}}%
\>[B]{}\mathbf{type\;}\;\Tyvarid{a}\;{}\<[16]%
\>[16]{}\Varid{method\char95 t}{}\<[29]%
\>[29]{}\mathrel{=}{}\<[29E]%
\>[32]{}\Conid{Method}\mathrel{\;:\;}(\Tyvarid{b},\,\Tyvarid{a})\;\Varid{method\char95 desc}\to \Tyvarid{a}\;\Varid{method\char95 t}{}\<[E]%
\\
\>[B]{}\mathbf{type\;}\;(\Tyvarid{b},\,\Tyvarid{a})\;{}\<[16]%
\>[16]{}\Varid{method\char95 desc}{}\<[29]%
\>[29]{}\mathrel{=}{}\<[29E]%
\>[32]{}\{\mskip1.5mu \Varid{name}{}\<[40]%
\>[40]{}\mathrel{\;:\;}{}\<[40E]%
\>[43]{}\Varid{string}\mathbin{;\;}\quad\Varid{send}{}\<[69]%
\>[69]{}\mathrel{\;:\;}{}\<[69E]%
\>[72]{}\Tyvarid{b}\to \Tyvarid{a}\mathbin{;\;}\quad\Varid{ty}{}\<[100]%
\>[100]{}\mathrel{\;:\;}{}\<[100E]%
\>[103]{}\Tyvarid{b}\;\Varid{ty}\mskip1.5mu\}{}\<[E]%
\ColumnHook
\end{hscode}\resethooks
For example, the point class:
\begin{hscode}\SaveRestoreHook
\column{B}{@{}>{\hspre}l<{\hspost}@{}}%
\column{21}{@{}>{\hspre}l<{\hspost}@{}}%
\column{24}{@{}>{\hspre}l<{\hspost}@{}}%
\column{39}{@{}>{\hspre}c<{\hspost}@{}}%
\column{39E}{@{}l@{}}%
\column{42}{@{}>{\hspre}l<{\hspost}@{}}%
\column{E}{@{}>{\hspre}l<{\hspost}@{}}%
\>[B]{}\Keyword{class}\;\Varid{point}\;\Varid{init}\mathrel{=}{}\<[21]%
\>[21]{}\Keyword{object}{}\<[E]%
\\
\>[21]{}\hsindent{3}{}\<[24]%
\>[24]{}\Keyword{val}\;\Keyword{mutable}\;\Varid{x}{}\<[39]%
\>[39]{}\mathrel{=}{}\<[39E]%
\>[42]{}\Varid{init}{}\<[E]%
\\
\>[21]{}\hsindent{3}{}\<[24]%
\>[24]{}\Keyword{method}\;\Varid{get\char95 x}{}\<[39]%
\>[39]{}\mathrel{=}{}\<[39E]%
\>[42]{}\Varid{x}{}\<[E]%
\\
\>[21]{}\hsindent{3}{}\<[24]%
\>[24]{}\Keyword{method}\;\Varid{move}\;\Varid{d}{}\<[39]%
\>[39]{}\mathrel{=}{}\<[39E]%
\>[42]{}\Varid{x}\leftarrow \Varid{x}\mathbin{+}\Varid{d}{}\<[E]%
\\
\>[21]{}\Keyword{end}{}\<[E]%
\ColumnHook
\end{hscode}\resethooks
is described by:
\begin{hscode}\SaveRestoreHook
\column{B}{@{}>{\hspre}l<{\hspost}@{}}%
\column{6}{@{}>{\hspre}l<{\hspost}@{}}%
\column{13}{@{}>{\hspre}c<{\hspost}@{}}%
\column{13E}{@{}l@{}}%
\column{16}{@{}>{\hspre}l<{\hspost}@{}}%
\column{35}{@{}>{\hspre}l<{\hspost}@{}}%
\column{63}{@{}>{\hspre}l<{\hspost}@{}}%
\column{E}{@{}>{\hspre}l<{\hspost}@{}}%
\>[B]{}\Keyword{let}\;{}\<[6]%
\>[6]{}\Varid{get\char95 x}{}\<[13]%
\>[13]{}\mathrel{=}{}\<[13E]%
\>[16]{}\{\mskip1.5mu \Varid{name}\mathrel{=}\text{\tt \char34 get\char95 x\char34}\mathbin{;\;}{}\<[35]%
\>[35]{}\Varid{send}\mathrel{=}(\Keyword{fun}\;\Varid{c}\to \Varid{c}\mathbin{\#}\Varid{get\char95 x})\mathbin{;\;}{}\<[63]%
\>[63]{}\Varid{ty}\mathrel{=}\Conid{Int}\mskip1.5mu\}{}\<[E]%
\\
\>[B]{}\Keyword{and}\;{}\<[6]%
\>[6]{}\Varid{move}{}\<[13]%
\>[13]{}\mathrel{=}{}\<[13E]%
\>[16]{}\{\mskip1.5mu \Varid{name}\mathrel{=}\text{\tt \char34 move\char34}\mathbin{;\;}{}\<[35]%
\>[35]{}\Varid{send}\mathrel{=}(\Keyword{fun}\;\Varid{c}\to \Varid{c}\mathbin{\#}\Varid{move})\mathbin{;\;}{}\<[63]%
\>[63]{}\Varid{ty}\mathrel{=}\Conid{Fun}\;(\Conid{Int},\,\Conid{Unit})\mskip1.5mu\}{}\<[E]%
\\
\>[B]{}\Keyword{in}\;\Conid{Class}\;\{\mskip1.5mu \Varid{name}\mathrel{=}\text{\tt \char34 Point\char34}\mathbin{;\;}\Varid{methods}\mathrel{=}[\mskip1.5mu \Conid{Method}\;\Varid{get\char95 x}\mathbin{;\;}\Conid{Method}\;\Varid{move}\mskip1.5mu]\mskip1.5mu\}{}\<[E]%
\ColumnHook
\end{hscode}\resethooks
\paragraph{Synonym}
The view for a type synonym \ensuremath{\mathbf{type\;}\;\Varid{s}\mathrel{=}\Varid{t}} consists of a type witness for \ensuremath{\Varid{t}} and a proof that the two types are equal. The equality type corresponds to the equivalence of type synonyms in \ocaml{}; it was introduced in section~\ref{sec:type-equality}.
\begin{hscode}\SaveRestoreHook
\column{B}{@{}>{\hspre}l<{\hspost}@{}}%
\column{18}{@{}>{\hspre}c<{\hspost}@{}}%
\column{18E}{@{}l@{}}%
\column{21}{@{}>{\hspre}l<{\hspost}@{}}%
\column{36}{@{}>{\hspre}c<{\hspost}@{}}%
\column{36E}{@{}l@{}}%
\column{39}{@{}>{\hspre}l<{\hspost}@{}}%
\column{E}{@{}>{\hspre}l<{\hspost}@{}}%
\>[B]{}\mathbf{type\;}\;\Tyvarid{a}\;\Varid{synonym}{}\<[18]%
\>[18]{}\mathrel{=}{}\<[18E]%
\>[21]{}\{\mskip1.5mu \Varid{name}{}\<[36]%
\>[36]{}\mathrel{\;:\;}{}\<[36E]%
\>[39]{}\Varid{string}{}\<[E]%
\\
\>[21]{}\mathbin{;\;}\Varid{module\char95 path}{}\<[36]%
\>[36]{}\mathrel{\;:\;}{}\<[36E]%
\>[39]{}\Varid{string}\;\Varid{list}{}\<[E]%
\\
\>[21]{}\mathbin{;\;}\Varid{ty}{}\<[36]%
\>[36]{}\mathrel{\;:\;}{}\<[36E]%
\>[39]{}\Tyvarid{b}\;\Varid{ty}{}\<[E]%
\\
\>[21]{}\mathbin{;\;}\Varid{eq}{}\<[36]%
\>[36]{}\mathrel{\;:\;}{}\<[36E]%
\>[39]{}(\Tyvarid{b},\,\Tyvarid{a})\;\Varid{equal}\mskip1.5mu\}{}\<[E]%
\ColumnHook
\end{hscode}\resethooks
\paragraph{Abstract}
No information is associated with an abstract type, except for its name, thus respecting the desire of the programmer to hide the concrete implementation of the type.
Still, we may run generic functions over abstract types if they have a public representation. This is explained in section~\ref{sec:abstract}.

\begin{hscode}\SaveRestoreHook
\column{B}{@{}>{\hspre}l<{\hspost}@{}}%
\column{30}{@{}>{\hspre}l<{\hspost}@{}}%
\column{65}{@{}>{\hspre}l<{\hspost}@{}}%
\column{E}{@{}>{\hspre}l<{\hspost}@{}}%
\>[B]{}\mathbf{type\;}\;\Tyvarid{a}\;\Varid{abstract}\mathrel{=}\{\mskip1.5mu \Varid{name}\mathrel{\;:\;}{}\<[30]%
\>[30]{}\Varid{string}\mathbin{;\;}\quad\Varid{module\char95 path}\mathrel{\;:\;}{}\<[65]%
\>[65]{}\Varid{string}\;\Varid{list}\mskip1.5mu\}{}\<[E]%
\ColumnHook
\end{hscode}\resethooks
\paragraph{NoDesc}
The constructor \ensuremath{\Conid{NoDesc}} is used to signify that a view is not yet, or cannot be, associated with a type. For instance a function type does not have a meaningful generic view.

\subsubsection{Objects and Polymorphic Variants}
\label{sec:view/objects-polyvariants}
Objects types and polymorphic variant types are structural types rather than nominal types: an object type is given by the set of his methods signatures, and a polymorphic variant type is given by the set of its constructors signatures.
Since there is no type name to be reflected in a type witness, one must instead provide a view as the type witness.
\begin{hscode}\SaveRestoreHook
\column{B}{@{}>{\hspre}l<{\hspost}@{}}%
\column{12}{@{}>{\hspre}c<{\hspost}@{}}%
\column{12E}{@{}l@{}}%
\column{16}{@{}>{\hspre}l<{\hspost}@{}}%
\column{29}{@{}>{\hspre}c<{\hspost}@{}}%
\column{29E}{@{}l@{}}%
\column{32}{@{}>{\hspre}l<{\hspost}@{}}%
\column{53}{@{}>{\hspre}l<{\hspost}@{}}%
\column{E}{@{}>{\hspre}l<{\hspost}@{}}%
\>[B]{}\mathbf{type\;}\;\anonymous \;\Varid{ty}{}\<[12]%
\>[12]{}\mathrel{+\!\!\!=}{}\<[12E]%
\>[16]{}\Conid{Object}{}\<[29]%
\>[29]{}\mathrel{\;:\;}{}\<[29E]%
\>[32]{}\Tyvarid{a}\;\Varid{object\char95 desc}{}\<[53]%
\>[53]{}\to \Tyvarid{a}\;\Varid{ty}{}\<[E]%
\\
\>[B]{}\mathbf{type\;}\;\anonymous \;\Varid{ty}{}\<[12]%
\>[12]{}\mathrel{+\!\!\!=}{}\<[12E]%
\>[16]{}\Conid{PolyVariant}{}\<[29]%
\>[29]{}\mathrel{\;:\;}{}\<[29E]%
\>[32]{}\Tyvarid{a}\;\Varid{polyvariant\char95 desc}{}\<[53]%
\>[53]{}\to \Tyvarid{a}\;\Varid{ty}{}\<[E]%
\ColumnHook
\end{hscode}\resethooks
To describe objects, the \ensuremath{\Keyword{method}} datatype of the previous section is reused:
\begin{hscode}\SaveRestoreHook
\column{B}{@{}>{\hspre}l<{\hspost}@{}}%
\column{E}{@{}>{\hspre}l<{\hspost}@{}}%
\>[B]{}\mathbf{type\;}\;\Tyvarid{a}\;\Varid{object\char95 desc}\mathrel{=}\Tyvarid{a}\;\Varid{method\char95 t}\;\Varid{list}{}\<[E]%
\ColumnHook
\end{hscode}\resethooks
Values of polymorphic variant types have a memory representation
different from values of normal variants, as such they need a distinct generic description.
Each data constructor of a polymorphic variant is associated to a hash value, thus we provide operations to compute that hash value, as well as operations to compute the data constructor corresponding to a given hash.

We define an abstract type \ensuremath{\Tyvarid{a}\;\Varid{poly\char95 variant}} representing the set of constructors of a polymorphic variant type \ensuremath{\Tyvarid{a}}, together with functions to create a \ensuremath{\Varid{poly\char95 variant}} and extract a constructor and compute its \ensuremath{\Varid{hash}} value.
\savecolumns
\begin{hscode}\SaveRestoreHook
\column{B}{@{}>{\hspre}l<{\hspost}@{}}%
\column{19}{@{}>{\hspre}c<{\hspost}@{}}%
\column{19E}{@{}l@{}}%
\column{22}{@{}>{\hspre}l<{\hspost}@{}}%
\column{E}{@{}>{\hspre}l<{\hspost}@{}}%
\>[B]{}\mathbf{type\;}\;\Tyvarid{a}\;\Varid{poly\char95 variant}{}\<[E]%
\\
\>[B]{}\Keyword{val}\;\Varid{poly\char95 variant}{}\<[19]%
\>[19]{}\mathrel{\;:\;}{}\<[19E]%
\>[22]{}\Tyvarid{a}\;\Varid{con}\;\Varid{list}\to \Tyvarid{a}\;\Varid{poly\char95 variant}{}\<[E]%
\\
\>[B]{}\Keyword{val}\;\Varid{hash}{}\<[19]%
\>[19]{}\mathrel{\;:\;}{}\<[19E]%
\>[22]{}\Tyvarid{a}\;\Varid{con}\to \Varid{int}{}\<[E]%
\\
\>[B]{}\Keyword{val}\;\Varid{find}{}\<[19]%
\>[19]{}\mathrel{\;:\;}{}\<[19E]%
\>[22]{}\Tyvarid{a}\;\Varid{poly\char95 variant}\to \Varid{int}\to \Tyvarid{a}\;\Varid{con}{}\<[E]%
\ColumnHook
\end{hscode}\resethooks
\ensuremath{\Varid{conap}} deconstructs a polymorphic variant value into its data constructor and its arguments (see the paragraph on variants, in section~\ref{sec:view/desc}).
\restorecolumns
\begin{hscode}\SaveRestoreHook
\column{B}{@{}>{\hspre}l<{\hspost}@{}}%
\column{19}{@{}>{\hspre}l<{\hspost}@{}}%
\column{E}{@{}>{\hspre}l<{\hspost}@{}}%
\>[B]{}\Keyword{val}\;\Varid{conap}{}\<[19]%
\>[19]{}\mathrel{\;:\;}\Tyvarid{a}\;\Varid{poly\char95 variant}\to \Tyvarid{a}\to \Tyvarid{a}\;\Varid{conap}{}\<[E]%
\ColumnHook
\end{hscode}\resethooks
Important note: the support for objects and polymorphic variant types is very fragile and should be considered experimental.
It is not obvious how they may be compared for equality. As a result
we may not yet extend a type-indexed function with a case for an object or polymorphic variant type.
It is however possible to define generic functions that work on them (through a generic view).
An example is the deserialisation function presented in section~\ref{sec:unmarshal}.

\subsubsection{List of Constructors}
The list of constructor view is similar to the underlying view of the RepLib Haskell library~\cite{weirich:06:replib}.
In a nutshell, the view sees all types as variants. Products and records are viewed as variants of a single constructor. The other categories of types do not fit well under that description, and are left as base cases with no constructors.

The view is similar to the sum-of-products view in the sense that each constructor is associated to a product and the type is the sum of those products.

We reuse the type of constructor descriptions \ensuremath{\Varid{con\char95 desc}} defined for the low level \ensuremath{\Varid{desc}} view. For instance, given the witness of the type parameter \ensuremath{\Varid{a}\mathrel{\;:\;}\Tyvarid{a}\;\Varid{ty}}, the list-of-constructors view for \ensuremath{\Tyvarid{a}\;\Varid{btree}} is:
\label{prog:cons-btree}
\begin{hscode}\SaveRestoreHook
\column{B}{@{}>{\hspre}l<{\hspost}@{}}%
\column{5}{@{}>{\hspre}l<{\hspost}@{}}%
\column{16}{@{}>{\hspre}l<{\hspost}@{}}%
\column{23}{@{}>{\hspre}l<{\hspost}@{}}%
\column{32}{@{}>{\hspre}l<{\hspost}@{}}%
\column{E}{@{}>{\hspre}l<{\hspost}@{}}%
\>[B]{}\Keyword{let}\;\Varid{cons\char95 btree}\;\Varid{a}\mathrel{=}{}\<[E]%
\\
\>[B]{}\hsindent{5}{}\<[5]%
\>[5]{}\Keyword{let}\;\Varid{empty}{}\<[16]%
\>[16]{}\mathrel{=}\Conid{Con}\;{}\<[23]%
\>[23]{}\{\mskip1.5mu \Varid{name}{}\<[32]%
\>[32]{}\mathrel{=}\text{\tt \char34 Empty\char34}{}\<[E]%
\\
\>[23]{}\mathbin{;\;}\Varid{args}{}\<[32]%
\>[32]{}\mathrel{=}\Conid{Nil}{}\<[E]%
\\
\>[23]{}\mathbin{;\;}\Varid{embed}{}\<[32]%
\>[32]{}\mathrel{=}(\Keyword{function}\;()\to \Conid{Empty}){}\<[E]%
\\
\>[23]{}\mathbin{;\;}\Varid{proj}{}\<[32]%
\>[32]{}\mathrel{=}(\Keyword{function}\;\Conid{Empty}\to \Conid{Some}\;()\mid \anonymous \to \Conid{None})\mskip1.5mu\}{}\<[E]%
\\
\>[B]{}\hsindent{5}{}\<[5]%
\>[5]{}\Keyword{and}\;\Varid{node}{}\<[16]%
\>[16]{}\mathrel{=}\Conid{Con}\;{}\<[23]%
\>[23]{}\{\mskip1.5mu \Varid{name}{}\<[32]%
\>[32]{}\mathrel{=}\text{\tt \char34 None\char34}{}\<[E]%
\\
\>[23]{}\mathbin{;\;}\Varid{args}{}\<[32]%
\>[32]{}\mathrel{=}\Varid{f3}\;(\Conid{Btree}\;\Varid{a})\;\Varid{a}\;(\Conid{Btree}\;\Varid{a}){}\<[E]%
\\
\>[23]{}\mathbin{;\;}\Varid{embed}{}\<[32]%
\>[32]{}\mathrel{=}(\Keyword{function}\;(\Varid{l},\,(\Varid{x},\,(\Varid{r},\,())))\to \Conid{None}\;(\Varid{l},\,\Varid{x},\,\Varid{r})){}\<[E]%
\\
\>[23]{}\mathbin{;\;}\Varid{proj}{}\<[32]%
\>[32]{}\mathrel{=}(\Keyword{function}\;\Conid{Node}\;(\Varid{l},\,\Varid{x},\,\Varid{r})\to \Conid{Some}\;(\Varid{l},\,(\Varid{x},\,(\Varid{r},\,())))\mid \anonymous \to \Conid{None})\mskip1.5mu\}{}\<[E]%
\\
\>[B]{}\hsindent{5}{}\<[5]%
\>[5]{}\Keyword{in}\;[\mskip1.5mu \Varid{empty}\mathbin{;\;}\Varid{node}\mskip1.5mu]{}\<[E]%
\ColumnHook
\end{hscode}\resethooks
\ensuremath{\Varid{f3}} computes a lists of three fields with empty labels.
\begin{hscode}\SaveRestoreHook
\column{B}{@{}>{\hspre}l<{\hspost}@{}}%
\column{E}{@{}>{\hspre}l<{\hspost}@{}}%
\>[B]{}\Keyword{val}\;\Varid{f3}\mathrel{\;:\;}\Tyvarid{a}\;\Varid{ty}\to \Tyvarid{b}\;\Varid{ty}\to \Tyvarid{c}\;\Varid{ty}\to (\Tyvarid{a}\mathbin{\times}(\Tyvarid{b}\mathbin{\times}(\Tyvarid{c}\mathbin{\times}\Varid{unit})),\,\Tyvarid{d})\;\Varid{fields}{}\<[E]%
\ColumnHook
\end{hscode}\resethooks
\paragraph{Equality}
Typically, functions using the list-of-constructors view are defined by three mutually recursive functions: the first works with type witnesses and delegates the work on the view, the second works on the view and iterates through the list of constructors to find the matching constructor, the third works on the product of arguments of a single constructor.
\begin{hscode}\SaveRestoreHook
\column{B}{@{}>{\hspre}l<{\hspost}@{}}%
\column{6}{@{}>{\hspre}l<{\hspost}@{}}%
\column{11}{@{}>{\hspre}l<{\hspost}@{}}%
\column{26}{@{}>{\hspre}c<{\hspost}@{}}%
\column{26E}{@{}l@{}}%
\column{29}{@{}>{\hspre}l<{\hspost}@{}}%
\column{39}{@{}>{\hspre}l<{\hspost}@{}}%
\column{44}{@{}>{\hspre}l<{\hspost}@{}}%
\column{45}{@{}>{\hspre}l<{\hspost}@{}}%
\column{60}{@{}>{\hspre}l<{\hspost}@{}}%
\column{67}{@{}>{\hspre}l<{\hspost}@{}}%
\column{68}{@{}>{\hspre}l<{\hspost}@{}}%
\column{E}{@{}>{\hspre}l<{\hspost}@{}}%
\>[B]{}\Keyword{let}\;{}\<[6]%
\>[6]{}\Keyword{rec}\;{}\<[11]%
\>[11]{}\Varid{equal}{}\<[26]%
\>[26]{}\mathrel{\;:\;}{}\<[26E]%
\>[29]{}\mathbf{type\;}\;\Varid{a}\mathbin{°.°}\Varid{a}\;\Varid{ty}\to \Varid{a}\to \Varid{a}\to \Varid{bool}{}\<[E]%
\\
\>[26]{}\mathrel{=}{}\<[26E]%
\>[29]{}\Keyword{fun}\;\Varid{t}\to {}\<[39]%
\>[39]{}\Keyword{match}\;\Varid{conlist}\;\Varid{t}\;\Keyword{with}{}\<[E]%
\\
\>[39]{}\mid [\mskip1.5mu \mskip1.5mu]{}\<[45]%
\>[45]{}\to (\mathrel{=}){}\<[67]%
\>[67]{}\mbox{\commentbegin  Base case (core types)  \commentend}{}\<[E]%
\\
\>[39]{}\mid \Varid{cs}{}\<[45]%
\>[45]{}\to \Varid{equal\char95 conlist}\;\Varid{cs}{}\<[67]%
\>[67]{}\mbox{\commentbegin  Generic case  \commentend}{}\<[E]%
\\[\blanklineskip]%
\>[6]{}\Keyword{and}\;{}\<[11]%
\>[11]{}\Varid{equal\char95 conlist}{}\<[26]%
\>[26]{}\mathrel{\;:\;}{}\<[26E]%
\>[29]{}\mathbf{type\;}\;\Varid{a}\mathbin{°.°}\Varid{a}\;\Varid{con}\;\Varid{list}\to \Varid{a}\to \Varid{a}\to \Varid{bool}{}\<[E]%
\\
\>[26]{}\mathrel{=}{}\<[26E]%
\>[29]{}\Keyword{fun}\;\Varid{cs}\;\Varid{x}\;\Varid{y}\to {}\<[44]%
\>[44]{}\Keyword{match}\;\Varid{conap}\;\Varid{cs}\;\Varid{x}\;\Keyword{with}{}\<[E]%
\\
\>[44]{}\mid \Conid{Conap}\;(\Varid{c},\,\Varid{x'}){}\<[60]%
\>[60]{}\to \Keyword{match}\;\Varid{c}.\Varid{proj}\;\Varid{y}\;\Keyword{with}{}\<[E]%
\\
\>[44]{}\mid \Conid{None}{}\<[60]%
\>[60]{}\to \Keyword{false}\;\mbox{\commentbegin  Not the same constructor  \commentend}{}\<[E]%
\\
\>[44]{}\mid \Conid{Some}\;\Varid{y'}{}\<[60]%
\>[60]{}\to \Varid{equal\char95 prod}\;(\Varid{product}\;\Varid{c})\;\Varid{x'}\;\Varid{y'}{}\<[E]%
\\[\blanklineskip]%
\>[B]{}\Keyword{and}\;\Varid{equal\char95 prod}{}\<[26]%
\>[26]{}\mathrel{\;:\;}{}\<[26E]%
\>[29]{}\mathbf{type\;}\;\Varid{p}.\Varid{p}\;\Varid{product}\to \Varid{p}\to \Varid{p}\to \Varid{bool}{}\<[E]%
\\
\>[26]{}\mathrel{=}{}\<[26E]%
\>[29]{}\Keyword{function}{}\<[E]%
\\
\>[29]{}\mid \Conid{Nil}{}\<[45]%
\>[45]{}\to \Keyword{fun}\;\anonymous \;{}\<[60]%
\>[60]{}\anonymous {}\<[68]%
\>[68]{}\to \Keyword{true}{}\<[E]%
\\
\>[29]{}\mid \Conid{Cons}\;(\Varid{t},\,\Varid{ts}){}\<[45]%
\>[45]{}\to \Keyword{fun}\;(\Varid{x},\,\Varid{xs})\;{}\<[60]%
\>[60]{}(\Varid{y},\,\Varid{ys}){}\<[68]%
\>[68]{}\to \Varid{equal}\;\Varid{t}\;\Varid{x}\;\Varid{y}\mathrel{\wedge}\Varid{equal\char95 prod}\;\Varid{ts}\;\Varid{xs}\;\Varid{ys}{}\<[E]%
\ColumnHook
\end{hscode}\resethooks
A particularly useful function is \ensuremath{\Varid{conap}} which deconstructs a value into a constructor and its arguments, it has the same semantics as the homonymous function on variants given earlier, but this one has a linear complexity since it must walk through its list argument in order to find a matching constructor.
\begin{hscode}\SaveRestoreHook
\column{B}{@{}>{\hspre}l<{\hspost}@{}}%
\column{E}{@{}>{\hspre}l<{\hspost}@{}}%
\>[B]{}\Keyword{val}\;\Varid{conap}\mathrel{\;:\;}\Tyvarid{a}\;\Varid{con}\;\Varid{list}\to \Tyvarid{a}\to \Tyvarid{a}\;\Varid{conap}{}\<[E]%
\ColumnHook
\end{hscode}\resethooks
\paragraph{Children}
The list-of-constructor view makes our job really easy here: \ensuremath{\Varid{conap}} computes the list of all children, whatever their types, and we only need to keep those that have the same type as the parent. The function \ensuremath{\Varid{child}} was defined at the end of section~\ref{prog:child}.
\begin{hscode}\SaveRestoreHook
\column{B}{@{}>{\hspre}l<{\hspost}@{}}%
\column{6}{@{}>{\hspre}l<{\hspost}@{}}%
\column{11}{@{}>{\hspre}l<{\hspost}@{}}%
\column{27}{@{}>{\hspre}c<{\hspost}@{}}%
\column{27E}{@{}l@{}}%
\column{30}{@{}>{\hspre}l<{\hspost}@{}}%
\column{47}{@{}>{\hspre}l<{\hspost}@{}}%
\column{58}{@{}>{\hspre}l<{\hspost}@{}}%
\column{E}{@{}>{\hspre}l<{\hspost}@{}}%
\>[B]{}\Keyword{let}\;{}\<[6]%
\>[6]{}\Keyword{rec}\;{}\<[11]%
\>[11]{}\Varid{filter\char95 child}\;\Varid{t}{}\<[27]%
\>[27]{}\mathrel{=}{}\<[27E]%
\>[30]{}\Keyword{function}{}\<[E]%
\\
\>[30]{}\mid \Conid{Nil}{}\<[47]%
\>[47]{},\,(){}\<[58]%
\>[58]{}\to [\mskip1.5mu \mskip1.5mu]{}\<[E]%
\\
\>[30]{}\mid \Conid{Cons}\;(\Varid{t'},\,\Varid{ts}){}\<[47]%
\>[47]{},\,(\Varid{x},\,\Varid{xs}){}\<[58]%
\>[58]{}\to \Varid{child}\;\Varid{t}\;\Varid{t'}\;\Varid{x}\mathbin{@}\Varid{filter\char95 child}\;\Varid{t}\;(\Varid{ts},\,\Varid{xs}){}\<[E]%
\\[\blanklineskip]%
\>[B]{}\Keyword{let}\;{}\<[11]%
\>[11]{}\Varid{children}\;\Varid{t}\;\Varid{x}{}\<[27]%
\>[27]{}\mathrel{=}{}\<[27E]%
\>[30]{}\Keyword{match}\;\Varid{conap}\;(\Varid{conlist}\;\Varid{t})\;\Varid{x}\;\Keyword{with}{}\<[E]%
\\
\>[30]{}\mid \Conid{Conap}\;(\Varid{c},\,\Varid{y})\to \Varid{filter\char95 child}\;\Varid{t}\;(\Varid{product}\;\Varid{c},\,\Varid{y}){}\<[E]%
\ColumnHook
\end{hscode}\resethooks

\subsubsection{Abstract Types}
\label{sec:abstract}

Abstract types are an essential element of modular programming. Separating the public interface from the concrete implementation allows us to change the implementation without consequences for the users of the module. Generic functions should respect the abstraction, therefore the concrete type structure of an abstract type should not be available through the generic views. This is why the low level view provides a constructor \ensuremath{\Conid{Abstract}} for abstract types that only exports their names. However, in order to compute anything useful, one needs a generic view to convert back and forth between the abstract type and a public representation on which the generic functions may act.

The view is given by a type \ensuremath{\Tyvarid{a}\;\Varid{repr}} that
specifies a representation for an abstract type \ensuremath{\Tyvarid{a}}, and a type-indexed function \ensuremath{\Varid{repr}} that returns the representation associated to a type witness:
\begin{hscode}\SaveRestoreHook
\column{B}{@{}>{\hspre}l<{\hspost}@{}}%
\column{E}{@{}>{\hspre}l<{\hspost}@{}}%
\>[B]{}\Keyword{val}\;\Varid{repr}\mathrel{\;:\;}\Tyvarid{a}\;\Varid{ty}\to \Tyvarid{a}\;\Varid{repr}{}\<[E]%
\ColumnHook
\end{hscode}\resethooks
The type \ensuremath{\Tyvarid{a}\;\Varid{repr}} is existentially quantified over the representation type \ensuremath{\Tyvarid{b}}:
\begin{hscode}\SaveRestoreHook
\column{B}{@{}>{\hspre}l<{\hspost}@{}}%
\column{E}{@{}>{\hspre}l<{\hspost}@{}}%
\>[B]{}\mathbf{type\;}\;\Tyvarid{a}\;\Varid{repr}\mathrel{=}\Conid{Repr}\mathrel{\;:\;}(\Tyvarid{a},\,\Tyvarid{b})\;\Varid{repr\char95 by}\to \Tyvarid{a}\;\Varid{repr}{}\<[E]%
\ColumnHook
\end{hscode}\resethooks
The type \ensuremath{(\Tyvarid{a},\,\Tyvarid{b})\;\Varid{repr\char95 by}} specifies how the abstract type \ensuremath{\Tyvarid{a}} is represented by the type \ensuremath{\Tyvarid{b}}. It is a record type whose fields we explain below:
\par
\vspace{5pt}
{\setlength{\tabcolsep}{0pt}
\begin{tabular}{l@{ : }l@{$\quad$}p{100mm}}
\ensuremath{\Varid{repr\char95 ty}} & \ensuremath{\Tyvarid{b}\;\Varid{ty}}
& Witness of the representation type.\\
\ensuremath{\Varid{to\char95 repr}} & \ensuremath{\Tyvarid{a}\to \Tyvarid{b}}
& Conversion from the abstract type to the representation.\\
\ensuremath{\Varid{from\char95 repr}} & \ensuremath{\Tyvarid{b}\to \Tyvarid{a}\;\Varid{option}}
& Partial conversion from representation to the abstract type. It may fail with \ensuremath{\Conid{None}} if the representation is not valid.
\end{tabular}
}
\vspace{5pt}



\paragraph{Example} An abstract type for natural numbers implemented as \ensuremath{\Varid{int}}.


The module signature hides the implementation of \ensuremath{\Varid{nat}}. The type witness \ensuremath{\Conid{Nat}} must be exported as well if we want to support generic programming.
However, the views \ensuremath{\Varid{desc}} and \ensuremath{\Varid{repr}} are extended as side effects and are not visible in the signature.
\savecolumns
\begin{hscode}\SaveRestoreHook
\column{B}{@{}>{\hspre}l<{\hspost}@{}}%
\column{5}{@{}>{\hspre}l<{\hspost}@{}}%
\column{17}{@{}>{\hspre}c<{\hspost}@{}}%
\column{17E}{@{}l@{}}%
\column{21}{@{}>{\hspre}l<{\hspost}@{}}%
\column{E}{@{}>{\hspre}l<{\hspost}@{}}%
\>[B]{}\Keyword{module}\;\Conid{Nat}\mathrel{\;:\;}\Keyword{sig}{}\<[E]%
\\
\>[B]{}\hsindent{5}{}\<[5]%
\>[5]{}\mathbf{type\;}\;\Varid{nat}{}\<[E]%
\\
\>[B]{}\hsindent{5}{}\<[5]%
\>[5]{}\mathbf{type\;}\;\anonymous \;\Varid{ty}\mathrel{+\!\!\!=}\Conid{Nat}\mathrel{\;:\;}\Varid{nat}\;\Varid{ty}{}\<[E]%
\\
\>[B]{}\Keyword{end}\mathrel{=}\Keyword{struct}{}\<[E]%
\\
\>[B]{}\hsindent{5}{}\<[5]%
\>[5]{}\mathbf{type\;}\;\Varid{nat}{}\<[17]%
\>[17]{}\mathrel{=}{}\<[17E]%
\>[21]{}\Varid{int}{}\<[E]%
\\
\>[B]{}\hsindent{5}{}\<[5]%
\>[5]{}\mathbf{type\;}\;\anonymous \;\Varid{ty}{}\<[17]%
\>[17]{}\mathrel{+\!\!\!=}{}\<[17E]%
\>[21]{}\Conid{Nat}\mathrel{\;:\;}\Varid{nat}\;\Varid{ty}{}\<[E]%
\ColumnHook
\end{hscode}\resethooks
We define \ensuremath{\Conid{Nat}} as \ensuremath{\Conid{Abstract}} in the low level view.
\restorecolumns
\begin{hscode}\SaveRestoreHook
\column{B}{@{}>{\hspre}l<{\hspost}@{}}%
\column{5}{@{}>{\hspre}l<{\hspost}@{}}%
\column{28}{@{}>{\hspre}l<{\hspost}@{}}%
\column{34}{@{}>{\hspre}l<{\hspost}@{}}%
\column{42}{@{}>{\hspre}l<{\hspost}@{}}%
\column{55}{@{}>{\hspre}c<{\hspost}@{}}%
\column{55E}{@{}l@{}}%
\column{58}{@{}>{\hspre}l<{\hspost}@{}}%
\column{71}{@{}>{\hspre}c<{\hspost}@{}}%
\column{71E}{@{}l@{}}%
\column{74}{@{}>{\hspre}l<{\hspost}@{}}%
\column{E}{@{}>{\hspre}l<{\hspost}@{}}%
\>[5]{}\Varid{\Conid{Desc\char95 fun}.ext}\;\Conid{Nat}\;\{\mskip1.5mu \Varid{f}\mathrel{=}{}\<[28]%
\>[28]{}\Keyword{fun}\;(\mathbf{type\;}\;\Varid{a})\;(\Varid{t}\mathrel{\;:\;}\Varid{a}\;\Varid{ty}){}\<[E]%
\\
\>[28]{}\to ({}\<[34]%
\>[34]{}\Keyword{match}\;\Varid{t}\;\Keyword{with}{}\<[E]%
\\
\>[34]{}\mid \Conid{Nat}{}\<[42]%
\>[42]{}\to \Conid{Abstract}\;{}\<[55]%
\>[55]{}\{\mskip1.5mu {}\<[55E]%
\>[58]{}\Varid{name}{}\<[71]%
\>[71]{}\mathrel{=}{}\<[71E]%
\>[74]{}\text{\tt \char34 nat\char34}{}\<[E]%
\\
\>[55]{}\mathbin{;\;}{}\<[55E]%
\>[58]{}\Varid{module\char95 path}{}\<[71]%
\>[71]{}\mathrel{=}{}\<[71E]%
\>[74]{}[\mskip1.5mu \text{\tt \char34 Test\char34}\mathbin{;\;}\text{\tt \char34 Nat\char34}\mskip1.5mu]\mskip1.5mu\}{}\<[E]%
\\
\>[34]{}\mid \anonymous {}\<[42]%
\>[42]{}\to \Keyword{assert}\;\Keyword{false}\mathrel{\;:\;}\Varid{a}\;\Varid{desc})\mskip1.5mu\}\mathbin{;\;}\!\!\!\mathbin{;\;}{}\<[E]%
\ColumnHook
\end{hscode}\resethooks
We define the representation using \ensuremath{\Varid{int}}
and making sure that negative integers are not converted to \ensuremath{\Varid{nat}}:
\restorecolumns
\begin{hscode}\SaveRestoreHook
\column{B}{@{}>{\hspre}l<{\hspost}@{}}%
\column{5}{@{}>{\hspre}l<{\hspost}@{}}%
\column{26}{@{}>{\hspre}l<{\hspost}@{}}%
\column{39}{@{}>{\hspre}l<{\hspost}@{}}%
\column{E}{@{}>{\hspre}l<{\hspost}@{}}%
\>[5]{}\Keyword{let}\;\Varid{nat\char95 repr}\mathrel{=}\Conid{Repr}\;{}\<[26]%
\>[26]{}\{\mskip1.5mu \Varid{repr\char95 ty}{}\<[39]%
\>[39]{}\mathrel{=}\Conid{Int}{}\<[E]%
\\
\>[26]{}\mathbin{;\;}\Varid{to\char95 repr}{}\<[39]%
\>[39]{}\mathrel{=}(\Keyword{fun}\;\Varid{x}\to \Varid{x}){}\<[E]%
\\
\>[26]{}\mathbin{;\;}\Varid{from\char95 repr}{}\<[39]%
\>[39]{}\mathrel{=}(\Keyword{fun}\;\Varid{x}\to \Keyword{if}\;\Varid{x}\geq \mathrm{0}\;\Keyword{then}\;\Conid{Some}\;\Varid{x}\;\Keyword{else}\;\Conid{None})\mskip1.5mu\}{}\<[E]%
\ColumnHook
\end{hscode}\resethooks
The abstract view \ensuremath{\Varid{repr}} must be extended manually.
\restorecolumns
\begin{hscode}\SaveRestoreHook
\column{B}{@{}>{\hspre}l<{\hspost}@{}}%
\column{5}{@{}>{\hspre}l<{\hspost}@{}}%
\column{54}{@{}>{\hspre}l<{\hspost}@{}}%
\column{68}{@{}>{\hspre}l<{\hspost}@{}}%
\column{75}{@{}>{\hspre}l<{\hspost}@{}}%
\column{E}{@{}>{\hspre}l<{\hspost}@{}}%
\>[5]{}\Varid{\Conid{Repr}.ext}\;\Conid{Nat}\;\{\mskip1.5mu \Varid{f}\mathrel{=}\Keyword{fun}\;(\mathbf{type\;}\;\Varid{a})\;(\Varid{t}\mathrel{\;:\;}\Varid{a}\;\Varid{ty})\to ({}\<[54]%
\>[54]{}\Keyword{match}\;\Varid{t}\;\Keyword{with}{}\<[68]%
\>[68]{}\mid \Conid{Nat}{}\<[75]%
\>[75]{}\to \Varid{nat\char95 repr}{}\<[E]%
\\
\>[68]{}\mid \anonymous {}\<[75]%
\>[75]{}\to \Keyword{assert}\;\Keyword{false}\mathrel{\;:\;}\Varid{a}\;\Varid{repr})\mskip1.5mu\}{}\<[E]%
\\
\>[B]{}\Keyword{end}\;\mbox{\commentbegin  end of Nat module  \commentend}{}\<[E]%
\ColumnHook
\end{hscode}\resethooks
OCaml uses modular abstraction, and, since module signatures may hide type aliases, type equality depends on the context: in the example above, \ensuremath{\Varid{nat}} and \ensuremath{\Varid{int}} are the same type inside the \ensuremath{\Conid{Nat}} module, and are distinct types outside. Generic operations on such types use different representations for the same type: internally, \ensuremath{\Varid{nat}} and \ensuremath{\Varid{int}} are type synonyms, and are represented by the type witness of \ensuremath{\Varid{int}}. On the other hand, externally, we may need to design a specific public representation for \ensuremath{\Varid{nat}}.

\subsection{Syntax Extensions}
\label{sec:ppx}
The library is compatible with \ocaml{} version 4.04. In order to provide support for a user type \ensuremath{\Tyvarid{t}}, one should add a corresponding type witness \ensuremath{\Tyvarid{t}\;\Varid{ty}}, and add a corresponding case to the low level view \ensuremath{\Tyvarid{t}\;\Varid{desc}}, as well as \ensuremath{\Tyvarid{t}\;\Varid{ty}\;\Varid{desc}}, and also extend the type equality function \ensuremath{\Varid{ty\char95 equal}} (section~\ref{sec:coercion}). This involves a lot of boilerplate which can be fully automated by using extension points (PPX).

When the structure item attribute \ensuremath{[\!\mathbin{@\!\!\!@}\Varid{reify}]} is associated with a type declaration, a type witness obtained by capitalising the type name is defined and the low level view is extended. The generated code is placed right after the type declaration.

Alternatively a global (floating) attribute \ensuremath{[\!\mathbin{@\!\!\!@\!\!\!@}\Varid{reify}\mathbin{-}\Varid{all}]}, placed at the top of the file, ensures that every type declaration is reified, unless the type declaration is marked with \ensuremath{[\!\mathbin{@\!\!\!@}\Varid{dont\char95 reify}]}.

The attribute \ensuremath{[\!\mathbin{@\!\!\!@}\Varid{abstract}]} ensures that the low level view for that type is \ensuremath{\Conid{Abstract}} and hides the concrete structure of the type. However, the \ensuremath{\Varid{repr}} view must still be extended manually.

The attribute \ensuremath{[\!\mathbin{@\!\!\!@}\Varid{no\char95 desc}]} may be used in case the user wants to provide his own implementation of the low level view for the type but still wants the type witness to be generated and a new case for the type equality function.

\paragraph{Potential Compile-Time Errors}
Name conflicts may arise from the generated type witnesses, which are new data constructors extending the type \ensuremath{\Varid{ty}}: constructor names are obtained from type names by capitalising them.

The user should make sure that all the necessary types and type witnesses are in scope. The PPX does not open any module.
In particular, it is usually necessary to open \ensuremath{\Conid{\Conid{Generic}.\Conid{Core}.\Conid{Ty}.T}} which exports the witnesses for the built-in types (bool, char, int, int32, int64, nativeint, float, bytes, string, array, exn, ref, option, list, ty, unit, and tuples up to decuples).

When reifying a type, the witnesses for all the types that are mentioned in the definition should be in scope (fields of records, constructors of variants, definitions of synonyms).

\paragraph{Reifying GADTs}
Currently the low level view for GADTs must be written by hand. The view that is derived by default for variant datatypes does not work with GADTs, one must use the attribute \ensuremath{[\!\mathbin{@\!\!\!@}\Varid{no\char95 desc}]} to prevent the generation of the view, or \ensuremath{[\!\mathbin{@}\Varid{abstract}]} if one wants to make the type abstract.

\paragraph{Reifying Classes}
Currently, classes are reified as abstract datatypes, the class representation (low level view) must be written by hand. In that case, one must use the attribute \ensuremath{[\!\mathbin{@\!\!\!@}\Varid{no\char95 desc}]}.

\paragraph{Extensible Type-Indexed Functions}
There is currently no syntax support for creating and extending type-indexed functions.
Extending by hand the low level view \ensuremath{\Varid{desc}} with a case for \ensuremath{\Varid{btree}} requires the following boiler plate:
\pagebreak
\begin{hscode}\SaveRestoreHook
\column{B}{@{}>{\hspre}l<{\hspost}@{}}%
\column{24}{@{}>{\hspre}l<{\hspost}@{}}%
\column{30}{@{}>{\hspre}l<{\hspost}@{}}%
\column{45}{@{}>{\hspre}c<{\hspost}@{}}%
\column{45E}{@{}l@{}}%
\column{48}{@{}>{\hspre}l<{\hspost}@{}}%
\column{57}{@{}>{\hspre}l<{\hspost}@{}}%
\column{70}{@{}>{\hspre}c<{\hspost}@{}}%
\column{70E}{@{}l@{}}%
\column{73}{@{}>{\hspre}l<{\hspost}@{}}%
\column{86}{@{}>{\hspre}c<{\hspost}@{}}%
\column{86E}{@{}l@{}}%
\column{89}{@{}>{\hspre}l<{\hspost}@{}}%
\column{E}{@{}>{\hspre}l<{\hspost}@{}}%
\>[B]{}\Varid{ext}\;(\Conid{Btree}\;\Conid{Any})\;\{\mskip1.5mu \Varid{f}\mathrel{=}{}\<[24]%
\>[24]{}\Keyword{fun}\;(\mathbf{type\;}\;\Varid{a})\;(\Varid{ty}\mathrel{\;:\;}\Varid{a}\;\Varid{ty}){}\<[E]%
\\
\>[24]{}\to ({}\<[30]%
\>[30]{}\Keyword{match}\;\Varid{ty}\;\Keyword{with}{}\<[45]%
\>[45]{}\mid {}\<[45E]%
\>[48]{}\Conid{Btree}\;\Varid{a}{}\<[57]%
\>[57]{}\to \Conid{Variant}\;{}\<[70]%
\>[70]{}\{\mskip1.5mu {}\<[70E]%
\>[73]{}\Varid{name}{}\<[86]%
\>[86]{}\mathrel{=}{}\<[86E]%
\>[89]{}\text{\tt \char34 btree\char34}{}\<[E]%
\\
\>[70]{}\mathbin{;\;}{}\<[70E]%
\>[73]{}\Varid{module\char95 path}{}\<[86]%
\>[86]{}\mathrel{=}{}\<[86E]%
\>[89]{}[\!\text{\tt \char34 Example\char34}]{}\<[E]%
\\
\>[70]{}\mathbin{;\;}{}\<[70E]%
\>[73]{}\Varid{cons}{}\<[86]%
\>[86]{}\mathrel{=}{}\<[86E]%
\>[89]{}\Varid{cons\char95 btree}\;\Varid{a}\mskip1.5mu\}{}\<[E]%
\\
\>[45]{}\mid {}\<[45E]%
\>[48]{}\anonymous {}\<[57]%
\>[57]{}\to \Keyword{assert}\;\Keyword{false}\mathrel{\;:\;}\Varid{a}\;\Varid{\Conid{Desc}.t})\mskip1.5mu\}\mathbin{;\;}\!\!\!\mathbin{;\;}{}\<[E]%
\ColumnHook
\end{hscode}\resethooks
where \ensuremath{\Varid{cons\char95 btree}} was in section~\ref{prog:cons-btree}.

\section{Boilerplate-less Generic Traversals}
\label{sec:traversals}
The examples in this section are adapted from Mitchell and Runciman~\cite{mitchell:Uniplate}.
Consider a simple expression language with constants, negation, addition, subtraction, variables, and bindings.
\begin{hscode}\SaveRestoreHook
\column{B}{@{}>{\hspre}l<{\hspost}@{}}%
\column{12}{@{}>{\hspre}c<{\hspost}@{}}%
\column{12E}{@{}l@{}}%
\column{15}{@{}>{\hspre}l<{\hspost}@{}}%
\column{E}{@{}>{\hspre}l<{\hspost}@{}}%
\>[B]{}\mathbf{type\;}\;\Varid{expr}{}\<[12]%
\>[12]{}\mathrel{=}{}\<[12E]%
\>[15]{}\Conid{Cst}\;\Keyword{of}\;\Varid{int}\mid \Conid{Neg}\;\Keyword{of}\;\Varid{expr}\mid \Conid{Add}\;\Keyword{of}\;\Varid{expr}\mathbin{\times}\Varid{expr}\mid \Conid{Sub}\;\Keyword{of}\;\Varid{expr}\mathbin{\times}\Varid{expr}{}\<[E]%
\\
\>[12]{}\mid {}\<[12E]%
\>[15]{}\Conid{Var}\;\Keyword{of}\;\Varid{string}\mid \Conid{Let}\;\Keyword{of}\;\Varid{string}\mathbin{\times}\Varid{expr}\mathbin{\times}\Varid{expr}\;[\!\mathbin{@\!\!\!@}\Varid{reify}]{}\<[E]%
\ColumnHook
\end{hscode}\resethooks
Let us compute the list of all constants occurring in an expression:
\begin{hscode}\SaveRestoreHook
\column{B}{@{}>{\hspre}l<{\hspost}@{}}%
\column{22}{@{}>{\hspre}l<{\hspost}@{}}%
\column{39}{@{}>{\hspre}l<{\hspost}@{}}%
\column{E}{@{}>{\hspre}l<{\hspost}@{}}%
\>[B]{}\Keyword{let}\;\Keyword{rec}\;\Varid{constants}\mathrel{=}{}\<[22]%
\>[22]{}\Keyword{function}{}\<[E]%
\\
\>[22]{}\mid \Conid{Cst}\;\Varid{x}{}\<[39]%
\>[39]{}\to [\mskip1.5mu \Varid{x}\mskip1.5mu]{}\<[E]%
\\
\>[22]{}\mid \Conid{Neg}\;\Varid{x}{}\<[39]%
\>[39]{}\to \Varid{constants}\;\Varid{x}{}\<[E]%
\\
\>[22]{}\mid \Conid{Add}\;(\Varid{x},\,\Varid{y}){}\<[39]%
\>[39]{}\to \Varid{constants}\;\Varid{x}\mathbin{@}\Varid{constants}\;\Varid{y}{}\<[E]%
\\
\>[22]{}\mid \Conid{Sub}\;(\Varid{x},\,\Varid{y}){}\<[39]%
\>[39]{}\to \Varid{constants}\;\Varid{x}\mathbin{@}\Varid{constants}\;\Varid{y}{}\<[E]%
\\
\>[22]{}\mid \Conid{Var}\;\Varid{n}{}\<[39]%
\>[39]{}\to [\mskip1.5mu \mskip1.5mu]{}\<[E]%
\\
\>[22]{}\mid \Conid{Let}\;(\Varid{n},\,\Varid{x},\,\Varid{y}){}\<[39]%
\>[39]{}\to \Varid{constants}\;\Varid{x}\mathbin{@}\Varid{constants}\;\Varid{y}{}\<[E]%
\ColumnHook
\end{hscode}\resethooks
This definition has the three characteristics of a boilerplate problem:
(1) adding a constructor to the type would require adding a new case to the function,
(2) most cases are repetitive and systematic, only one case here---Cst---is really specific, and
(3) the code is tied to a particular operation and cannot be shared.

In real world compilers, AST have many more constructors and those issues become all the more frustrating.
Generic traversals are the answer.

With a generic function \ensuremath{\Varid{family}\mathrel{\;:\;}\Tyvarid{a}\;\Varid{ty}\to \Tyvarid{a}\to \Tyvarid{a}\;\Varid{list}} that returns the list of all the sub-expressions of an expression, the previous example may be written:
\begin{hscode}\SaveRestoreHook
\column{B}{@{}>{\hspre}l<{\hspost}@{}}%
\column{18}{@{}>{\hspre}c<{\hspost}@{}}%
\column{18E}{@{}l@{}}%
\column{21}{@{}>{\hspre}l<{\hspost}@{}}%
\column{31}{@{}>{\hspre}c<{\hspost}@{}}%
\column{31E}{@{}l@{}}%
\column{34}{@{}>{\hspre}l<{\hspost}@{}}%
\column{41}{@{}>{\hspre}l<{\hspost}@{}}%
\column{E}{@{}>{\hspre}l<{\hspost}@{}}%
\>[B]{}\Keyword{let}\;\Varid{is\char95 cst}{}\<[18]%
\>[18]{}\mathrel{=}{}\<[18E]%
\>[21]{}\Keyword{function}{}\<[31]%
\>[31]{}\mid {}\<[31E]%
\>[34]{}\Conid{Cst}\;\Varid{k}{}\<[41]%
\>[41]{}\to [\mskip1.5mu \Varid{k}\mskip1.5mu]{}\<[E]%
\\
\>[31]{}\mid {}\<[31E]%
\>[34]{}\anonymous {}\<[41]%
\>[41]{}\to [\mskip1.5mu \mskip1.5mu]{}\<[E]%
\\
\>[B]{}\Keyword{let}\;\Varid{constants}\;\Varid{e}{}\<[18]%
\>[18]{}\mathrel{=}{}\<[18E]%
\>[21]{}\Varid{\Conid{List}.concat}\;(\Varid{\Conid{List}.map}\;\Varid{is\char95 cst}\;(\Varid{family}\;\Conid{Expr}\;\Varid{e})){}\<[E]%
\ColumnHook
\end{hscode}\resethooks
Notice that (1) \ensuremath{\Varid{is\char95 cst}} only mentions the constructor \ensuremath{\Conid{Cst}}, therefore adding new constructors to the type would not break the behaviour of \ensuremath{\Varid{constants}}, (2) the repetitive cases have disappeared, (3) the traversal code is shared in the library function \ensuremath{\Varid{family}}.

A few libraries for Haskell offer a similar functionality, of which Uniplate and Multiplate were our main inspiration. The term ``Uniplate'' is a contraction of ``uniform boilerplate'' used by Mitchell and Runciman in \cite{mitchell:Uniplate} as the name of their library. The term ``Multiplate'' was introduced by O'Commor in \cite{Multiplate} as the name of his own library for generic programming on mutually recursive data types.

\subsection{Uniplate}
\label{sec:Uniplate}
Our whole Uniplate library\footnote{The combinators have been renamed for consistency with the Multiplate library, a correspondence is given in section~\ref{sec:comparison}.} relies on a single generic function \ensuremath{\Varid{scrap}} which may easily be implemented using the spine view, or the list-of-constructors view.
\begin{hscode}\SaveRestoreHook
\column{B}{@{}>{\hspre}l<{\hspost}@{}}%
\column{E}{@{}>{\hspre}l<{\hspost}@{}}%
\>[B]{}\Keyword{val}\;\Varid{scrap}\mathrel{\;:\;}\Tyvarid{a}\;\Varid{ty}\to \Tyvarid{a}\to \Tyvarid{a}\;\Varid{list}\mathbin{\times}(\Tyvarid{a}\;\Varid{list}\to \Tyvarid{a}){}\<[E]%
\ColumnHook
\end{hscode}\resethooks
\ensuremath{\Varid{scrap}\;\Varid{a}\;\Varid{v}} returns the list of children of a value of type \ensuremath{\Varid{a}} and a function to replace the children. By children, we mean the maximal substructures of the same type. For instance, the tail of a list is the only child of a non-empty list.

\subsubsection{Children, Descendents, Family}
From \ensuremath{\Varid{scrap}} we can of course derive \ensuremath{\Varid{children}} and \ensuremath{\Varid{replace\char95 children}} which are simply the first and second components of the result:

\begin{hscode}\SaveRestoreHook
\column{B}{@{}>{\hspre}l<{\hspost}@{}}%
\column{23}{@{}>{\hspre}l<{\hspost}@{}}%
\column{E}{@{}>{\hspre}l<{\hspost}@{}}%
\>[B]{}\Keyword{val}\;\Varid{children}{}\<[23]%
\>[23]{}\mathrel{\;:\;}\Tyvarid{a}\;\Varid{ty}\to \Tyvarid{a}\to \Tyvarid{a}\;\Varid{list}{}\<[E]%
\\
\>[B]{}\Keyword{val}\;\Varid{replace\char95 children}{}\<[23]%
\>[23]{}\mathrel{\;:\;}\Tyvarid{a}\;\Varid{ty}\to \Tyvarid{a}\to \Tyvarid{a}\;\Varid{list}\to \Tyvarid{a}{}\<[E]%
\ColumnHook
\end{hscode}\resethooks
Note that \ensuremath{\Varid{replace\char95 children}} expects a list of the same size as the one returned by \ensuremath{\Varid{children}}, that property is only checked at runtime. And \ensuremath{(\Varid{replace\char95 children}\;\Varid{t}\;\Varid{x}\;(\Varid{children}\;\Varid{t}\;\Varid{x}))} is the same  as \ensuremath{\Varid{x}}.

Let us define a \emph{descendent} of a value as either the value itself or a descendent of one of its children. The \emph{family} is the set of all the descendents of a value.
\begin{hscode}\SaveRestoreHook
\column{B}{@{}>{\hspre}l<{\hspost}@{}}%
\column{E}{@{}>{\hspre}l<{\hspost}@{}}%
\>[B]{}\Keyword{let}\;\Keyword{rec}\;\Varid{family}\;\Varid{a}\;\Varid{x}\mathrel{=}\Varid{x}\mathbin{::}\Varid{\Conid{List}.concat}\;(\Varid{\Conid{List}.map}\;(\Varid{family}\;\Varid{a})\;(\Varid{children}\;\Varid{a}\;\Varid{x})){}\<[E]%
\ColumnHook
\end{hscode}\resethooks
Most applications of \ensuremath{\Varid{family}} consist in filtering the descendents and extracting some information.

\subsubsection{Transformation and Queries}
A transformation is modifying a value and has some type \ensuremath{\Tyvarid{a}\to \Tyvarid{a}}, whereas a query is extracting some information: its type is \ensuremath{\Tyvarid{a}\to \Tyvarid{b}}.
The generic traversals in Uniplate are higher-order functions that take a transformation or a query to compute a more complex transformation or query.
For instance, one can define a non-recursive transformation to rename a variable and use the combinator \ensuremath{\Varid{map\char95 family}} to apply it recursively on an AST, with the effect of changing every variable of an expression.

\subsubsection{Paramorphisms}
A paramorphism is a bottom-up recursive function whose inductive step may also depend on the initial value in addition to the recursive results~\cite{Meertens1992}.
Accordingly, we express the inductive step of our \ensuremath{\Varid{para}} operator as a function with type \ensuremath{\Tyvarid{a}\to \Tyvarid{r}\;\Varid{list}\to \Tyvarid{r}}, that takes the initial value and the list of children's results. The \ensuremath{\Varid{para}} combinator then recursively applies the inductive step to compute a result for the whole expression.
\begin{hscode}\SaveRestoreHook
\column{B}{@{}>{\hspre}l<{\hspost}@{}}%
\column{10}{@{}>{\hspre}l<{\hspost}@{}}%
\column{16}{@{}>{\hspre}c<{\hspost}@{}}%
\column{16E}{@{}l@{}}%
\column{19}{@{}>{\hspre}l<{\hspost}@{}}%
\column{33}{@{}>{\hspre}l<{\hspost}@{}}%
\column{E}{@{}>{\hspre}l<{\hspost}@{}}%
\>[B]{}\Keyword{let}\;\Keyword{rec}\;{}\<[10]%
\>[10]{}\Varid{para}{}\<[16]%
\>[16]{}\mathrel{\;:\;}{}\<[16E]%
\>[19]{}\Tyvarid{a}\;\Varid{ty}\to (\Tyvarid{a}\to \Tyvarid{r}\;\Varid{list}\to \Tyvarid{r})\to \Tyvarid{a}\to \Tyvarid{r}{}\<[E]%
\\
\>[16]{}\mathrel{=}{}\<[16E]%
\>[19]{}\Keyword{fun}\;\Varid{a}\;\Varid{f}\;\Varid{x}\to {}\<[33]%
\>[33]{}\Varid{f}\;\Varid{x}\;(\Varid{\Conid{List}.map}\;(\Varid{para}\;\Varid{a}\;\Varid{f})\;(\Varid{children}\;\Varid{a}\;\Varid{x})){}\<[E]%
\ColumnHook
\end{hscode}\resethooks
As an example, the \ensuremath{\Varid{family}} function could be expressed as a paramorphism:
\begin{hscode}\SaveRestoreHook
\column{B}{@{}>{\hspre}l<{\hspost}@{}}%
\column{E}{@{}>{\hspre}l<{\hspost}@{}}%
\>[B]{}\Keyword{let}\;\Varid{family}\;\Varid{a}\mathrel{=}\Varid{para}\;\Varid{a}\mathbin{@\!\!\!@}\Keyword{fun}\;\Varid{x}\;\Varid{xs}\to \Varid{x}\mathbin{::}\Varid{\Conid{List}.concat}\;\Varid{xs}{}\<[E]%
\ColumnHook
\end{hscode}\resethooks
The \ensuremath{\Varid{height}} function from the introduction may also be computed directly as a paramorphism:
\begin{hscode}\SaveRestoreHook
\column{B}{@{}>{\hspre}l<{\hspost}@{}}%
\column{27}{@{}>{\hspre}l<{\hspost}@{}}%
\column{46}{@{}>{\hspre}l<{\hspost}@{}}%
\column{57}{@{}>{\hspre}l<{\hspost}@{}}%
\column{E}{@{}>{\hspre}l<{\hspost}@{}}%
\>[B]{}\Keyword{let}\;\Varid{height}\;\Varid{a}\mathrel{=}\Varid{para}\;\Varid{a}\mathbin{@\!\!\!@}{}\<[27]%
\>[27]{}\Keyword{fun}\;\anonymous \to \Keyword{function}{}\<[46]%
\>[46]{}\mid [\mskip1.5mu \mskip1.5mu]{}\<[57]%
\>[57]{}\to \mathrm{0}{}\<[E]%
\\
\>[46]{}\mid \Varid{h}\mathbin{::}\Varid{hs}{}\<[57]%
\>[57]{}\to \mathrm{1}\mathbin{+}\Varid{\Conid{List}.fold\char95 left}\;\Varid{max}\;\Varid{h}\;\Varid{hs}{}\<[E]%
\ColumnHook
\end{hscode}\resethooks
\subsubsection{Top-Down Transformations}
\ensuremath{\Varid{map\char95 children}} rewrites each child of the root using a given transformation.
\begin{hscode}\SaveRestoreHook
\column{B}{@{}>{\hspre}l<{\hspost}@{}}%
\column{19}{@{}>{\hspre}l<{\hspost}@{}}%
\column{22}{@{}>{\hspre}l<{\hspost}@{}}%
\column{35}{@{}>{\hspre}l<{\hspost}@{}}%
\column{E}{@{}>{\hspre}l<{\hspost}@{}}%
\>[B]{}\Keyword{let}\;\Varid{map\char95 children}{}\<[19]%
\>[19]{}\mathrel{\;:\;}{}\<[22]%
\>[22]{}\Tyvarid{a}\;\Varid{ty}\to (\Tyvarid{a}\to \Tyvarid{a})\to (\Tyvarid{a}\to \Tyvarid{a}){}\<[E]%
\\
\>[19]{}\mathrel{=}\Keyword{fun}\;\Varid{a}\;\Varid{f}\;\Varid{x}\to {}\<[35]%
\>[35]{}\Keyword{let}\;(\Varid{children},\,\Varid{replace})\mathrel{=}\Varid{scrap}\;\Varid{a}\;\Varid{x}{}\<[E]%
\\
\>[35]{}\Keyword{in}\;\Varid{replace}\;(\Varid{\Conid{List}.map}\;\Varid{f}\;\Varid{children}){}\<[E]%
\ColumnHook
\end{hscode}\resethooks
Let us define substitution for our expressions. First we define a finite map for our environments:
\begin{hscode}\SaveRestoreHook
\column{B}{@{}>{\hspre}l<{\hspost}@{}}%
\column{E}{@{}>{\hspre}l<{\hspost}@{}}%
\>[B]{}\Keyword{module}\;\Conid{Env}\mathrel{=}\Conid{\Conid{Map}.Make}\;(\Conid{String}){}\<[E]%
\\
\>[B]{}\mathbf{type\;}\;\Varid{env}\mathrel{=}\Varid{expr}\;\Varid{\Conid{Env}.t}{}\<[E]%
\ColumnHook
\end{hscode}\resethooks
Substitution is only really concerned with two cases, \ensuremath{\Conid{Let}} and \ensuremath{\Conid{Var}}, the rest of the cases involve recursing on the children (and rebuilding the term).
\begin{hscode}\SaveRestoreHook
\column{B}{@{}>{\hspre}l<{\hspost}@{}}%
\column{16}{@{}>{\hspre}c<{\hspost}@{}}%
\column{16E}{@{}l@{}}%
\column{19}{@{}>{\hspre}l<{\hspost}@{}}%
\column{43}{@{}>{\hspre}c<{\hspost}@{}}%
\column{43E}{@{}l@{}}%
\column{47}{@{}>{\hspre}l<{\hspost}@{}}%
\column{E}{@{}>{\hspre}l<{\hspost}@{}}%
\>[B]{}\Keyword{let}\;\Keyword{rec}\;\Varid{subst}{}\<[16]%
\>[16]{}\mathrel{\;:\;}{}\<[16E]%
\>[19]{}\Varid{env}\to \Varid{expr}\to \Varid{expr}{}\<[E]%
\\
\>[16]{}\mathrel{=}{}\<[16E]%
\>[19]{}\Keyword{fun}\;\Varid{env}\to \Keyword{let}\;\Keyword{open}\;\Conid{Env}\;\Keyword{in}\;\Keyword{function}{}\<[E]%
\\
\>[19]{}\mid \Conid{Let}\;(\Varid{n},\,\Varid{x},\,\Varid{y}){}\<[43]%
\>[43]{}\to {}\<[43E]%
\>[47]{}\Keyword{let}\;\Varid{env'}\mathrel{=}\Varid{remove}\;\Varid{n}\;\Varid{env}{}\<[E]%
\\
\>[47]{}\Keyword{in}\;\Conid{Let}\;(\Varid{n},\,\Varid{subst}\;\Varid{env}\;\Varid{x},\,\Varid{subst}\;\Varid{env'}\;\Varid{y}){}\<[E]%
\\
\>[19]{}\mid \Conid{Var}\;\Varid{n}\;\Keyword{when}\;\Varid{mem}\;\Varid{n}\;\Varid{env}{}\<[43]%
\>[43]{}\to {}\<[43E]%
\>[47]{}\Varid{find}\;\Varid{n}\;\Varid{env}{}\<[E]%
\\
\>[19]{}\mid \Varid{x}{}\<[43]%
\>[43]{}\to {}\<[43E]%
\>[47]{}\Varid{map\char95 children}\;\Conid{Expr}\;(\Varid{subst}\;\Varid{env})\;\Varid{x}{}\<[E]%
\ColumnHook
\end{hscode}\resethooks
\subsubsection{Bottom-Up Transformations}\mbox{}
\ensuremath{\Varid{map\char95 family}} recursively applies a transformation in a bottom-up traversal.
\begin{hscode}\SaveRestoreHook
\column{B}{@{}>{\hspre}l<{\hspost}@{}}%
\column{E}{@{}>{\hspre}l<{\hspost}@{}}%
\>[B]{}\Keyword{val}\;\Varid{map\char95 family}\mathrel{\;:\;}\Tyvarid{a}\;\Varid{ty}\to (\Tyvarid{a}\to \Tyvarid{a})\to (\Tyvarid{a}\to \Tyvarid{a}){}\<[E]%
\ColumnHook
\end{hscode}\resethooks
For instance, on a list \ensuremath{[\mskip1.5mu \Varid{x}\mathbin{;\;}\Varid{y}\mathbin{;\;}\Varid{z}\mskip1.5mu]} the transformation is applied along the spine of the list.
\begin{hscode}\SaveRestoreHook
\column{B}{@{}>{\hspre}l<{\hspost}@{}}%
\column{45}{@{}>{\hspre}l<{\hspost}@{}}%
\column{E}{@{}>{\hspre}l<{\hspost}@{}}%
\>[B]{}\Varid{map\char95 family}\;(\Conid{List}\;\Varid{a})\;\Varid{f}\;[\mskip1.5mu \Varid{x}\mathbin{;\;}\Varid{y}\mathbin{;\;}\Varid{z}\mskip1.5mu]\equiv \Varid{f}\;(\Varid{x}\mathbin{::}{}\<[45]%
\>[45]{}\Varid{f}\;(\Varid{y}\mathbin{::}\Varid{f}\;(\Varid{z}\mathbin{::}\Varid{f}\;[\mskip1.5mu \mskip1.5mu]))){}\<[E]%
\ColumnHook
\end{hscode}\resethooks
The families of the children are transformed before the value itself:
\begin{hscode}\SaveRestoreHook
\column{B}{@{}>{\hspre}l<{\hspost}@{}}%
\column{E}{@{}>{\hspre}l<{\hspost}@{}}%
\>[B]{}\Keyword{let}\;\Keyword{rec}\;\Varid{map\char95 family}\;\Varid{a}\;\Varid{f}\;\Varid{x}\mathrel{=}\Varid{f}\;(\Varid{map\char95 children}\;\Varid{a}\;(\Varid{map\char95 family}\;\Varid{a}\;\Varid{f})\;\Varid{x}){}\<[E]%
\ColumnHook
\end{hscode}\resethooks
For instance, we may remove double negations. The one-step transformation is applied bottom-up, ensuring that all double negations are removed.
\begin{hscode}\SaveRestoreHook
\column{B}{@{}>{\hspre}l<{\hspost}@{}}%
\column{36}{@{}>{\hspre}l<{\hspost}@{}}%
\column{46}{@{}>{\hspre}l<{\hspost}@{}}%
\column{61}{@{}>{\hspre}l<{\hspost}@{}}%
\column{E}{@{}>{\hspre}l<{\hspost}@{}}%
\>[B]{}\Keyword{let}\;\Varid{simplify}\mathrel{=}\Varid{map\char95 family}\;\Conid{Expr}\mathbin{@\!\!\!@}{}\<[36]%
\>[36]{}\Keyword{function}{}\<[46]%
\>[46]{}\mid \Conid{Neg}\;(\Conid{Neg}\;\Varid{x}){}\<[61]%
\>[61]{}\to \Varid{x}{}\<[E]%
\\
\>[46]{}\mid \Varid{x}{}\<[61]%
\>[61]{}\to \Varid{x}{}\<[E]%
\ColumnHook
\end{hscode}\resethooks
We may implement constant folding, i.e. evaluate the sub-expressions involving only constants:
\begin{hscode}\SaveRestoreHook
\column{B}{@{}>{\hspre}l<{\hspost}@{}}%
\column{47}{@{}>{\hspre}l<{\hspost}@{}}%
\column{69}{@{}>{\hspre}l<{\hspost}@{}}%
\column{E}{@{}>{\hspre}l<{\hspost}@{}}%
\>[B]{}\Keyword{let}\;\Varid{const\char95 fold}\mathrel{=}\Varid{map\char95 family}\;\Conid{Expr}\mathbin{@\!\!\!@}\Keyword{function}{}\<[47]%
\>[47]{}\mid \Conid{Add}\;(\Conid{Cst}\;\Varid{x},\,\Conid{Cst}\;\Varid{y}){}\<[69]%
\>[69]{}\to \Conid{Cst}\;(\Varid{x}\mathbin{+}\Varid{y}){}\<[E]%
\\
\>[47]{}\mid \Conid{Sub}\;(\Conid{Cst}\;\Varid{x},\,\Conid{Cst}\;\Varid{y}){}\<[69]%
\>[69]{}\to \Conid{Cst}\;(\Varid{x}\mathbin{-}\Varid{y}){}\<[E]%
\\
\>[47]{}\mid \Conid{Neg}\;(\Conid{Cst}\;\Varid{x}){}\<[69]%
\>[69]{}\to \Conid{Cst}\;(\mathbin{-}\Varid{x}){}\<[E]%
\\
\>[47]{}\mid \Varid{x}{}\<[69]%
\>[69]{}\to \Varid{x}{}\<[E]%
\ColumnHook
\end{hscode}\resethooks
\subsubsection{Normal Forms}
In some cases,  we want to apply a rewriting rule exhaustively until a normal form is reached. The rewriting rule is given as a function of type \ensuremath{\Tyvarid{a}\to \Tyvarid{a}\;\Varid{option}} which returns \ensuremath{\Conid{None}} when its argument is in normal form and otherwise performs one reduction step.
\begin{hscode}\SaveRestoreHook
\column{B}{@{}>{\hspre}l<{\hspost}@{}}%
\column{E}{@{}>{\hspre}l<{\hspost}@{}}%
\>[B]{}\Keyword{val}\;\Varid{reduce\char95 family}\mathrel{\;:\;}\Tyvarid{a}\;\Varid{ty}\to (\Tyvarid{a}\to \Tyvarid{a}\;\Varid{option})\to \Tyvarid{a}\to \Tyvarid{a}{}\<[E]%
\ColumnHook
\end{hscode}\resethooks
\ensuremath{\Varid{reduce\char95 family}} applies the rewriting rule until it returns \ensuremath{\Conid{None}} for all the descendents of the result.
\begin{hscode}\SaveRestoreHook
\column{B}{@{}>{\hspre}l<{\hspost}@{}}%
\column{32}{@{}>{\hspre}l<{\hspost}@{}}%
\column{62}{@{}>{\hspre}l<{\hspost}@{}}%
\column{72}{@{}>{\hspre}l<{\hspost}@{}}%
\column{E}{@{}>{\hspre}l<{\hspost}@{}}%
\>[B]{}\Keyword{let}\;\Keyword{rec}\;\Varid{reduce\char95 family}\;\Varid{a}\;\Varid{f}\;\Varid{x}\mathrel{=}{}\<[32]%
\>[32]{}\Keyword{let}\;\Keyword{rec}\;\Varid{g}\;\Varid{x}\mathrel{=}\Keyword{match}\;\Varid{f}\;\Varid{x}\;\Keyword{with}{}\<[62]%
\>[62]{}\mid \Conid{None}{}\<[72]%
\>[72]{}\to \Varid{x}{}\<[E]%
\\
\>[62]{}\mid \Conid{Some}\;\Varid{y}{}\<[72]%
\>[72]{}\to \Varid{map\char95 family}\;\Varid{a}\;\Varid{g}\;\Varid{y}{}\<[E]%
\\
\>[32]{}\Keyword{in}\;\Varid{map\char95 family}\;\Varid{a}\;\Varid{g}\;\Varid{x}{}\<[E]%
\ColumnHook
\end{hscode}\resethooks
We may extend our previous example with another rewrite rule to remove the use of subtraction from our expressions:
\begin{hscode}\SaveRestoreHook
\column{B}{@{}>{\hspre}l<{\hspost}@{}}%
\column{53}{@{}>{\hspre}l<{\hspost}@{}}%
\column{68}{@{}>{\hspre}l<{\hspost}@{}}%
\column{E}{@{}>{\hspre}l<{\hspost}@{}}%
\>[B]{}\Keyword{let}\;\Varid{simplify\char95 more}\mathrel{=}\Varid{reduce\char95 family}\;\Conid{Expr}\mathbin{@\!\!\!@}\Keyword{function}{}\<[53]%
\>[53]{}\mid \Conid{Neg}\;(\Conid{Neg}\;\Varid{x}){}\<[68]%
\>[68]{}\to \Conid{Some}\;\Varid{x}{}\<[E]%
\\
\>[53]{}\mid \Conid{Sub}\;(\Varid{x},\,\Varid{y}){}\<[68]%
\>[68]{}\to \Conid{Some}\;(\Conid{Add}\;(\Varid{x},\,\Conid{Neg}\;\Varid{y})){}\<[E]%
\\
\>[53]{}\mid \anonymous {}\<[68]%
\>[68]{}\to \Conid{None}{}\<[E]%
\ColumnHook
\end{hscode}\resethooks
The rewrite rule for \ensuremath{\Conid{Sub}} introduces a \ensuremath{\Conid{Neg}} constructor, which is itself on the left hand side of a rewrite rule, this may create a new rewriting opportunity: for instance \ensuremath{\Conid{Sub}\;\Varid{x}\;(\Conid{Neg}\;\Varid{y})} rewrites to \ensuremath{\Conid{Add}\;(\Varid{x},\,\Conid{Neg}\;(\Conid{Neg}\;\Varid{y}))} which in turns rewrites to \ensuremath{\Conid{Add}\;(\Varid{x},\,\Varid{y})}.
Using \ensuremath{\Varid{reduce\char95 family}} ensures that no rewriting opportunity is missed.

\subsubsection{Effectful Transformations}
Finally, all the traversals combinators have an effectful counterpart that threads the effects of an effectful transformation. We used the encoding of higher-kinded polymorphism explained in section~\ref{app} to parametrise the functions over applicative functors and monads.
\begin{hscode}\SaveRestoreHook
\column{B}{@{}>{\hspre}l<{\hspost}@{}}%
\column{24}{@{}>{\hspre}l<{\hspost}@{}}%
\column{42}{@{}>{\hspre}l<{\hspost}@{}}%
\column{74}{@{}>{\hspre}l<{\hspost}@{}}%
\column{E}{@{}>{\hspre}l<{\hspost}@{}}%
\>[B]{}\Keyword{val}\;\Varid{traverse\char95 children}{}\<[24]%
\>[24]{}\mathrel{\;:\;}\Tyvarid{f}\;\Varid{applicative}{}\<[42]%
\>[42]{}\to \Tyvarid{a}\;\Varid{ty}\to (\Tyvarid{a}\to (\Tyvarid{a},\,{}\<[74]%
\>[74]{}\Tyvarid{f})\;\Varid{app})\to (\Tyvarid{a}\to (\Tyvarid{a},\,\Tyvarid{f})\;\Varid{app}){}\<[E]%
\\
\>[B]{}\Keyword{val}\;\Varid{traverse\char95 family}{}\<[24]%
\>[24]{}\mathrel{\;:\;}\Tyvarid{f}\;\Varid{monad}{}\<[42]%
\>[42]{}\to \Tyvarid{a}\;\Varid{ty}\to (\Tyvarid{a}\to (\Tyvarid{a},\,{}\<[74]%
\>[74]{}\Tyvarid{f})\;\Varid{app})\to (\Tyvarid{a}\to (\Tyvarid{a},\,\Tyvarid{f})\;\Varid{app}){}\<[E]%
\\
\>[B]{}\Keyword{val}\;\Varid{mreduce\char95 family}{}\<[24]%
\>[24]{}\mathrel{\;:\;}\Tyvarid{f}\;\Varid{monad}{}\<[42]%
\>[42]{}\to \Tyvarid{a}\;\Varid{ty}\to (\Tyvarid{a}\to (\Tyvarid{a}\;\Varid{option},\,{}\<[74]%
\>[74]{}\Tyvarid{f})\;\Varid{app})\to (\Tyvarid{a}\to (\Tyvarid{a},\,\Tyvarid{f})\;\Varid{app}){}\<[E]%
\ColumnHook
\end{hscode}\resethooks
We will show later how one could use \ensuremath{\Varid{traverse\char95 family}} with a state monad to rename each variable to be unique. Beforehand, we must introduce some definitions.

\paragraph{Functors, Applicative, Monad, Monoid}
Using our encoding of higher-kinded polymorphism, we define the operations of functorial, applicative and monadic types. We assume the reader knows about those operations, they have been extensively discussed in the literature~\cite{Wadler92theessence, mcbride:applicative}
\begin{hscode}\SaveRestoreHook
\column{B}{@{}>{\hspre}l<{\hspost}@{}}%
\column{22}{@{}>{\hspre}c<{\hspost}@{}}%
\column{22E}{@{}l@{}}%
\column{25}{@{}>{\hspre}c<{\hspost}@{}}%
\column{25E}{@{}l@{}}%
\column{28}{@{}>{\hspre}l<{\hspost}@{}}%
\column{36}{@{}>{\hspre}l<{\hspost}@{}}%
\column{E}{@{}>{\hspre}l<{\hspost}@{}}%
\>[B]{}\mathbf{type\;}\;\Tyvarid{f}\;\Varid{functorial}{}\<[22]%
\>[22]{}\mathrel{=}{}\<[22E]%
\>[25]{}\{\mskip1.5mu {}\<[25E]%
\>[28]{}\Varid{fmap}{}\<[36]%
\>[36]{}\mathrel{\;:\;}\forall\!\;\Tyvarid{a}\;\Tyvarid{b}\mathbin{°.°}(\Tyvarid{a}\to \Tyvarid{b})\to (\Tyvarid{a},\,\Tyvarid{f})\;\Varid{app}\to (\Tyvarid{b},\,\Tyvarid{f})\;\Varid{app}\mskip1.5mu\}{}\<[E]%
\\
\>[B]{}\mathbf{type\;}\;\Tyvarid{f}\;\Varid{applicative}{}\<[22]%
\>[22]{}\mathrel{=}{}\<[22E]%
\>[25]{}\{\mskip1.5mu {}\<[25E]%
\>[28]{}\Varid{pure}{}\<[36]%
\>[36]{}\mathrel{\;:\;}\forall\!\;\Tyvarid{a}\mathbin{°.°}\Tyvarid{a}\to (\Tyvarid{a},\,\Tyvarid{f})\;\Varid{app}{}\<[E]%
\\
\>[25]{}\mathbin{;\;}{}\<[25E]%
\>[28]{}\Varid{apply}{}\<[36]%
\>[36]{}\mathrel{\;:\;}\forall\!\;\Tyvarid{a}\;\Tyvarid{b}\mathbin{°.°}(\Tyvarid{a}\to \Tyvarid{b},\,\Tyvarid{f})\;\Varid{app}\to (\Tyvarid{a},\,\Tyvarid{f})\;\Varid{app}\to (\Tyvarid{b},\,\Tyvarid{f})\;\Varid{app}\mskip1.5mu\}{}\<[E]%
\\
\>[B]{}\mathbf{type\;}\;\Tyvarid{f}\;\Varid{monad}{}\<[22]%
\>[22]{}\mathrel{=}{}\<[22E]%
\>[25]{}\{\mskip1.5mu {}\<[25E]%
\>[28]{}\Varid{return}{}\<[36]%
\>[36]{}\mathrel{\;:\;}\forall\!\;\Tyvarid{a}\mathbin{°.°}\Tyvarid{a}\to (\Tyvarid{a},\,\Tyvarid{f})\;\Varid{app}{}\<[E]%
\\
\>[25]{}\mathbin{;\;}{}\<[25E]%
\>[28]{}\Varid{bind}{}\<[36]%
\>[36]{}\mathrel{\;:\;}\forall\!\;\Tyvarid{a}\;\Tyvarid{b}\mathbin{°.°}(\Tyvarid{a},\,\Tyvarid{f})\;\Varid{app}\to (\Tyvarid{a}\to (\Tyvarid{b},\,\Tyvarid{f})\;\Varid{app})\to (\Tyvarid{b},\,\Tyvarid{f})\;\Varid{app}\mskip1.5mu\}{}\<[E]%
\ColumnHook
\end{hscode}\resethooks
Applicative functors and monads are functors:
\savecolumns
\begin{hscode}\SaveRestoreHook
\column{B}{@{}>{\hspre}l<{\hspost}@{}}%
\column{6}{@{}>{\hspre}l<{\hspost}@{}}%
\column{18}{@{}>{\hspre}l<{\hspost}@{}}%
\column{21}{@{}>{\hspre}c<{\hspost}@{}}%
\column{21E}{@{}l@{}}%
\column{24}{@{}>{\hspre}c<{\hspost}@{}}%
\column{24E}{@{}l@{}}%
\column{27}{@{}>{\hspre}l<{\hspost}@{}}%
\column{34}{@{}>{\hspre}c<{\hspost}@{}}%
\column{34E}{@{}l@{}}%
\column{37}{@{}>{\hspre}l<{\hspost}@{}}%
\column{47}{@{}>{\hspre}l<{\hspost}@{}}%
\column{E}{@{}>{\hspre}l<{\hspost}@{}}%
\>[B]{}\Keyword{let}\;{}\<[6]%
\>[6]{}\Varid{fun\char95 of\char95 app}\;{}\<[18]%
\>[18]{}\Varid{a}{}\<[21]%
\>[21]{}\mathrel{=}{}\<[21E]%
\>[24]{}\{\mskip1.5mu {}\<[24E]%
\>[27]{}\Varid{fmap}{}\<[34]%
\>[34]{}\mathrel{=}{}\<[34E]%
\>[37]{}\Keyword{fun}\;\Varid{f}{}\<[47]%
\>[47]{}\to \Varid{a}.\Varid{apply}\;(\Varid{a}.\Varid{pure}\;\Varid{f})\mskip1.5mu\}{}\<[E]%
\\
\>[B]{}\Keyword{let}\;{}\<[6]%
\>[6]{}\Varid{fun\char95 of\char95 mon}\;{}\<[18]%
\>[18]{}\Varid{m}{}\<[21]%
\>[21]{}\mathrel{=}{}\<[21E]%
\>[24]{}\{\mskip1.5mu {}\<[24E]%
\>[27]{}\Varid{fmap}{}\<[34]%
\>[34]{}\mathrel{=}{}\<[34E]%
\>[37]{}\Keyword{fun}\;\Varid{f}\;\Varid{kx}{}\<[47]%
\>[47]{}\to \Varid{m}.\Varid{bind}\;\Varid{kx}\;(\Keyword{fun}\;\Varid{x}\to \Varid{m}.\Varid{return}\;(\Varid{f}\;\Varid{x}))\mskip1.5mu\}{}\<[E]%
\ColumnHook
\end{hscode}\resethooks
Monads are applicative functors:
\restorecolumns
\begin{hscode}\SaveRestoreHook
\column{B}{@{}>{\hspre}l<{\hspost}@{}}%
\column{6}{@{}>{\hspre}l<{\hspost}@{}}%
\column{18}{@{}>{\hspre}l<{\hspost}@{}}%
\column{21}{@{}>{\hspre}c<{\hspost}@{}}%
\column{21E}{@{}l@{}}%
\column{24}{@{}>{\hspre}c<{\hspost}@{}}%
\column{24E}{@{}l@{}}%
\column{27}{@{}>{\hspre}l<{\hspost}@{}}%
\column{34}{@{}>{\hspre}c<{\hspost}@{}}%
\column{34E}{@{}l@{}}%
\column{37}{@{}>{\hspre}l<{\hspost}@{}}%
\column{E}{@{}>{\hspre}l<{\hspost}@{}}%
\>[B]{}\Keyword{let}\;{}\<[6]%
\>[6]{}\Varid{app\char95 of\char95 mon}\;{}\<[18]%
\>[18]{}\Varid{m}{}\<[21]%
\>[21]{}\mathrel{=}{}\<[21E]%
\>[24]{}\{\mskip1.5mu {}\<[24E]%
\>[27]{}\Varid{pure}{}\<[34]%
\>[34]{}\mathrel{=}{}\<[34E]%
\>[37]{}\Varid{m}.\Varid{return}{}\<[E]%
\\
\>[24]{}\mathbin{;\;}{}\<[24E]%
\>[27]{}\Varid{apply}{}\<[34]%
\>[34]{}\mathrel{=}{}\<[34E]%
\>[37]{}\Keyword{fun}\;\Varid{kf}\;\Varid{kx}\to \Varid{m}.\Varid{bind}\;\Varid{kf}\;(\Keyword{fun}\;\Varid{f}\to (\Varid{fun\char95 of\char95 mon}\;\Varid{m}).\Varid{fmap}\;\Varid{f}\;\Varid{kx})\mskip1.5mu\}{}\<[E]%
\ColumnHook
\end{hscode}\resethooks
Pure functions may be lifted to an applicative functor or a monad, with functions \ensuremath{\Varid{liftA}},..\ensuremath{\Varid{liftA4}}, and \ensuremath{\Varid{liftM}} variants. For instance, \ensuremath{\Varid{liftA2}} lifts a binary function:
\begin{hscode}\SaveRestoreHook
\column{B}{@{}>{\hspre}l<{\hspost}@{}}%
\column{E}{@{}>{\hspre}l<{\hspost}@{}}%
\>[B]{}\Keyword{val}\;\Varid{liftA2}\mathrel{\;:\;}\Tyvarid{f}\;\Varid{applicative}\to (\Tyvarid{a}\to \Tyvarid{b}\to \Tyvarid{c})\to (\Tyvarid{a},\,\Tyvarid{f})\;\Varid{app}\to (\Tyvarid{b},\,\Tyvarid{f})\;\Varid{app}\to (\Tyvarid{c},\,\Tyvarid{f})\;\Varid{app}{}\<[E]%
\ColumnHook
\end{hscode}\resethooks
We may traverse a list, executing an effectful function on each element.
\label{prog:traverse-list}
\begin{hscode}\SaveRestoreHook
\column{B}{@{}>{\hspre}l<{\hspost}@{}}%
\column{19}{@{}>{\hspre}c<{\hspost}@{}}%
\column{19E}{@{}l@{}}%
\column{22}{@{}>{\hspre}l<{\hspost}@{}}%
\column{43}{@{}>{\hspre}l<{\hspost}@{}}%
\column{53}{@{}>{\hspre}l<{\hspost}@{}}%
\column{E}{@{}>{\hspre}l<{\hspost}@{}}%
\>[B]{}\Keyword{let}\;\Keyword{rec}\;\Varid{traverse}{}\<[19]%
\>[19]{}\mathrel{\;:\;}{}\<[19E]%
\>[22]{}\Tyvarid{f}\;\Varid{applicative}\to (\Tyvarid{a}\to (\Tyvarid{b},\,\Tyvarid{f})\;\Varid{app})\to \Tyvarid{a}\;\Varid{list}\to (\Tyvarid{b}\;\Varid{list},\,\Tyvarid{f})\;\Varid{app}{}\<[E]%
\\
\>[19]{}\mathrel{=}{}\<[19E]%
\>[22]{}\Keyword{fun}\;\Varid{a}\;\Varid{f}\to \Keyword{function}{}\<[43]%
\>[43]{}\mid [\mskip1.5mu \mskip1.5mu]{}\<[53]%
\>[53]{}\to \Varid{a}.\Varid{pure}\;[\mskip1.5mu \mskip1.5mu]{}\<[E]%
\\
\>[43]{}\mid \Varid{h}\mathbin{::}\Varid{t}{}\<[53]%
\>[53]{}\to \Varid{liftA2}\;\Varid{a}\;\Varid{cons}\;(\Varid{f}\;\Varid{h})\;(\Varid{traverse}\;\Varid{a}\;\Varid{f}\;\Varid{t}){}\<[E]%
\ColumnHook
\end{hscode}\resethooks
A specific case of traversing is when the list contains effectful elements.
We may sequence the effects of the element and obtain an effectful list of pure elements. The list functor and the applicative functor commute.
\begin{hscode}\SaveRestoreHook
\column{B}{@{}>{\hspre}l<{\hspost}@{}}%
\column{6}{@{}>{\hspre}l<{\hspost}@{}}%
\column{16}{@{}>{\hspre}c<{\hspost}@{}}%
\column{16E}{@{}l@{}}%
\column{19}{@{}>{\hspre}l<{\hspost}@{}}%
\column{26}{@{}>{\hspre}c<{\hspost}@{}}%
\column{26E}{@{}l@{}}%
\column{30}{@{}>{\hspre}l<{\hspost}@{}}%
\column{E}{@{}>{\hspre}l<{\hspost}@{}}%
\>[B]{}\Keyword{let}\;{}\<[6]%
\>[6]{}\Varid{sequence}{}\<[16]%
\>[16]{}\mathrel{\;:\;}{}\<[16E]%
\>[19]{}\Tyvarid{f}\;\Varid{applicative}\to (\Tyvarid{a},\,\Tyvarid{f})\;\Varid{app}\;\Varid{list}\to (\Tyvarid{a}\;\Varid{list},\,\Tyvarid{f})\;\Varid{app}{}\<[E]%
\\
\>[16]{}\mathrel{=}{}\<[16E]%
\>[19]{}\Keyword{fun}\;\Varid{a}{}\<[26]%
\>[26]{}\to {}\<[26E]%
\>[30]{}\Varid{traverse}\;\Varid{a}\;(\Keyword{fun}\;\Varid{x}\to \Varid{x}){}\<[E]%
\ColumnHook
\end{hscode}\resethooks
We may derive monadic versions of \ensuremath{\Varid{traverse}} and \ensuremath{\Varid{sequence}}:
\begin{hscode}\SaveRestoreHook
\column{B}{@{}>{\hspre}l<{\hspost}@{}}%
\column{6}{@{}>{\hspre}l<{\hspost}@{}}%
\column{17}{@{}>{\hspre}l<{\hspost}@{}}%
\column{20}{@{}>{\hspre}c<{\hspost}@{}}%
\column{20E}{@{}l@{}}%
\column{23}{@{}>{\hspre}l<{\hspost}@{}}%
\column{33}{@{}>{\hspre}l<{\hspost}@{}}%
\column{E}{@{}>{\hspre}l<{\hspost}@{}}%
\>[B]{}\Keyword{let}\;{}\<[6]%
\>[6]{}\Varid{traverseM}\;{}\<[17]%
\>[17]{}\Varid{m}{}\<[20]%
\>[20]{}\mathrel{=}{}\<[20E]%
\>[23]{}\Varid{traverse}\;{}\<[33]%
\>[33]{}(\Varid{app\char95 of\char95 mon}\;\Varid{m}){}\<[E]%
\\
\>[B]{}\Keyword{let}\;{}\<[6]%
\>[6]{}\Varid{sequenceM}\;{}\<[17]%
\>[17]{}\Varid{m}{}\<[20]%
\>[20]{}\mathrel{=}{}\<[20E]%
\>[23]{}\Varid{sequence}\;{}\<[33]%
\>[33]{}(\Varid{app\char95 of\char95 mon}\;\Varid{m}){}\<[E]%
\ColumnHook
\end{hscode}\resethooks
\paragraph{Reader Monad}
The reader monad is parametrised by the type of an environment \ensuremath{\Tyvarid{b}} which may be read as a side effect of a monadic computation.
A value in the reader monad is a function \ensuremath{\Tyvarid{b}\to \Tyvarid{a}} from the environment to a result.
\begin{hscode}\SaveRestoreHook
\column{B}{@{}>{\hspre}l<{\hspost}@{}}%
\column{7}{@{}>{\hspre}l<{\hspost}@{}}%
\column{19}{@{}>{\hspre}c<{\hspost}@{}}%
\column{19E}{@{}l@{}}%
\column{23}{@{}>{\hspre}l<{\hspost}@{}}%
\column{33}{@{}>{\hspre}l<{\hspost}@{}}%
\column{45}{@{}>{\hspre}l<{\hspost}@{}}%
\column{E}{@{}>{\hspre}l<{\hspost}@{}}%
\>[B]{}\mathbf{type\;}\;{}\<[7]%
\>[7]{}\Tyvarid{b}\;\Varid{reader}{}\<[19]%
\>[19]{}\mathrel{=}{}\<[19E]%
\>[23]{}\Conid{READER}{}\<[E]%
\\
\>[B]{}\mathbf{type\;}\;{}\<[7]%
\>[7]{}(\anonymous ,\,\anonymous )\;\Varid{app}{}\<[19]%
\>[19]{}\mathrel{+\!\!\!=}{}\<[19E]%
\>[23]{}\Conid{Reader}\mathrel{\;:\;}(\Tyvarid{b}\to \Tyvarid{a})\to (\Tyvarid{a},\,\Tyvarid{b}\;\Varid{reader})\;\Varid{app}{}\<[E]%
\\
\>[B]{}\Keyword{let}\;{}\<[7]%
\>[7]{}\Varid{run\char95 reader}{}\<[19]%
\>[19]{}\mathrel{=}{}\<[19E]%
\>[23]{}\Keyword{function}{}\<[33]%
\>[33]{}\mid \Conid{Reader}\;\Varid{f}{}\<[45]%
\>[45]{}\to \Varid{f}{}\<[E]%
\\
\>[33]{}\mid \anonymous {}\<[45]%
\>[45]{}\to \Keyword{assert}\;\Keyword{false}{}\<[E]%
\ColumnHook
\end{hscode}\resethooks
\ensuremath{\Varid{return}} brings a pure value into the reader monad, the environment is ignored.
\ensuremath{\Varid{bind}\;\Varid{x}\;\Varid{f}} passes the environment to both \ensuremath{\Varid{x}} and \ensuremath{\Varid{f}}.
\begin{hscode}\SaveRestoreHook
\column{B}{@{}>{\hspre}l<{\hspost}@{}}%
\column{15}{@{}>{\hspre}c<{\hspost}@{}}%
\column{15E}{@{}l@{}}%
\column{18}{@{}>{\hspre}l<{\hspost}@{}}%
\column{26}{@{}>{\hspre}l<{\hspost}@{}}%
\column{38}{@{}>{\hspre}l<{\hspost}@{}}%
\column{58}{@{}>{\hspre}l<{\hspost}@{}}%
\column{62}{@{}>{\hspre}l<{\hspost}@{}}%
\column{E}{@{}>{\hspre}l<{\hspost}@{}}%
\>[B]{}\Keyword{let}\;\Varid{reader}\mathrel{=}{}\<[15]%
\>[15]{}\{\mskip1.5mu {}\<[15E]%
\>[18]{}\Varid{return}{}\<[26]%
\>[26]{}\mathrel{=}(\Keyword{fun}\;\Varid{x}{}\<[38]%
\>[38]{}\to \Conid{Reader}\;(\Keyword{fun}\;\Varid{env}{}\<[58]%
\>[58]{}\to \Varid{x})){}\<[E]%
\\
\>[15]{}\mathbin{;\;}{}\<[15E]%
\>[18]{}\Varid{bind}{}\<[26]%
\>[26]{}\mathrel{=}(\Keyword{fun}\;\Varid{x}\;\Varid{f}{}\<[38]%
\>[38]{}\to \Conid{Reader}\;(\Keyword{fun}\;\Varid{env}{}\<[58]%
\>[58]{}\to {}\<[62]%
\>[62]{}\Keyword{let}\;\Varid{y}\mathrel{=}\Varid{run\char95 reader}\;\Varid{x}\;\Varid{env}{}\<[E]%
\\
\>[62]{}\Keyword{in}\;\Varid{run\char95 reader}\;(\Varid{f}\;\Varid{y})\;\Varid{env}))\mskip1.5mu\}{}\<[E]%
\ColumnHook
\end{hscode}\resethooks
In addition to the monad primitives, the reader monad has a primitive \ensuremath{\Varid{ask}} to access the environment and \ensuremath{\Varid{local}} to run a reader action in a modified environment.
\begin{hscode}\SaveRestoreHook
\column{B}{@{}>{\hspre}l<{\hspost}@{}}%
\column{6}{@{}>{\hspre}l<{\hspost}@{}}%
\column{13}{@{}>{\hspre}c<{\hspost}@{}}%
\column{13E}{@{}l@{}}%
\column{16}{@{}>{\hspre}l<{\hspost}@{}}%
\column{33}{@{}>{\hspre}l<{\hspost}@{}}%
\column{49}{@{}>{\hspre}l<{\hspost}@{}}%
\column{E}{@{}>{\hspre}l<{\hspost}@{}}%
\>[B]{}\Keyword{let}\;{}\<[6]%
\>[6]{}\Varid{ask}{}\<[13]%
\>[13]{}\mathrel{\;:\;}{}\<[13E]%
\>[16]{}(\Tyvarid{a},\,\Tyvarid{a}\;\Varid{reader})\;\Varid{app}{}\<[E]%
\\
\>[13]{}\mathrel{=}{}\<[13E]%
\>[16]{}\Conid{Reader}\;(\Keyword{fun}\;\Varid{x}{}\<[33]%
\>[33]{}\to \Varid{x}){}\<[E]%
\\
\>[B]{}\Keyword{let}\;{}\<[6]%
\>[6]{}\Varid{local}{}\<[13]%
\>[13]{}\mathrel{\;:\;}{}\<[13E]%
\>[16]{}(\Tyvarid{a}\to \Tyvarid{b})\to (\Tyvarid{c},\,\Tyvarid{b}\;\Varid{reader})\;\Varid{app}\to (\Tyvarid{c},\,\Tyvarid{a}\;\Varid{reader})\;\Varid{app}{}\<[E]%
\\
\>[13]{}\mathrel{=}{}\<[13E]%
\>[16]{}\Keyword{fun}\;\Varid{modify}\;\Varid{r}\to \Conid{Reader}\;(\Keyword{fun}\;\Varid{env}{}\<[49]%
\>[49]{}\to \Varid{run\char95 reader}\;\Varid{r}\;(\Varid{modify}\;\Varid{env})){}\<[E]%
\ColumnHook
\end{hscode}\resethooks
\paragraph{State Monad}
The state monad is parametrised by the type of a state \ensuremath{\Tyvarid{b}} and allows the threading of a state as a side effect of a monadic computation.
A value in the state monad is a function \ensuremath{\Tyvarid{b}\to \Tyvarid{a}\mathbin{\times}\Tyvarid{b}} from an initial state to a result and a new state.
\begin{hscode}\SaveRestoreHook
\column{B}{@{}>{\hspre}l<{\hspost}@{}}%
\column{7}{@{}>{\hspre}l<{\hspost}@{}}%
\column{18}{@{}>{\hspre}c<{\hspost}@{}}%
\column{18E}{@{}l@{}}%
\column{22}{@{}>{\hspre}l<{\hspost}@{}}%
\column{32}{@{}>{\hspre}l<{\hspost}@{}}%
\column{43}{@{}>{\hspre}l<{\hspost}@{}}%
\column{E}{@{}>{\hspre}l<{\hspost}@{}}%
\>[B]{}\mathbf{type\;}\;{}\<[7]%
\>[7]{}\Tyvarid{b}\;\Varid{state}{}\<[18]%
\>[18]{}\mathrel{=}{}\<[18E]%
\>[22]{}\Conid{STATE}{}\<[E]%
\\
\>[B]{}\mathbf{type\;}\;{}\<[7]%
\>[7]{}(\anonymous ,\,\anonymous )\;\Varid{app}{}\<[18]%
\>[18]{}\mathrel{+\!\!\!=}{}\<[18E]%
\>[22]{}\Conid{State}\mathrel{\;:\;}(\Tyvarid{b}\to \Tyvarid{a}\mathbin{\times}\Tyvarid{b})\to (\Tyvarid{a},\,\Tyvarid{b}\;\Varid{state})\;\Varid{app}{}\<[E]%
\\
\>[B]{}\Keyword{let}\;{}\<[7]%
\>[7]{}\Varid{run\char95 state}{}\<[18]%
\>[18]{}\mathrel{=}{}\<[18E]%
\>[22]{}\Keyword{function}{}\<[32]%
\>[32]{}\mid \Conid{State}\;\Varid{f}{}\<[43]%
\>[43]{}\to \Varid{f}{}\<[E]%
\\
\>[32]{}\mid \anonymous {}\<[43]%
\>[43]{}\to \Keyword{assert}\;\Keyword{false}{}\<[E]%
\ColumnHook
\end{hscode}\resethooks
\ensuremath{\Varid{return}} brings a pure value into the state monad. The state is left untouched.
\ensuremath{\Varid{bind}\;\Varid{x}\;\Varid{f}} runs the stateful \ensuremath{\Varid{x}} with a state \ensuremath{\Varid{s}} obtaining a result \ensuremath{\Varid{y}} and a new state \ensuremath{\Varid{s'}}; then \ensuremath{\Varid{f}} is applied to \ensuremath{\Varid{y}} yielding a stateful computation which is run in the new state \ensuremath{\Varid{s'}}.
\begin{hscode}\SaveRestoreHook
\column{B}{@{}>{\hspre}l<{\hspost}@{}}%
\column{14}{@{}>{\hspre}c<{\hspost}@{}}%
\column{14E}{@{}l@{}}%
\column{17}{@{}>{\hspre}l<{\hspost}@{}}%
\column{25}{@{}>{\hspre}l<{\hspost}@{}}%
\column{37}{@{}>{\hspre}l<{\hspost}@{}}%
\column{57}{@{}>{\hspre}l<{\hspost}@{}}%
\column{E}{@{}>{\hspre}l<{\hspost}@{}}%
\>[B]{}\Keyword{let}\;\Varid{state}\mathrel{=}{}\<[14]%
\>[14]{}\{\mskip1.5mu {}\<[14E]%
\>[17]{}\Varid{return}{}\<[25]%
\>[25]{}\mathrel{=}(\Keyword{fun}\;\Varid{x}{}\<[37]%
\>[37]{}\to \Conid{State}\;(\Keyword{fun}\;\Varid{s}\to (\Varid{x},\,\Varid{s}))){}\<[E]%
\\
\>[14]{}\mathbin{;\;}{}\<[14E]%
\>[17]{}\Varid{bind}{}\<[25]%
\>[25]{}\mathrel{=}(\Keyword{fun}\;\Varid{x}\;\Varid{f}{}\<[37]%
\>[37]{}\to \Conid{State}\;(\Keyword{fun}\;\Varid{s}\to {}\<[57]%
\>[57]{}\Keyword{let}\;(\Varid{y},\,\Varid{s'})\mathrel{=}\Varid{run\char95 state}\;\Varid{x}\;\Varid{s}{}\<[E]%
\\
\>[57]{}\Keyword{in}\;\Varid{run\char95 state}\;(\Varid{f}\;\Varid{y})\;\Varid{s'}))\mskip1.5mu\}{}\<[E]%
\ColumnHook
\end{hscode}\resethooks
In addition to the monad primitives, the state monad has a primitive \ensuremath{\Varid{get}} to access the state and \ensuremath{\Varid{set}} to replace the state with a new one.
\begin{hscode}\SaveRestoreHook
\column{B}{@{}>{\hspre}l<{\hspost}@{}}%
\column{6}{@{}>{\hspre}l<{\hspost}@{}}%
\column{11}{@{}>{\hspre}c<{\hspost}@{}}%
\column{11E}{@{}l@{}}%
\column{14}{@{}>{\hspre}l<{\hspost}@{}}%
\column{43}{@{}>{\hspre}c<{\hspost}@{}}%
\column{43E}{@{}l@{}}%
\column{46}{@{}>{\hspre}l<{\hspost}@{}}%
\column{53}{@{}>{\hspre}c<{\hspost}@{}}%
\column{53E}{@{}l@{}}%
\column{57}{@{}>{\hspre}l<{\hspost}@{}}%
\column{60}{@{}>{\hspre}l<{\hspost}@{}}%
\column{71}{@{}>{\hspre}l<{\hspost}@{}}%
\column{E}{@{}>{\hspre}l<{\hspost}@{}}%
\>[B]{}\Keyword{let}\;{}\<[6]%
\>[6]{}\Varid{get}{}\<[11]%
\>[11]{}\mathrel{\;:\;}{}\<[11E]%
\>[14]{}(\Tyvarid{a},\,\Tyvarid{a}\;\Varid{state})\;\Varid{app}{}\<[43]%
\>[43]{}\mathrel{=}{}\<[43E]%
\>[46]{}\Conid{State}\;(\Keyword{fun}\;\Varid{s}{}\<[60]%
\>[60]{}\to (\Varid{s},\,\Varid{s})){}\<[E]%
\\
\>[B]{}\Keyword{let}\;{}\<[6]%
\>[6]{}\Varid{put}{}\<[11]%
\>[11]{}\mathrel{\;:\;}{}\<[11E]%
\>[14]{}\Tyvarid{a}\to (\Varid{unit},\,\Tyvarid{a}\;\Varid{state})\;\Varid{app}{}\<[43]%
\>[43]{}\mathrel{=}{}\<[43E]%
\>[46]{}\Keyword{fun}\;\Varid{s}{}\<[53]%
\>[53]{}\to {}\<[53E]%
\>[57]{}\Conid{State}\;(\Keyword{fun}\;\anonymous {}\<[71]%
\>[71]{}\to ((),\,\Varid{s})){}\<[E]%
\ColumnHook
\end{hscode}\resethooks
\paragraph{Abstracting over Constants}
Consider the task of replacing all constants in an expression with unique variables. We will use the state monad to hold a counter.
Let us write a function that increments the counter and returns the last value:
\begin{hscode}\SaveRestoreHook
\column{B}{@{}>{\hspre}l<{\hspost}@{}}%
\column{6}{@{}>{\hspre}l<{\hspost}@{}}%
\column{12}{@{}>{\hspre}c<{\hspost}@{}}%
\column{12E}{@{}l@{}}%
\column{15}{@{}>{\hspre}l<{\hspost}@{}}%
\column{19}{@{}>{\hspre}l<{\hspost}@{}}%
\column{30}{@{}>{\hspre}l<{\hspost}@{}}%
\column{42}{@{}>{\hspre}c<{\hspost}@{}}%
\column{42E}{@{}l@{}}%
\column{E}{@{}>{\hspre}l<{\hspost}@{}}%
\>[B]{}\Keyword{val}\;{}\<[6]%
\>[6]{}\Varid{incr}{}\<[12]%
\>[12]{}\mathrel{\;:\;}{}\<[12E]%
\>[15]{}(\Varid{int},\,\Varid{int}\;\Varid{state})\;\Varid{app}{}\<[E]%
\\
\>[B]{}\Keyword{let}\;{}\<[6]%
\>[6]{}\Varid{incr}{}\<[12]%
\>[12]{}\mathrel{=}{}\<[12E]%
\>[15]{}\Keyword{let}\;(\bind )\mathrel{=}\Varid{state}.\Varid{bind}\;\Keyword{and}\;\Varid{return}\mathrel{=}\Varid{state}.\Varid{return}{}\<[E]%
\\
\>[15]{}\Keyword{in}\;{}\<[19]%
\>[19]{}\Varid{get}{}\<[30]%
\>[30]{}\bind \Keyword{fun}\;\Varid{i}{}\<[42]%
\>[42]{}\to {}\<[42E]%
\\
\>[19]{}\Varid{put}\;(\Varid{i}\mathbin{+}\mathrm{1}){}\<[30]%
\>[30]{}\bind \Keyword{fun}\;(){}\<[42]%
\>[42]{}\to {}\<[42E]%
\\
\>[19]{}\Varid{return}\;\Varid{i}{}\<[E]%
\ColumnHook
\end{hscode}\resethooks
The core of the program uses the effectful traversal combinator \ensuremath{\Varid{traverse\char95 family}} to recursively apply the transformation in a bottom-up traversal of the expression.
\begin{hscode}\SaveRestoreHook
\column{B}{@{}>{\hspre}l<{\hspost}@{}}%
\column{21}{@{}>{\hspre}c<{\hspost}@{}}%
\column{21E}{@{}l@{}}%
\column{24}{@{}>{\hspre}l<{\hspost}@{}}%
\column{31}{@{}>{\hspre}l<{\hspost}@{}}%
\column{E}{@{}>{\hspre}l<{\hspost}@{}}%
\>[B]{}\Keyword{let}\;\Varid{abstract\char95 state}{}\<[21]%
\>[21]{}\mathrel{=}{}\<[21E]%
\>[24]{}\Varid{traverse\char95 family}\;\Varid{state}\;\Conid{Expr}\mathbin{@\!\!\!@}\Keyword{function}{}\<[E]%
\\
\>[21]{}\mid {}\<[21E]%
\>[24]{}\Conid{Cst}\;\anonymous {}\<[31]%
\>[31]{}\to \Varid{liftM}\;\Varid{state}\;(\Keyword{fun}\;\Varid{i}\to \Conid{Var}\;(\text{\tt \char34 x\char34}\mathbin{\mbox{}^\wedge}\Varid{string\char95 of\char95 int}\;\Varid{i}))\;\Varid{incr}{}\<[E]%
\\
\>[21]{}\mid {}\<[21E]%
\>[24]{}\Varid{x}{}\<[31]%
\>[31]{}\to \Varid{state}.\Varid{return}\;\Varid{x}{}\<[E]%
\ColumnHook
\end{hscode}\resethooks
The main function runs the stateful action with an initial counter value.
\begin{hscode}\SaveRestoreHook
\column{B}{@{}>{\hspre}l<{\hspost}@{}}%
\column{6}{@{}>{\hspre}l<{\hspost}@{}}%
\column{18}{@{}>{\hspre}c<{\hspost}@{}}%
\column{18E}{@{}l@{}}%
\column{21}{@{}>{\hspre}l<{\hspost}@{}}%
\column{E}{@{}>{\hspre}l<{\hspost}@{}}%
\>[B]{}\Keyword{val}\;{}\<[6]%
\>[6]{}\Varid{abstract}{}\<[18]%
\>[18]{}\mathrel{\;:\;}{}\<[18E]%
\>[21]{}\Varid{expr}\to \Varid{expr}{}\<[E]%
\\
\>[B]{}\Keyword{let}\;{}\<[6]%
\>[6]{}\Varid{abstract}\;\Varid{x}{}\<[18]%
\>[18]{}\mathrel{=}{}\<[18E]%
\>[21]{}\Varid{fst}\;(\Varid{run\char95 state}\;(\Varid{abstract\char95 state}\;\Varid{x})\;\mathrm{0}){}\<[E]%
\ColumnHook
\end{hscode}\resethooks
\paragraph{Free Variables\protect\footnote{This example is adapted from Sebastian Fischer's blog-post \url{http://www-ps.informatik.uni-kiel.de/~sebf/projects/traversal.html}}}
To collect the free variables of an expression, we will use the reader monad to keep track of the variables in scope. The reader environment is the list of variables in scope.
\begin{hscode}\SaveRestoreHook
\column{B}{@{}>{\hspre}l<{\hspost}@{}}%
\column{E}{@{}>{\hspre}l<{\hspost}@{}}%
\>[B]{}\mathbf{type\;}\;\Varid{scoped}\mathrel{=}\Varid{string}\;\Varid{list}\;\Varid{reader}{}\<[E]%
\ColumnHook
\end{hscode}\resethooks
We may define the function \ensuremath{\Varid{in\char95 scope}} that checks if a variable is in the environment:
\begin{hscode}\SaveRestoreHook
\column{B}{@{}>{\hspre}l<{\hspost}@{}}%
\column{E}{@{}>{\hspre}l<{\hspost}@{}}%
\>[B]{}\Keyword{let}\;\Varid{in\char95 scope}\;\Varid{n}\mathrel{=}\Conid{Reader}\;(\Varid{\Conid{List}.mem}\;\Varid{n}){}\<[E]%
\ColumnHook
\end{hscode}\resethooks
We also need to extend the scope with a new bound variable. \ensuremath{\Varid{extend\char95 scope}\;\Varid{n}\;\Varid{c}} runs the scoped computation \ensuremath{\Varid{c}} in the scope extended with \ensuremath{\Varid{n}}.
\begin{hscode}\SaveRestoreHook
\column{B}{@{}>{\hspre}l<{\hspost}@{}}%
\column{6}{@{}>{\hspre}l<{\hspost}@{}}%
\column{22}{@{}>{\hspre}c<{\hspost}@{}}%
\column{22E}{@{}l@{}}%
\column{25}{@{}>{\hspre}l<{\hspost}@{}}%
\column{E}{@{}>{\hspre}l<{\hspost}@{}}%
\>[B]{}\Keyword{val}\;{}\<[6]%
\>[6]{}\Varid{extend\char95 scope}{}\<[22]%
\>[22]{}\mathrel{\;:\;}{}\<[22E]%
\>[25]{}\Varid{string}\to (\Tyvarid{a},\,\Varid{scoped})\;\Varid{app}\to (\Tyvarid{a},\,\Varid{scoped})\;\Varid{app}{}\<[E]%
\\
\>[B]{}\Keyword{let}\;{}\<[6]%
\>[6]{}\Varid{extend\char95 scope}\;\Varid{n}{}\<[22]%
\>[22]{}\mathrel{=}{}\<[22E]%
\>[25]{}\Varid{local}\;(\Keyword{fun}\;\Varid{ns}\to \Varid{n}\mathbin{::}\Varid{ns}){}\<[E]%
\ColumnHook
\end{hscode}\resethooks
The function \ensuremath{\Varid{free\char95 vars}} is run in an initially empty scope:
\begin{hscode}\SaveRestoreHook
\column{B}{@{}>{\hspre}l<{\hspost}@{}}%
\column{6}{@{}>{\hspre}l<{\hspost}@{}}%
\column{19}{@{}>{\hspre}c<{\hspost}@{}}%
\column{19E}{@{}l@{}}%
\column{22}{@{}>{\hspre}l<{\hspost}@{}}%
\column{E}{@{}>{\hspre}l<{\hspost}@{}}%
\>[B]{}\Keyword{val}\;{}\<[6]%
\>[6]{}\Varid{free\char95 vars}{}\<[19]%
\>[19]{}\mathrel{\;:\;}{}\<[19E]%
\>[22]{}\Varid{expr}\to \Varid{string}\;\Varid{list}{}\<[E]%
\\
\>[B]{}\Keyword{let}\;{}\<[6]%
\>[6]{}\Varid{free\char95 vars}\;\Varid{x}{}\<[19]%
\>[19]{}\mathrel{=}{}\<[19E]%
\>[22]{}\Varid{run\char95 reader}\;(\Varid{free\char95 vars\char95 scoped}\;\Varid{x})\;[\mskip1.5mu \mskip1.5mu]{}\<[E]%
\ColumnHook
\end{hscode}\resethooks
The core of the algorithm uses the \ensuremath{\Varid{para}} combinator to recursively apply an inductive step. The step takes a list of scoped lists of free variables \ensuremath{\Varid{rs}} which are sequenced and concatenated into a scoped list of free variables \ensuremath{\Varid{r}}. Only two cases are significant: when the expression is a variable, we check if it is bound or free before returning either the empty list or a singleton; when an expression is a let binding, we run the scoped result in an extended scope including the new bound variable.
\begin{hscode}\SaveRestoreHook
\column{B}{@{}>{\hspre}l<{\hspost}@{}}%
\column{6}{@{}>{\hspre}l<{\hspost}@{}}%
\column{24}{@{}>{\hspre}c<{\hspost}@{}}%
\column{24E}{@{}l@{}}%
\column{27}{@{}>{\hspre}l<{\hspost}@{}}%
\column{31}{@{}>{\hspre}l<{\hspost}@{}}%
\column{35}{@{}>{\hspre}l<{\hspost}@{}}%
\column{52}{@{}>{\hspre}c<{\hspost}@{}}%
\column{52E}{@{}l@{}}%
\column{56}{@{}>{\hspre}l<{\hspost}@{}}%
\column{E}{@{}>{\hspre}l<{\hspost}@{}}%
\>[B]{}\Keyword{let}\;{}\<[6]%
\>[6]{}\Varid{free\char95 vars\char95 scoped}{}\<[24]%
\>[24]{}\mathrel{\;:\;}{}\<[24E]%
\>[27]{}\Varid{expr}\to (\Varid{string}\;\Varid{list},\,\Varid{scoped})\;\Varid{app}{}\<[E]%
\\
\>[24]{}\mathrel{=}{}\<[24E]%
\>[27]{}\Varid{para}\;\Conid{Expr}\mathbin{@\!\!\!@}\Keyword{fun}\;\Varid{expr}\;\Varid{rs}\to {}\<[E]%
\\
\>[27]{}\hsindent{4}{}\<[31]%
\>[31]{}\Keyword{let}\;\Varid{r}\mathrel{=}\Varid{liftM}\;\Varid{reader}\;\Varid{\Conid{List}.concat}\;(\Varid{sequenceM}\;\Varid{reader}\;\Varid{rs}){}\<[E]%
\\
\>[27]{}\hsindent{4}{}\<[31]%
\>[31]{}\Keyword{in}\;{}\<[35]%
\>[35]{}\Keyword{match}\;\Varid{expr}\;\Keyword{with}{}\<[E]%
\\
\>[35]{}\mid \Conid{Var}\;\Varid{n}{}\<[52]%
\>[52]{}\to {}\<[52E]%
\>[56]{}\Varid{reader}.\Varid{bind}\;(\Varid{in\char95 scope}\;\Varid{n})\;(\Keyword{fun}\;\Varid{is\char95 in\char95 scope}\to {}\<[E]%
\\
\>[56]{}\Varid{reader}.\Varid{return}\;(\Keyword{if}\;\Varid{is\char95 in\char95 scope}\;\Keyword{then}\;[\mskip1.5mu \mskip1.5mu]\;\Keyword{else}\;[\mskip1.5mu \Varid{n}\mskip1.5mu])){}\<[E]%
\\
\>[35]{}\mid \Conid{Let}\;(\Varid{n},\,\anonymous ,\,\anonymous ){}\<[52]%
\>[52]{}\to {}\<[52E]%
\>[56]{}\Varid{extend\char95 scope}\;\Varid{n}\;\Varid{r}{}\<[E]%
\\
\>[35]{}\mid \anonymous {}\<[52]%
\>[52]{}\to {}\<[52E]%
\>[56]{}\Varid{r}{}\<[E]%
\ColumnHook
\end{hscode}\resethooks

\subsection{Multiplate}
\label{sec:Multiplate}
One of the design goal of the Uniplate library was the simplicity of its types~\cite{mitchell:Uniplate}. However, this came at the cost of a loss of generality: the traversals are only expressed in terms of a single recursive type. The library had a slight generalisation to two mutually recursive types called \emph{biplate}. However, yet another generalisation called \emph{Multiplate} made it possible to deal with any number of mutually defined types~\cite{Multiplate}.

Multiplate is very similar to Uniplate: it has the same combinators that take a simple transformation and apply it to the children or recursively to the whole family of descendents. However that transformation, rather than being on a single type \ensuremath{\Tyvarid{a}\to \Tyvarid{a}}, is a type indexed transformation that can transform any type \ensuremath{\forall\!\;\Tyvarid{a}\mathbin{°.°}\Tyvarid{a}\;\Varid{ty}\to \Tyvarid{a}\to \Tyvarid{a}}.
Accordingly, the notion of children in Multiplate is more general: now children can have any type. The children of a variant value are all the arguments of its constructor. The children of a record are all of its fields. We refer the reader to \cite{Multiplate} for a complete description of O'Connors' Multiplate library.

\subsubsection{Deconstructing a value}
Multiplate generalises the function \ensuremath{\Varid{scrap}}. Since the children have different types, we cannot use lists anymore. The concrete type for the tuple of children is captured by the \ensuremath{\Varid{product}} GADT already seen in section~\ref{prog:product}.
\begin{hscode}\SaveRestoreHook
\column{B}{@{}>{\hspre}l<{\hspost}@{}}%
\column{E}{@{}>{\hspre}l<{\hspost}@{}}%
\>[B]{}\mathbf{type\;}\;\Tyvarid{a}\;\Varid{scrapped}\mathrel{=}\Conid{Scrapped}\mathrel{\;:\;}\Tyvarid{b}\;\Varid{product}\mathbin{\times}\Tyvarid{b}\mathbin{\times}(\Tyvarid{b}\to \Tyvarid{a})\to \Tyvarid{a}\;\Varid{scrapped}{}\<[E]%
\ColumnHook
\end{hscode}\resethooks
The function \ensuremath{\Varid{scrap}\mathrel{\;:\;}\Tyvarid{a}\;\Varid{ty}\to \Tyvarid{a}\to \Tyvarid{a}\;\Varid{scrapped}} is quite simple to define using the list-of-constructors view:
\begin{hscode}\SaveRestoreHook
\column{B}{@{}>{\hspre}l<{\hspost}@{}}%
\column{12}{@{}>{\hspre}c<{\hspost}@{}}%
\column{12E}{@{}l@{}}%
\column{15}{@{}>{\hspre}l<{\hspost}@{}}%
\column{20}{@{}>{\hspre}c<{\hspost}@{}}%
\column{20E}{@{}l@{}}%
\column{23}{@{}>{\hspre}l<{\hspost}@{}}%
\column{25}{@{}>{\hspre}c<{\hspost}@{}}%
\column{25E}{@{}l@{}}%
\column{29}{@{}>{\hspre}l<{\hspost}@{}}%
\column{33}{@{}>{\hspre}c<{\hspost}@{}}%
\column{33E}{@{}l@{}}%
\column{35}{@{}>{\hspre}l<{\hspost}@{}}%
\column{37}{@{}>{\hspre}l<{\hspost}@{}}%
\column{E}{@{}>{\hspre}l<{\hspost}@{}}%
\>[B]{}\Keyword{let}\;\Varid{scrap\char95 conlist}{}\<[20]%
\>[20]{}\mathrel{\;:\;}{}\<[20E]%
\>[23]{}\Tyvarid{a}\;\Varid{\Conid{Conlist}.t}\to \Tyvarid{a}\to \Tyvarid{a}\;\Varid{scrapped}{}\<[E]%
\\
\>[20]{}\mathrel{=}{}\<[20E]%
\>[23]{}\Keyword{fun}\;\Varid{cs}\;\Varid{x}{}\<[33]%
\>[33]{}\to {}\<[33E]%
\>[37]{}\Keyword{match}\;\Varid{\Conid{Conlist}.conap}\;\Varid{cs}\;\Varid{x}\;\Keyword{with}{}\<[E]%
\\
\>[37]{}\mid \Conid{Conap}\;(\Varid{c},\,\Varid{y})\to \Conid{Scrapped}\;(\Varid{product}\;\Varid{c},\,\Varid{y},\,\Varid{c}.\Varid{embed}){}\<[E]%
\\
\>[B]{}\Keyword{let}\;\Varid{scrap}{}\<[12]%
\>[12]{}\mathrel{\;:\;}{}\<[12E]%
\>[15]{}\Tyvarid{a}\;\Varid{ty}\to \Tyvarid{a}\to \Tyvarid{a}\;\Varid{scrapped}{}\<[E]%
\\
\>[12]{}\mathrel{=}{}\<[12E]%
\>[15]{}\Keyword{fun}\;\Varid{t}\;\Varid{x}{}\<[25]%
\>[25]{}\to {}\<[25E]%
\>[29]{}\Keyword{match}\;\Varid{\Conid{Conlist}.view}\;\Varid{t}\;\Keyword{with}{}\<[E]%
\\
\>[29]{}\mid [\mskip1.5mu \mskip1.5mu]{}\<[35]%
\>[35]{}\to \Conid{Scrapped}\;(\Conid{Nil},\,(),\,\Varid{const}\;\Varid{x}){}\<[E]%
\\
\>[29]{}\mid \Varid{cs}{}\<[35]%
\>[35]{}\to \Varid{scrap\char95 conlist}\;\Varid{cs}\;\Varid{x}{}\<[E]%
\ColumnHook
\end{hscode}\resethooks
where \ensuremath{\Varid{product}\mathrel{\;:\;}(\Tyvarid{b},\,\Tyvarid{a})\;\Varid{con}\to \Tyvarid{b}\;\Varid{product}}.

\subsubsection{Plates}
\noindent In Multiplate terminology, an effectful transformation is called a \emph{plate}
\begin{hscode}\SaveRestoreHook
\column{B}{@{}>{\hspre}l<{\hspost}@{}}%
\column{E}{@{}>{\hspre}l<{\hspost}@{}}%
\>[B]{}\mathbf{type\;}\;\Tyvarid{f}\;\Varid{plate}\mathrel{=}\{\mskip1.5mu \Varid{plate}\mathrel{\;:\;}\forall\!\;\Tyvarid{a}\mathbin{°.°}\Tyvarid{a}\;\Varid{ty}\to \Tyvarid{a}\to (\Tyvarid{a},\,\Tyvarid{f})\;\Varid{app}\mskip1.5mu\}\mbox{\commentbegin  for some applicative functor   \ensuremath{\Tyvarid{f}}  \commentend}{}\<[E]%
\ColumnHook
\end{hscode}\resethooks
We also specialise \ensuremath{\Varid{plate}} for the identity and the constant functors, obtaining the types of pure transformations and queries:
\begin{hscode}\SaveRestoreHook
\column{B}{@{}>{\hspre}l<{\hspost}@{}}%
\column{22}{@{}>{\hspre}c<{\hspost}@{}}%
\column{22E}{@{}l@{}}%
\column{25}{@{}>{\hspre}l<{\hspost}@{}}%
\column{39}{@{}>{\hspre}l<{\hspost}@{}}%
\column{E}{@{}>{\hspre}l<{\hspost}@{}}%
\>[B]{}\mathbf{type\;}\;\Varid{id\char95 plate}{}\<[22]%
\>[22]{}\mathrel{=}{}\<[22E]%
\>[25]{}\{\mskip1.5mu \Varid{id\char95 plate}{}\<[39]%
\>[39]{}\mathrel{\;:\;}\forall\!\;\Tyvarid{a}\mathbin{°.°}\Tyvarid{a}\;\Varid{ty}\to \Tyvarid{a}\to \Tyvarid{a}\mskip1.5mu\}{}\<[E]%
\\
\>[B]{}\mathbf{type\;}\;\Tyvarid{b}\;\Varid{const\char95 plate}{}\<[22]%
\>[22]{}\mathrel{=}{}\<[22E]%
\>[25]{}\{\mskip1.5mu \Varid{const\char95 plate}{}\<[39]%
\>[39]{}\mathrel{\;:\;}\forall\!\;\Tyvarid{a}\mathbin{°.°}\Tyvarid{a}\;\Varid{ty}\to \Tyvarid{a}\to \Tyvarid{b}\mskip1.5mu\}{}\<[E]%
\ColumnHook
\end{hscode}\resethooks
\subsubsection{Applying an Effectful Transformation to the Children}
All of Multiplate's combinators may be derived from a single combinator, called \emph{Multiplate} in the original article~\cite{Multiplate} but that we renamed \ensuremath{\Varid{traverse\char95 children}} for consistency with our presentation of Uniplate.
\begin{hscode}\SaveRestoreHook
\column{B}{@{}>{\hspre}l<{\hspost}@{}}%
\column{E}{@{}>{\hspre}l<{\hspost}@{}}%
\>[B]{}\Keyword{val}\;\Varid{traverse\char95 children\char95 p}\mathrel{\;:\;}\Tyvarid{f}\;\Varid{applicative}\to \Tyvarid{f}\;\Varid{plate}\to \Tyvarid{f}\;\Varid{plate}{}\<[E]%
\ColumnHook
\end{hscode}\resethooks
Thinking about the corresponding Uniplate function, \ensuremath{\Varid{traverse\char95 children\char95 p}} modifies the children of a value using a given effectful transformation. We also provide a version where the plate is inlined:
\begin{hscode}\SaveRestoreHook
\column{B}{@{}>{\hspre}l<{\hspost}@{}}%
\column{E}{@{}>{\hspre}l<{\hspost}@{}}%
\>[B]{}\Keyword{val}\;\Varid{traverse\char95 children}\mathrel{\;:\;}\Tyvarid{f}\;\Varid{applicative}\to \Tyvarid{f}\;\Varid{plate}\to \Tyvarid{a}\;\Varid{ty}\to \Tyvarid{a}\to (\Tyvarid{a},\,\Tyvarid{f})\;\Varid{app}{}\<[E]%
\ColumnHook
\end{hscode}\resethooks
The implementation is strikingly similar to the Uniplate version.
\begin{hscode}\SaveRestoreHook
\column{B}{@{}>{\hspre}l<{\hspost}@{}}%
\column{30}{@{}>{\hspre}c<{\hspost}@{}}%
\column{30E}{@{}l@{}}%
\column{34}{@{}>{\hspre}l<{\hspost}@{}}%
\column{56}{@{}>{\hspre}l<{\hspost}@{}}%
\column{E}{@{}>{\hspre}l<{\hspost}@{}}%
\>[B]{}\Keyword{let}\;\Varid{traverse\char95 children\char95 p}\;\Varid{a}\;\Varid{f}{}\<[30]%
\>[30]{}\mathrel{=}{}\<[30E]%
\>[34]{}\{\mskip1.5mu \Varid{plate}\mathrel{=}\Keyword{fun}\;\Varid{t}\;\Varid{x}\to {}\<[56]%
\>[56]{}\Keyword{let}\;\Conid{Scrapped}\;(\Varid{p},\,\Varid{cs},\,\Varid{rep})\mathrel{=}\Varid{scrap}\;\Varid{t}\;\Varid{x}{}\<[E]%
\\
\>[56]{}\Keyword{in}\;(\Varid{fun\char95 of\char95 app}\;\Varid{a}).\Varid{fmap}\;\Varid{rep}\;(\Varid{traverse}\;\Varid{a}\;\Varid{f}\;\Varid{p}\;\Varid{cs})\mskip1.5mu\}{}\<[E]%
\\
\>[B]{}\Keyword{let}\;\Varid{traverse\char95 children}\;\Varid{a}\;\Varid{f}{}\<[30]%
\>[30]{}\mathrel{=}{}\<[30E]%
\>[34]{}(\Varid{traverse\char95 children\char95 p}\;\Varid{a}\;\Varid{f}).\Varid{plate}{}\<[E]%
\ColumnHook
\end{hscode}\resethooks
The function \ensuremath{\Varid{traverse}} used above generalises the homonymous function on lists (section \ref{prog:traverse-list}) to tuples. It applies an effectful transformation to each component of a tuple from left to right, returning the modified tuple in an effectful context.
\begin{hscode}\SaveRestoreHook
\column{B}{@{}>{\hspre}l<{\hspost}@{}}%
\column{3}{@{}>{\hspre}l<{\hspost}@{}}%
\column{21}{@{}>{\hspre}l<{\hspost}@{}}%
\column{38}{@{}>{\hspre}l<{\hspost}@{}}%
\column{49}{@{}>{\hspre}c<{\hspost}@{}}%
\column{49E}{@{}l@{}}%
\column{53}{@{}>{\hspre}l<{\hspost}@{}}%
\column{E}{@{}>{\hspre}l<{\hspost}@{}}%
\>[B]{}\Keyword{let}\;\Keyword{rec}\;\Varid{traverse}\mathrel{\;:\;}\mathbf{type\;}\;\Varid{x}\mathbin{°.°}\Tyvarid{f}\;\Varid{applicative}\to \Tyvarid{f}\;\Varid{plate}\to \Varid{x}\;\Varid{product}\to \Varid{x}\to (\Varid{x},\,\Tyvarid{f})\;\Varid{app}{}\<[E]%
\\
\>[B]{}\hsindent{3}{}\<[3]%
\>[3]{}\mathrel{=}\Keyword{fun}\;\Varid{a}\;\Varid{f}\;\Varid{p}\;\Varid{x}\to {}\<[21]%
\>[21]{}\Keyword{match}\;(\Varid{p},\,\Varid{x})\;\Keyword{with}{}\<[E]%
\\
\>[21]{}\mid \Conid{Nil}{}\<[38]%
\>[38]{},\,(){}\<[49]%
\>[49]{}\to {}\<[49E]%
\>[53]{}\Varid{a}.\Varid{pure}\;(){}\<[E]%
\\
\>[21]{}\mid \Conid{Cons}\;(\Varid{t},\,\Varid{ts}){}\<[38]%
\>[38]{},\,(\Varid{x},\,\Varid{xs}){}\<[49]%
\>[49]{}\to {}\<[49E]%
\>[53]{}\Keyword{let}\;\Varid{pair}\;\Varid{a}\;\Varid{b}\mathrel{=}(\Varid{a},\,\Varid{b}){}\<[E]%
\\
\>[53]{}\Keyword{in}\;\Varid{liftA2}\;\Varid{a}\;\Varid{pair}\;(\Varid{f}.\Varid{plate}\;\Varid{t}\;\Varid{x})\;(\Varid{traverse}\;\Varid{a}\;\Varid{f}\;\Varid{ts}\;\Varid{xs}){}\<[E]%
\ColumnHook
\end{hscode}\resethooks
\subsubsection{Module Interface}
All the Uniplate functions can be generalised. Their type signature in Multiplate are:
\begin{hscode}\SaveRestoreHook
\column{B}{@{}>{\hspre}l<{\hspost}@{}}%
\column{26}{@{}>{\hspre}c<{\hspost}@{}}%
\column{26E}{@{}l@{}}%
\column{31}{@{}>{\hspre}l<{\hspost}@{}}%
\column{E}{@{}>{\hspre}l<{\hspost}@{}}%
\>[B]{}\Keyword{val}\;\Varid{children}{}\<[26]%
\>[26]{}\mathrel{\;:\;}{}\<[26E]%
\>[31]{}\Tyvarid{a}\;\Varid{ty}\to \Tyvarid{a}\to \Varid{dyn}\;\Varid{list}{}\<[E]%
\\
\>[B]{}\Keyword{val}\;\Varid{family}{}\<[26]%
\>[26]{}\mathrel{\;:\;}{}\<[26E]%
\>[31]{}\Tyvarid{a}\;\Varid{ty}\to \Tyvarid{a}\to \Varid{dyn}\;\Varid{list}{}\<[E]%
\\[\blanklineskip]%
\>[B]{}\Keyword{val}\;\Varid{traverse\char95 children\char95 p}{}\<[26]%
\>[26]{}\mathrel{\;:\;}{}\<[26E]%
\>[31]{}\Tyvarid{f}\;\Varid{applicative}\to \Tyvarid{f}\;\Varid{plate}\to \Tyvarid{f}\;\Varid{plate}{}\<[E]%
\\
\>[B]{}\Keyword{val}\;\Varid{map\char95 children\char95 p}{}\<[26]%
\>[26]{}\mathrel{\;:\;}{}\<[26E]%
\>[31]{}\Varid{id\char95 plate}\to \Varid{id\char95 plate}{}\<[E]%
\\
\>[B]{}\Keyword{val}\;\Varid{fold\char95 children\char95 p}{}\<[26]%
\>[26]{}\mathrel{\;:\;}{}\<[26E]%
\>[31]{}\Tyvarid{t}\;\Varid{monoid}\to \Tyvarid{t}\;\Varid{const\char95 plate}\to \Tyvarid{t}\;\Varid{const\char95 plate}{}\<[E]%
\\[\blanklineskip]%
\>[B]{}\Keyword{val}\;\Varid{traverse\char95 family\char95 p}{}\<[26]%
\>[26]{}\mathrel{\;:\;}{}\<[26E]%
\>[31]{}\Tyvarid{f}\;\Varid{monad}\to \Tyvarid{f}\;\Varid{plate}\to \Tyvarid{f}\;\Varid{plate}{}\<[E]%
\\
\>[B]{}\Keyword{val}\;\Varid{map\char95 family\char95 p}{}\<[26]%
\>[26]{}\mathrel{\;:\;}{}\<[26E]%
\>[31]{}\Varid{id\char95 plate}\to \Varid{id\char95 plate}{}\<[E]%
\\
\>[B]{}\Keyword{val}\;\Varid{pre\char95 fold\char95 p}{}\<[26]%
\>[26]{}\mathrel{\;:\;}{}\<[26E]%
\>[31]{}\Tyvarid{t}\;\Varid{monoid}\to \Tyvarid{t}\;\Varid{const\char95 plate}\to \Tyvarid{t}\;\Varid{const\char95 plate}{}\<[E]%
\\
\>[B]{}\Keyword{val}\;\Varid{post\char95 fold\char95 p}{}\<[26]%
\>[26]{}\mathrel{\;:\;}{}\<[26E]%
\>[31]{}\Tyvarid{t}\;\Varid{monoid}\to \Tyvarid{t}\;\Varid{const\char95 plate}\to \Tyvarid{t}\;\Varid{const\char95 plate}{}\<[E]%
\\[\blanklineskip]%
\>[B]{}\Keyword{val}\;\Varid{para\char95 p}{}\<[26]%
\>[26]{}\mathrel{\;:\;}{}\<[26E]%
\>[31]{}(\Tyvarid{r}\;\Varid{list}\to \Tyvarid{r})\;\Varid{const\char95 plate}\to \Tyvarid{r}\;\Varid{const\char95 plate}{}\<[E]%
\ColumnHook
\end{hscode}\resethooks
where \ensuremath{\Varid{dyn}} is the type of {\em dynamic} values, that is, values paired with their own type representation~\cite{dyn:amber:cardelli,dyn:abadi,dyn:leroy}.
The \ensuremath{\Varid{map\char95 }} functions are specialisations of the corresponding \ensuremath{\Varid{traverse\char95 }} functions to the identity functor. The function \ensuremath{\Varid{fold\char95 children}} is a specialisation of \ensuremath{\Varid{traverse\char95 children}} to the constant functor.

\subsubsection{Open Recursion}
In the \ocaml{} compiler libraries, one can find the modules \ensuremath{\Conid{Ast\char95 mapper}} and \ensuremath{\Conid{Ast\char95 iterator}} whose purpose is to ease the definition of traversals over the \ensuremath{\Conid{Parsetree}} mutually recursive data types.
Both \ensuremath{\Conid{Ast\char95 mapper}} and \ensuremath{\Conid{Ast\char95 iterator}} implement open recursion which takes the form of a large record \ensuremath{\Varid{mapper}} (\emph{resp.}\ \ensuremath{\Varid{iterator}}) where each field corresponds to one of the recursively defined types and is a function that takes a \ensuremath{\Varid{mapper}} (\emph{resp.}\ \ensuremath{\Varid{iterator}}) and a value of the corresponding type, and outputs a value of the same type (\emph{resp.}\ \ensuremath{\Varid{unit}}).
Defining mappers (\emph{resp.}\ iterators) is usually done by modifying a default record implementing the identity (\emph{resp.}\ a traversal of the tree without side effect). Only specific fields of interest need to be modified, the rest being taken care of by the default behaviour.

The behaviour of \ensuremath{\Conid{Ast\char95 mapper}} and \ensuremath{\Conid{Ast\char95 iterator}} can be implemented using Multiplate by defining a recursive type:
\begin{hscode}\SaveRestoreHook
\column{B}{@{}>{\hspre}l<{\hspost}@{}}%
\column{E}{@{}>{\hspre}l<{\hspost}@{}}%
\>[B]{}\mathbf{type\;}\;\Tyvarid{f}\;\Varid{openrec}\mathrel{=}\{\mskip1.5mu \Varid{run}\mathrel{\;:\;}\Tyvarid{f}\;\Varid{openrec}\to \Tyvarid{f}\;\Varid{plate}\mskip1.5mu\}{}\<[E]%
\ColumnHook
\end{hscode}\resethooks
The default records of \ensuremath{\Conid{Ast\char95 mapper}} and \ensuremath{\Conid{Ast\char95 iterator}} correspond to the \ensuremath{\Varid{openrec}} function \ensuremath{\Varid{default}} which uses the \ensuremath{\Varid{openrec}} parameter to continue the recursion below the immediate children of a value.
\begin{hscode}\SaveRestoreHook
\column{B}{@{}>{\hspre}l<{\hspost}@{}}%
\column{E}{@{}>{\hspre}l<{\hspost}@{}}%
\>[B]{}\Keyword{let}\;\Varid{default}\;\Varid{a}\mathrel{=}\{\mskip1.5mu \Varid{run}\mathrel{=}\Keyword{fun}\;\Varid{r}\to \Varid{traverse\char95 children\char95 p}\;\Varid{a}\;(\Varid{r}.\Varid{run}\;\Varid{r})\mskip1.5mu\}{}\<[E]%
\ColumnHook
\end{hscode}\resethooks
\ensuremath{\Conid{Ast\char95 mapper}} corresponds to the specialisation of \ensuremath{\Varid{openrec}} to the identity functor, while \ensuremath{\Conid{Ast\char95 iterator}} corresponds to a use of \ensuremath{\Varid{openrec}} with the IO monad to embed \ocaml{} effectful computations:
\begin{hscode}\SaveRestoreHook
\column{B}{@{}>{\hspre}l<{\hspost}@{}}%
\column{E}{@{}>{\hspre}l<{\hspost}@{}}%
\>[B]{}\mathbf{type\;}\;\Varid{io}{}\<[E]%
\\
\>[B]{}\Keyword{val}\;\Varid{io}\mathrel{\;:\;}\Varid{io}\;\Varid{monad}{}\<[E]%
\ColumnHook
\end{hscode}\resethooks
Effectful computations may be embedded in the IO monad with \ensuremath{\Varid{embed\char95 io}}. They are functions from unit to some result type \ensuremath{\Tyvarid{a}} which may carry side effects when evaluated.
\begin{hscode}\SaveRestoreHook
\column{B}{@{}>{\hspre}l<{\hspost}@{}}%
\column{E}{@{}>{\hspre}l<{\hspost}@{}}%
\>[B]{}\Keyword{val}\;\Varid{embed\char95 io}\mathrel{\;:\;}(\Varid{unit}\to \Tyvarid{a})\to (\Tyvarid{a},\,\Varid{io})\;\Varid{app}{}\<[E]%
\ColumnHook
\end{hscode}\resethooks
IO computations may be executed with \ensuremath{\Varid{run\char95 io}}.
\begin{hscode}\SaveRestoreHook
\column{B}{@{}>{\hspre}l<{\hspost}@{}}%
\column{E}{@{}>{\hspre}l<{\hspost}@{}}%
\>[B]{}\Keyword{val}\;\Varid{run\char95 io}\mathrel{\;:\;}(\Tyvarid{a},\,\Varid{io})\;\Varid{app}\to \Tyvarid{a}{}\<[E]%
\ColumnHook
\end{hscode}\resethooks
\paragraph{Discussion}
\ensuremath{\Conid{Ast\char95 mapper}} and \ensuremath{\Conid{Ast\char95 iterator}} are long pieces of boilerplate that cannot be reused for other AST types, and do not allow easy extension or addition of new operations.
In contrast, Multiplate implements the same functionality (and more) in a concise implementation that exploits the theoretical properties of applicative functors and monads. Built on top of the generic library, Multiplate may be used with any type with no boilerplate.

\section{A Case Study: Safe Deserialisation}
\label{sec:unmarshal}

Serialisation and deserialisation in \ocaml{} are provided by the standard library module Marshal.  They are generic functions defined in C that rely on the concrete structure of the runtime values of \ocaml{} programs.
There is a serious safety issue with the use of the deserialisation primitive: since it must be able to reconstruct values of any type from an input channel, it is polymorphic in its return type:
\begin{hscode}\SaveRestoreHook
\column{B}{@{}>{\hspre}l<{\hspost}@{}}%
\column{E}{@{}>{\hspre}l<{\hspost}@{}}%
\>[B]{}\Keyword{val}\;\Varid{from\char95 channel}\mathrel{\;:\;}\Varid{in\char95 channel}\to \Tyvarid{a}{}\<[E]%
\ColumnHook
\end{hscode}\resethooks
This may easily cause a segmentation fault if the deserialised value is used with the wrong type.
This problem may be solved using the generic library, constraining the return type by its witness:
\begin{hscode}\SaveRestoreHook
\column{B}{@{}>{\hspre}l<{\hspost}@{}}%
\column{E}{@{}>{\hspre}l<{\hspost}@{}}%
\>[B]{}\Keyword{val}\;\Varid{from\char95 channel}\mathrel{\;:\;}\Tyvarid{a}\;\Varid{ty}\to \Varid{in\char95 channel}\to \Tyvarid{a}{}\<[E]%
\ColumnHook
\end{hscode}\resethooks
The low level generic view gives us the type structure to guide the deserialisation process.

\subsection{Outline of the Algorithm}
The algorithm follows the same approach as the work of Henry~{\em et al.}~\cite{henry:12:icfp} to check that the result of the standard deserialisation function is compatible with a given type.
A substantial difference is that our implementation also deals with abstract types. This involves converting the abstract values to a public representation before serialisation, and converting back the representation to the abstract type after deserialisation. Therefore, the heart of the program is a function
\ensuremath{\Varid{convert}} that not only deals with such conversions but also does the compatibility check.
The function \ensuremath{\Varid{convert}} takes a \ensuremath{\Varid{direction}} argument (\emph{to} or \emph{from} the public representation), a type witness and converts directly the runtime values using the \ensuremath{\Conid{Obj}} module, which gives an API to the memory representation of \ocaml{} values.
\begin{hscode}\SaveRestoreHook
\column{B}{@{}>{\hspre}l<{\hspost}@{}}%
\column{E}{@{}>{\hspre}l<{\hspost}@{}}%
\>[B]{}\mathbf{type\;}\;\Varid{direction}\mathrel{=}\Conid{To}\mid \Conid{From}{}\<[E]%
\\
\>[B]{}\Keyword{val}\;\Varid{convert}\mathrel{\;:\;}\Varid{direction}\to \Tyvarid{a}\;\Varid{ty}\to \Varid{obj}\to \Varid{obj}{}\<[E]%
\ColumnHook
\end{hscode}\resethooks
\ensuremath{\Varid{convert}} is private to the module, the type \ensuremath{\Varid{obj}} is not shown to the user of the library.
The module exports type safe functions that call \ensuremath{\Varid{convert}} internally.
\begin{hscode}\SaveRestoreHook
\column{B}{@{}>{\hspre}l<{\hspost}@{}}%
\column{18}{@{}>{\hspre}c<{\hspost}@{}}%
\column{18E}{@{}l@{}}%
\column{21}{@{}>{\hspre}l<{\hspost}@{}}%
\column{E}{@{}>{\hspre}l<{\hspost}@{}}%
\>[B]{}\Keyword{val}\;\Varid{to\char95 string}{}\<[18]%
\>[18]{}\mathrel{\;:\;}{}\<[18E]%
\>[21]{}\Tyvarid{a}\;\Varid{ty}\to \Tyvarid{a}\to \Varid{string}{}\<[E]%
\\
\>[B]{}\Keyword{val}\;\Varid{from\char95 string}{}\<[18]%
\>[18]{}\mathrel{\;:\;}{}\<[18E]%
\>[21]{}\Tyvarid{a}\;\Varid{ty}\to \Varid{string}\to \Tyvarid{a}{}\<[E]%
\ColumnHook
\end{hscode}\resethooks
\ensuremath{\Varid{from\char95 string}} can safely cast the result of \ensuremath{\Varid{convert}} from \ensuremath{\Varid{obj}} to \ensuremath{\Tyvarid{a}}, because the compatibility check ensures that the value of type \ensuremath{\Varid{obj}} is also a valid value of type \ensuremath{\Tyvarid{a}}.

\subsection{Type Compatibility}
\ocaml{} runtime values are either immediate values taking a word of memory minus one bit, or a pointer to a block of memory allocated in the heap, which has a header containing the size of the block and a tag indicating how the block is structured.
Our goal is to check that such a runtime value is compatible with a certain type.
The structure of runtime values is accessible through the standard module \ensuremath{\Conid{Obj}} and the structure of the type is described by the low-level generic view \ensuremath{\Varid{desc}} presented in section~\ref{sec:view/desc}.

The algorithm is recursive. The base cases are immediate values: to check that a value is compatible with an integer value for instance, we simply check that it is an immediate value.
Checking a record involves checking that the value is a block of tag 0, that it has a number of fields corresponding to that of the record type, and recursively that each field is compatible with its corresponding type.
To check that a value is compatible with a variant type, we must check that: either it is an immediate value and corresponds to one of the constant constructors, or it is a block whose tag corresponds to one of the non-constant constructors and the fields must be recursively checked with the types of the constructor arguments.

\subsection{Sharing and Cycles}
The structure of runtime values is a directed graph where the vertices are memory blocks and the edges are the pointers that may be stored in the fields of a block. Sharing and cycles in the graph raise two questions: (1) could we avoid checking again a value that has already been checked?
(2) how do we ensure that the algorithm terminates?
In a monomorphic setting, both questions may be answered by storing the addresses of blocks together with the witness of their expected type when they are visited for the first time, and upon subsequent visits, simply check that the new witness is equal to the stored one.
However both sharing and cycles can be polymorphic.
Here is a contrived example of an infinite tree with polymorphic recursion, whose representation is a (finite) cyclic graph:
\begin{hscode}\SaveRestoreHook
\column{B}{@{}>{\hspre}l<{\hspost}@{}}%
\column{3}{@{}>{\hspre}l<{\hspost}@{}}%
\column{7}{@{}>{\hspre}l<{\hspost}@{}}%
\column{17}{@{}>{\hspre}c<{\hspost}@{}}%
\column{17E}{@{}l@{}}%
\column{E}{@{}>{\hspre}l<{\hspost}@{}}%
\>[B]{}\mathbf{type\;}\;{}\<[7]%
\>[7]{}\Tyvarid{a}\;\Varid{t}\mathrel{=}\Conid{Leaf}\;\mathbf{of\;}\;\Varid{int}\mid \Conid{Node}\;\mathbf{of\;}\;\Tyvarid{a}\;\Varid{t}\mathbin{\times}(\Tyvarid{a}\mathbin{\times}\Tyvarid{a})\;\Varid{t}{}\<[E]%
\\
\>[B]{}\Keyword{let}\;\Varid{poly\char95 cycle}{}\<[17]%
\>[17]{}\mathrel{=}{}\<[17E]%
\\
\>[B]{}\hsindent{3}{}\<[3]%
\>[3]{}\Keyword{let}\;\Keyword{rec}\;\Varid{go}\mathrel{\;:\;}\forall\!\;\Tyvarid{a}\mathbin{°.°}\Tyvarid{a}\;\Varid{t}\mathrel{=}\Conid{Node}\;(\Conid{Leaf}\;\mathrm{0},\,\Varid{go})\;\Keyword{in}{}\<[E]%
\\
\>[B]{}\hsindent{3}{}\<[3]%
\>[3]{}\Varid{go}{}\<[E]%
\ColumnHook
\end{hscode}\resethooks
The type parameter of the tree is not ever used, which makes it a bit pointless but this illustrates perfectly the sort of complex situations that our compatibility checker must deal with.
In this context, the previous solution does not work anymore and
non-termination becomes an issue because the number of types to check is infinite.

If an \ocaml{} value admits many types, then they must all be instances of a more general type scheme. In the previous example, the value \ensuremath{\Varid{go}} admits the types
\begin{itemize}[itemsep=0pt, label=\mbox{}, leftmargin=\parindent]
\item \ensuremath{\forall\!\;\Tyvarid{a}\mathbin{°.°}\Tyvarid{a}\;\Varid{t}}
\item \ensuremath{\forall\!\;\Tyvarid{a}\mathbin{°.°}(\Tyvarid{a}\mathbin{\times}\Tyvarid{a})\;\Varid{t}}
\item \ensuremath{\forall\!\;\Tyvarid{a}\mathbin{°.°}((\Tyvarid{a}\mathbin{\times}\Tyvarid{a})\mathbin{\times}(\Tyvarid{a}\mathbin{\times}\Tyvarid{a}))\;\Varid{t}}
\item \ldots
\end{itemize}
and so on, of which the first is the most general.

The \emph{anti-unifier} of a set of types is the most precise type that is more general than each of them. For instance,
the anti-unifier of \ensuremath{\Varid{int}} and \ensuremath{\Varid{bool}} is \ensuremath{\forall\!\;\Tyvarid{a}\mathbin{°.°}\Tyvarid{a}}, and the anti-unifier of \ensuremath{\Varid{int}\;\Varid{list}} and \ensuremath{\Varid{bool}\;\Varid{list}} is \ensuremath{\forall\!\;\Tyvarid{a}\mathbin{°.°}\Tyvarid{a}\;\Varid{list}}.

Now, we may keep track of the anti-unifier so far associated with a block.
The sequence of updated anti-unifiers is sure to reach a fixed-point in which case we no longer need to visit that block.
When checking the \ensuremath{\Varid{poly\char95 cycle}} example, the first visit would already be checking \ensuremath{\Varid{go}} with its most general type \ensuremath{\forall\!\;\Tyvarid{a}\mathbin{°.°}\Tyvarid{a}\;\Varid{t}}, hence no other visit would be performed.

Now the algorithm terminates, but it may be improved: when there is no cycle, it is faster to traverse the graph in topological order. During the traversal, all the expected types of a block may be collected before that block is visited. The anti-unifier of a block's set of expected types may then be computed and the block needs only be visited once. In the presence of cycles, one may still apply that strategy on the strongly connected components of a graph and compute a lower bound for the anti-unifier of the roots of each strongly connected component before traversing it.

The interested reader is referred to Henry {\em et al.}~\cite{henry:12:icfp} for the theoretical background and justifications.
\subsection{Abstract types}
The standard (de)-serialisation functions from the module \ensuremath{\Conid{Marshal}} break type abstraction since it becomes possible to inspect an abstract value through its serialisation and to cast a value to an abstract type.
In order to resolve that issue and respect the abstraction, we must serialise a public representation of the abstract value. This is the purpose of the datatype \ensuremath{\Varid{repr}} given in section~\ref{sec:abstract}.
The partial application \ensuremath{(\Varid{convert}\;\Conid{To})} uses the field \ensuremath{\Varid{to\char95 repr}} to serialise the representation,
while \ensuremath{(\Varid{convert}\;\Conid{From})} uses the field \ensuremath{\Varid{from\char95 repr}} on the deserialised value.

Abstract types introduce some complexity.
Instead of \ensuremath{\Varid{check}\mathrel{\;:\;}\Tyvarid{a}\;\Varid{ty}\to \Varid{obj}\to \Varid{bool}} that returns a boolean when a value is compatible with a type, we define \ensuremath{\Varid{convert}\mathrel{\;:\;}\Varid{direction}\to \Tyvarid{a}\;\Varid{ty}\to \Varid{obj}\to \Varid{obj}} that computes a new value where all the blocks of abstract types have been converted (in one direction or the other). A call to \ensuremath{\Varid{convert}} fails with an exception when the input value is not compatible with the type witness.

Note that when we serialise a value, we first convert it by recursively computing its representation.
The result of that conversion usually does not have a corresponding \ocaml{} type in the program. However when converting back after deserializing a value, we rebuild a value of a valid type by recursively computing its components.

The main challenge is that the conversion should preserve the graph structure in recursively transforming its sub-graphs.

\paragraph{Sharing}
Sharing in an acyclic graph does not cause too much trouble. In addition to keeping track of visited blocks and their most general type so far, we also memoise the function \ensuremath{\Varid{convert}}. The result of converting a block is stored, so that when the same block is visited again and the expected type is not more general, the previous result may be retrieved directly.

\paragraph{Cycles}
Cycles in memory graphs on the other hand require a lot of care.
When visiting a block, the presence of cycles means that we may visit the same block again {\em before} the first visit is completed, therefore before we have been able to store the converted block. The solution is to introduce an indirection, a reference to an \ensuremath{\Varid{option}} which is set to \ensuremath{\Conid{None}} upon visiting a block for the first time, and updated with the result when the function returns. If during the visit, the same block is checked, the reference is already known even if the content is not.  In a final traversal of the graph, we may remove all the indirections.

\section{Related Work}
\label{sec:comparison}

Generic programming is a very rich topic that we have barely touched in this article, the reader may consult the following tutorials for a deeper understanding~\cite{jansson:99:gp-intro,hinze:03:gh-theory,hinze:03:gh-applications}, mostly in the context of the Haskell type system and programming language. More fundamental ideas and higher levels of generality can be obtained using dependent types, as for instance in \cite{Altenkirch:GPDT, Altenkirch:IC}. However, we only compare here our library to those that have been developed in Haskell or ML-like languages, whose expressive power and usage is similar to \ocaml.

Generic libraries have blossomed in the past twenty years, and the many different approaches have been compared extensively~\cite{rodriguez:08:comparing-gp-libraries,hinze:2006:comparing-gp-approaches,generic-programming-in-3d}.
Our design with separate type witnesses and generic view
was directly influenced by Hinze and Löh~\cite{generic-programming-in-3d}.

\subsection{Views}
The open design of the library enables the user to define his own views.
In addition to the low level view, we have included the sum-of-products view underlying the LIGD library~\cite{Cheney:02:LIGD} and Instant-Generics~\cite{chakravarty09instantgenerics,Magalhaes:11:GPID},
the spine view underlying the SYB library~\cite{hinze:06:syb-reloaded} and the list-of-constructors view underlying RepLib~\cite{weirich:06:replib}.

Adapting other libraries is possible when their underlying type representation is first order---where closed types are reflected.

\subsection{Type Representation}
The SYB library relies on a type reflection using a non-parametrised type \ensuremath{\Conid{TypRep}} and an unsafe coercion operation~\cite{lammel:03:SYB}. In contrast, our type witness GADT reflecting its type parameter makes it possible to define a safe coerce (section~\ref{prog:coerce}).

\ensuremath{\Conid{TypRep}} is similar to our \ensuremath{\Tyvarid{a}\;\Varid{ty}} in that it captures an open universe of types. However, \ensuremath{\Tyvarid{a}\;\Varid{ty}} does so with an extensible variant ensuring strong type guarantees, whereas \ensuremath{\Conid{TypRep}} does so with a unique integer tag.

LIGD~\cite{Cheney:02:LIGD} and Replib~\cite{weirich:06:replib} use a GADT for type representation in the same way as we did with \ensuremath{\Tyvarid{a}\;\Varid{ty}}, however their representation is closed and is fused with the view. Note that open variants are not natively supported in Haskell.

With TypeCase~\cite{Oliveira:2005:typecase}, a GADT is made implicit through by using type classes to implement catamorphisms over the GADT. That technique may be used to make a Haskell-98 compatible library, since GADT are not valid Haskell-98. LIGD and PolyP have been adapted using TypeCase.

In Instant-Generics~\cite{chakravarty09instantgenerics}, the representation is given by a type family---a GHC extension which allows to define type functions.


\subsection{Higher-Order Kinded Types}
The choice of a type representation determines the universe of types that can be represented. Our library represents types with first-order kind.

In a first order representation, the list type constructor is represented as a data constructor of arity one:
\begin{hscode}\SaveRestoreHook
\column{B}{@{}>{\hspre}l<{\hspost}@{}}%
\column{E}{@{}>{\hspre}l<{\hspost}@{}}%
\>[B]{}\Conid{List}\mathrel{\;:\;}\Tyvarid{a}\;\Varid{ty}\to \Tyvarid{a}\;\Varid{list}\;\Varid{ty}{}\<[E]%
\ColumnHook
\end{hscode}\resethooks
Whereas in a second order type representation, it would be represented as a constant data constructor:
\begin{hscode}\SaveRestoreHook
\column{B}{@{}>{\hspre}l<{\hspost}@{}}%
\column{E}{@{}>{\hspre}l<{\hspost}@{}}%
\>[B]{}\Conid{List}\mathrel{\;:\;}\Varid{list'}\;\Varid{ty'}{}\<[E]%
\ColumnHook
\end{hscode}\resethooks
Where \ensuremath{\Varid{list'}} corresponds to the unapplied list constructor in our encoding of higher-order kinded type variables.

The latter approach allows us to define generic functions that work on type constructors, like a generic map:
\begin{hscode}\SaveRestoreHook
\column{B}{@{}>{\hspre}l<{\hspost}@{}}%
\column{E}{@{}>{\hspre}l<{\hspost}@{}}%
\>[B]{}\Keyword{val}\;\Varid{gmap}\mathrel{\;:\;}\Tyvarid{f}\;\Varid{ty'}\to (\Tyvarid{a}\to \Tyvarid{b})\to (\Tyvarid{a},\,\Tyvarid{f})\;\Varid{app}\to (\Tyvarid{b},\,\Tyvarid{f})\;\Varid{app}{}\<[E]%
\ColumnHook
\end{hscode}\resethooks
Ours was a choice of simplicity since \ocaml{} does not have a native support for higher order kind type variables.  A higher order kind generic library is possible in \ocaml{} using the encoding presented in this article. More flexibility is obtained at the cost of more complexity.

In Haskell, representing higher kinded types is possible~\cite{generic-programming-in-3d}. For instance,
PolyP~\cite{jansson:97:polyp} and its library implementation~\cite{norell:03:polytypic} represent parametrised datatypes as fixed-points of functors.
Generic Deriving~\cite{Magalhaes:10:generic-deriving}, another Haskell library and GHC extensions, allows users to define generic instances of type classes. It has two type representations: one for closed types and one for parametric types (of one parameter). More representations could be defined in the same way, and would allow users to derive class-instances for types of the corresponding kind.

\subsection{Type Indexed Functions}
The first version of SYB relied on operations \ensuremath{\Varid{mkT}}, \ensuremath{\Varid{extT}}, \ensuremath{\Varid{mkQ}}, \ensuremath{\Varid{extQ}}, \ensuremath{\Varid{mkM}}, \ensuremath{\Varid{extM}} to define extensible type-indexed functions.
Their implementation suffered from the same shortcomings as the simple implementation given in section~\ref{sec:gp/ext}, and furthermore once a generic function was defined, no more ad-hoc cases could be added.
A latter version of SYB~\cite{lammel:05:syb-with-class} resolved this issue with a clever use of type classes, requiring some extensions to the class system. With our explicit use of a type witness to define type-indexed functions, we face no such difficulties.

Most Haskell generic libraries rely on the powerful type class system to implement type-indexed functions. Usually, they have a \ensuremath{\Conid{Rep}} class that builds the representation. A generic function is usually implemented by a class with a default instance depending on a \ensuremath{\Conid{Rep}} constraint to implement the generic behaviour. Ad-hoc behaviour may be defined by implementing instances of the class for specific types.

\subsection{Language Extensions}
Library implementations of genericity must fight with the limits set by the programming language. Extending the language with direct support for generic programming through a dedicated syntax and semantics gives much more freedom to the designer.

PolyP~\cite{jansson:97:polyp} represent parametrised datatypes as fixed-points of functors, which makes it possible to define a generic map.
That extension initially implemented as a preprocessor was subsequently implemented as a library using extensions to the type class system~\cite{norell:03:polytypic}.

Generic Haskell~\cite{hinze:03:gh-theory} is the most expressive of all generic systems so far. In Generic Haskell the type contains types of any kinds. Functions defined by induction on the structure of types have a type that is defined by induction on the structure of kinds. This allows a truly generic \ensuremath{\Varid{map}} that works on types of any kinds.

The level of expressivity of Generic Haskell may be achieved by a library in \ocaml{}, but at such a cost in readability that one may wonder if that would be useful.

\subsection{Generic Traversals}
Our implementation of Uniplate and Multiplate follows directly from the work on the homonymous Haskell libraries~\cite{mitchell:Uniplate,Multiplate}, but also on Fischer's implementation of Uniplate for his naming convention.
The use of applicative functors for generic traversals has a small history~\cite{mcbride:applicative,gibbons:iterator,bringert:06:compos,Multiplate}.

Other approaches to generic traversals include Compos~\cite{bringert:06:compos} and SYB~\cite{lammel:03:SYB} which are equivalent to Multiplate in expressivity. The fundamental mechanism underlying Compos and Multiplate is the same. They differ in their Haskell embodiment by the way that type classes are used. In our \ocaml{} implementation, the type classes are replaced with explicit type-indexed functions. In fact, the missing link between SYB, Compos and Multiplate, is yet another variation called
Traverse-with-class~\cite{cheplyaka:traversal}. It is the closest to what we have implemented: in essence, our single type-indexed function \ensuremath{\Varid{traverse\char95 children}} corresponds to its single type class \ensuremath{\Conid{Gtraverse}}:
\begin{hscode}\SaveRestoreHook
\column{B}{@{}>{\hspre}l<{\hspost}@{}}%
\column{5}{@{}>{\hspre}l<{\hspost}@{}}%
\column{E}{@{}>{\hspre}l<{\hspost}@{}}%
\>[B]{}\Keyword{class}\;\Conid{Gtraverse}\;\mathbf{where}{}\<[E]%
\\
\>[B]{}\hsindent{5}{}\<[5]%
\>[5]{}\Varid{gtraverse}\mathbin{::}\Conid{Applicative}\;\Varid{c}\Rightarrow (\Varid{forall}\;\Varid{d}.\Conid{GTraversable}\;\Varid{d}\Rightarrow \Varid{d}\to \Varid{c}\;\Varid{d})\to \Varid{a}\to \Varid{c}\;\Varid{a}{}\<[E]%
\ColumnHook
\end{hscode}\resethooks
The SYB primitive \ensuremath{\Varid{gfold}} corresponds to a catamorphism over the spine view~\cite{hinze:06:syb-reloaded}.

Compos, Uniplate, Multiplate, and Traverse-with-class are independent on a particular type representation, their necessary class instances may be either written manually or derived using Template Haskell or using a Generic Deriving mechanism, or using the SYB Typeable or even Data class.

%
%
%

\subsection{Generic Libraries and Extensions in ML}
Generics for the Working ML'er~\cite{karvonen:2007:generics-ml} is a library for SML implementing Generics for the Masses~\cite{hinze04generic-for-the-masses} which is a variation of LIGD that uses a type class instead of a GADT for the type representation. In the SML implementation, a module is used instead.
\begin{hscode}\SaveRestoreHook
\column{B}{@{}>{\hspre}l<{\hspost}@{}}%
\column{5}{@{}>{\hspre}l<{\hspost}@{}}%
\column{11}{@{}>{\hspre}l<{\hspost}@{}}%
\column{17}{@{}>{\hspre}l<{\hspost}@{}}%
\column{E}{@{}>{\hspre}l<{\hspost}@{}}%
\>[B]{}\Keyword{module}\;\mathbf{type\;}\;\Conid{Rep}\mathrel{=}\Keyword{sig}{}\<[E]%
\\
\>[B]{}\hsindent{5}{}\<[5]%
\>[5]{}\mathbf{type\;}\;{}\<[11]%
\>[11]{}\Tyvarid{a}\;\Varid{ty}{}\<[E]%
\\
\>[B]{}\hsindent{5}{}\<[5]%
\>[5]{}\Keyword{val}\;{}\<[11]%
\>[11]{}\Varid{int}{}\<[17]%
\>[17]{}\mathrel{\;:\;}\Varid{int}\;\Varid{ty}{}\<[E]%
\\
\>[B]{}\hsindent{5}{}\<[5]%
\>[5]{}\Keyword{val}\;{}\<[11]%
\>[11]{}\Varid{list}{}\<[17]%
\>[17]{}\mathrel{\;:\;}\Tyvarid{a}\;\Varid{ty}\to \Tyvarid{a}\;\Varid{list}\;\Varid{ty}{}\<[E]%
\\
\>[B]{}\hsindent{5}{}\<[5]%
\>[5]{}\mathbin{...}{}\<[E]%
\\
\>[B]{}\Keyword{end}{}\<[E]%
\ColumnHook
\end{hscode}\resethooks
Generic functions are all modules of the same signature, whose functions correspond to the constructors of the GADT, and in the context of the library, they correspond to the type representation constructors. For instance, let us write a generic show function:
\begin{hscode}\SaveRestoreHook
\column{B}{@{}>{\hspre}l<{\hspost}@{}}%
\column{5}{@{}>{\hspre}l<{\hspost}@{}}%
\column{11}{@{}>{\hspre}l<{\hspost}@{}}%
\column{19}{@{}>{\hspre}l<{\hspost}@{}}%
\column{E}{@{}>{\hspre}l<{\hspost}@{}}%
\>[B]{}\Keyword{module}\;\Conid{Show}\mathrel{\;:\;}\Conid{Rep}\mathrel{=}\Keyword{struct}{}\<[E]%
\\
\>[B]{}\hsindent{5}{}\<[5]%
\>[5]{}\mathbf{type\;}\;{}\<[11]%
\>[11]{}\Tyvarid{a}\;\Varid{ty}{}\<[19]%
\>[19]{}\mathrel{=}\Tyvarid{a}\to \Varid{string}{}\<[E]%
\\
\>[B]{}\hsindent{5}{}\<[5]%
\>[5]{}\Keyword{let}\;{}\<[11]%
\>[11]{}\Varid{int}{}\<[19]%
\>[19]{}\mathrel{=}\Varid{string\char95 of\char95 int}{}\<[E]%
\\
\>[B]{}\hsindent{5}{}\<[5]%
\>[5]{}\Keyword{let}\;{}\<[11]%
\>[11]{}\Varid{list}{}\<[19]%
\>[19]{}\mathrel{=}\Keyword{fun}\;\Varid{show\char95 x}\;\Varid{xs}\to \text{\tt \char34 [\char34}\mathbin{\mbox{}^\wedge}\Varid{\Conid{String}.concat}\;\text{\tt \char34 ;~\char34}\;(\Varid{\Conid{List}.map}\;\Varid{show\char95 x}\;\Varid{xs})\mathbin{\mbox{}^\wedge}\text{\tt \char34 ]\char34}{}\<[E]%
\\
\>[B]{}\hsindent{5}{}\<[5]%
\>[5]{}\mathbin{...}{}\<[E]%
\\
\>[B]{}\Keyword{end}{}\<[E]%
\ColumnHook
\end{hscode}\resethooks
There is an inherent problem with this approach: the type representations are not unique as they must be instantiated with the module of the generic function that is called.

Deriving~\cite{Yallop:2007:Deriving} is an extension to \ocaml{} implemented using the preprocessor Camlp4. Generic functions are defined over the structure of types definitions, using a special syntax. The extension is used to implement a safe deserialisation function which supports sharing and cycles, and allows the user to override the default behaviour for specific types.

SYB was implemented in MetaOCaml extended with modular implicits and was shown to perform on par with manually written traversals~\cite{Yallop:2016:staging-generic-programming}. The implementation uses an extensible GADT for type witnesses, like in our library. There is no support for extensible type indexed functions. The spine view is implicit much like in the Haskell implementation. The Haskell \ensuremath{\Conid{Typeable}} and \ensuremath{\Conid{Data}} type classes are directly translated as a modules, using the correspondence explained in~\cite{modular-implicits}.

The addition of GADTs to \ocaml{} made it possible to reflect types as a basis for generic programming.

\section{Conclusion}
\label{sec:conclusion}
We have presented a library for generic programming in \ocaml{},
it is built modularly around three main ingredients: (1) an extensible GADT that reflects the names of types; (2) an implementation of extensible type-indexed functions, suitable to define ad-hoc polymorphic functions; (3) generic views which reflect the structure of types. Views are implemented as type-indexed functions, and new views can be added by the user. The built-in view is automatically derived by a PPX for the types marked with an attribute \ensuremath{\Varid{reify}}.
Abstract types are supported by means of a public representation.

On top of the library, we implemented a library for generic traversals that removes a lot of the boilerplate normally associated with the functions on mutually defined recursive types with a large number of constructors.
The library was seamlessly adapted from a Haskell library.

Finally we presented a complex generic function that fixed some of the shortcomings of the built-in deserialisation: not only is our function type-safe, it also respects abstract types by serialising their public representation.

\bibliographystyle{eptcs}
\bibliography{generic}

\begin{thebibliography}{10}
\providecommand{\bibitemdeclare}[2]{}
\providecommand{\surnamestart}{}
\providecommand{\surnameend}{}
\providecommand{\urlprefix}{Available at }
\providecommand{\url}[1]{\texttt{#1}}
\providecommand{\href}[2]{\texttt{#2}}
\providecommand{\urlalt}[2]{\href{#1}{#2}}
\providecommand{\doi}[1]{doi:\urlalt{http://dx.doi.org/#1}{#1}}
\providecommand{\bibinfo}[2]{#2}

\bibitemdeclare{article}{dyn:abadi}
\bibitem{dyn:abadi}
\bibinfo{author}{Mart{\'{\i}}n \surnamestart Abadi\surnameend},
  \bibinfo{author}{Luca \surnamestart Cardelli\surnameend},
  \bibinfo{author}{Benjamin~C. \surnamestart Pierce\surnameend} \&
  \bibinfo{author}{Gordon~D. \surnamestart Plotkin\surnameend}
  (\bibinfo{year}{1991}): \emph{\bibinfo{title}{Dynamic Typing in a Statically
  Typed Language}}.
\newblock {\sl \bibinfo{journal}{{ACM} Trans. Program. Lang. Syst.}}
  \bibinfo{volume}{13}(\bibinfo{number}{2}), pp. \bibinfo{pages}{237--268},
  \doi{10.1145/103135.103138}.

\bibitemdeclare{article}{Altenkirch:IC}
\bibitem{Altenkirch:IC}
\bibinfo{author}{Thorsten \surnamestart Altenkirch\surnameend},
  \bibinfo{author}{Neil \surnamestart Ghani\surnameend}, \bibinfo{author}{Peter
  \surnamestart Hancock\surnameend}, \bibinfo{author}{Conor \surnamestart
  McBride\surnameend} \& \bibinfo{author}{Peter \surnamestart
  Morris\surnameend} (\bibinfo{year}{2015}): \emph{\bibinfo{title}{Indexed
  containers}}.
\newblock {\sl \bibinfo{journal}{Journal of Functional Programming}}
  \bibinfo{volume}{25}, p.~\bibinfo{pages}{e5},
  \doi{10.1017/S095679681500009X}.

\bibitemdeclare{inproceedings}{Altenkirch:GPDT}
\bibitem{Altenkirch:GPDT}
\bibinfo{author}{Thorsten \surnamestart Altenkirch\surnameend},
  \bibinfo{author}{Conor \surnamestart Mcbride\surnameend} \&
  \bibinfo{author}{Peter \surnamestart Morris\surnameend}
  (\bibinfo{year}{2007}): \emph{\bibinfo{title}{Generic Programming with
  Dependent Types}}.
\newblock In: {\sl \bibinfo{booktitle}{Proceedings of the 2006 International
  Conference on Datatype-generic Programming}}, \bibinfo{series}{SSDGP'06},
  \bibinfo{publisher}{Springer-Verlag}, \bibinfo{address}{Berlin, Heidelberg},
  pp. \bibinfo{pages}{209--257}, \doi{10.1007/978-3-540-76786-2\_4}.
\newblock \urlprefix\url{http://dl.acm.org/citation.cfm?id=1782894.1782898}.

\bibitemdeclare{inproceedings}{bahr:11:compositional-data-types}
\bibitem{bahr:11:compositional-data-types}
\bibinfo{author}{Patrick \surnamestart Bahr\surnameend} \& \bibinfo{author}{Tom
  \surnamestart Hvitved\surnameend} (\bibinfo{year}{2011}):
  \emph{\bibinfo{title}{Compositional Data Types}}.
\newblock In: {\sl \bibinfo{booktitle}{Proceedings of the Seventh ACM SIGPLAN
  Workshop on Generic Programming}}, \bibinfo{series}{WGP '11},
  \bibinfo{publisher}{ACM}, \bibinfo{address}{New York, NY, USA}, pp.
  \bibinfo{pages}{83--94}, \doi{10.1145/2036918.2036930}.

\bibitemdeclare{article}{bringert:06:compos}
\bibitem{bringert:06:compos}
\bibinfo{author}{Bj\"{o}rn \surnamestart Bringert\surnameend} \&
  \bibinfo{author}{Aarne \surnamestart Ranta\surnameend}
  (\bibinfo{year}{2006}): \emph{\bibinfo{title}{A Pattern for Almost
  Compositional Functions}}.
\newblock {\sl \bibinfo{journal}{SIGPLAN Not.}}
  \bibinfo{volume}{41}(\bibinfo{number}{9}), pp. \bibinfo{pages}{216--226},
  \doi{10.1145/1160074.1159834}.

\bibitemdeclare{inproceedings}{dyn:amber:cardelli}
\bibitem{dyn:amber:cardelli}
\bibinfo{author}{Luca \surnamestart Cardelli\surnameend}
  (\bibinfo{year}{1985}): \emph{\bibinfo{title}{Amber}}.
\newblock In: {\sl \bibinfo{booktitle}{Combinators and Functional Programming
  Languages, Thirteenth Spring School of the LITP, Val d'Ajol, France, May
  6-10, 1985, Proceedings}}, pp. \bibinfo{pages}{21--47},
  \doi{10.1007/3-540-17184-3\_38}.

\bibitemdeclare{unpublished}{chakravarty09instantgenerics}
\bibitem{chakravarty09instantgenerics}
\bibinfo{author}{Manuel M.~T. \surnamestart Chakravarty\surnameend},
  \bibinfo{author}{Gabriel~C. \surnamestart Ditu\surnameend} \&
  \bibinfo{author}{Roman \surnamestart Leshchinskiy\surnameend}
  (\bibinfo{year}{2009}): \emph{\bibinfo{title}{Instant Generics: Fast and
  Easy}}.
\newblock
  \urlprefix\url{https://pdfs.semanticscholar.org/0e4b/bbf2738c6d91ce81c7f22973071eefe935e2.pdf}.
\newblock \bibinfo{note}{Draft}.

\bibitemdeclare{inproceedings}{Cheney:02:LIGD}
\bibitem{Cheney:02:LIGD}
\bibinfo{author}{James \surnamestart Cheney\surnameend} \&
  \bibinfo{author}{Ralf \surnamestart Hinze\surnameend} (\bibinfo{year}{2002}):
  \emph{\bibinfo{title}{A Lightweight Implementation of Generics and
  Dynamics}}.
\newblock In: {\sl \bibinfo{booktitle}{Proceedings of the 2002 ACM SIGPLAN
  Workshop on Haskell}}, \bibinfo{series}{Haskell '02},
  \bibinfo{publisher}{ACM}, \bibinfo{address}{New York, NY, USA}, pp.
  \bibinfo{pages}{90--104}, \doi{10.1145/581690.581698}.

\bibitemdeclare{misc}{cheplyaka:traversal}
\bibitem{cheplyaka:traversal}
\bibinfo{author}{Roman \surnamestart Cheplyaka\surnameend}
  (\bibinfo{year}{2013}): \emph{\bibinfo{title}{Generalizing generic fold}}.
\newblock
  \bibinfo{howpublished}{\url{https://ro-che.info/articles/2013-03-11-generalizing-gfoldl},
  \url{http://hackage.haskell.org/package/traverse-with-class}}.
\newblock \bibinfo{note}{Blog article}.

\bibitemdeclare{article}{gibbons:iterator}
\bibitem{gibbons:iterator}
\bibinfo{author}{Jeremy \surnamestart Gibbons\surnameend} \&
  \bibinfo{author}{Bruno~César \surnamestart dos Santos~Oliveira\surnameend}
  (\bibinfo{year}{2009}): \emph{\bibinfo{title}{The Essence of the Iterator
  Pattern}}.
\newblock {\sl \bibinfo{journal}{Journal of Functional Programming}}
  \bibinfo{volume}{19}(\bibinfo{number}{34}), pp. \bibinfo{pages}{377--402},
  \doi{10.1017/S0956796809007291}.
\newblock
  \urlprefix\url{http://www.comlab.ox.ac.uk/jeremy.gibbons/publications/iterator.pdf}.
\newblock \bibinfo{note}{Revised version of \cite{gibbons:iterator}}.

\bibitemdeclare{inproceedings}{henry:12:icfp}
\bibitem{henry:12:icfp}
\bibinfo{author}{Gr{\'e}goire \surnamestart Henry\surnameend},
  \bibinfo{author}{Michel \surnamestart Mauny\surnameend},
  \bibinfo{author}{Emmanuel \surnamestart Chailloux\surnameend} \&
  \bibinfo{author}{Pascal \surnamestart Manoury\surnameend}
  (\bibinfo{year}{2012}): \emph{\bibinfo{title}{Typing Unmarshalling Without
  Marshalling Types}}.
\newblock In: {\sl \bibinfo{booktitle}{Proceedings of the 17th ACM SIGPLAN
  International Conference on Functional Programming}}, \bibinfo{series}{ICFP
  '12}, \bibinfo{publisher}{ACM}, \bibinfo{address}{New York, NY, USA}, pp.
  \bibinfo{pages}{287--298}, \doi{10.1145/2364527.2364569}.

\bibitemdeclare{inproceedings}{hinze04generic-for-the-masses}
\bibitem{hinze04generic-for-the-masses}
\bibinfo{author}{Ralf \surnamestart Hinze\surnameend} (\bibinfo{year}{2004}):
  \emph{\bibinfo{title}{Generics for the masses}}.
\newblock In \bibinfo{editor}{Kathleen \surnamestart Fisher\surnameend},
  editor: {\sl \bibinfo{booktitle}{Proceedings of the ninth ACM SIGPLAN
  international conference on Functional Programming (ICFP '04)}},
  \bibinfo{publisher}{ACM}, \bibinfo{address}{New York, NY, USA}, pp.
  \bibinfo{pages}{236--243}, \doi{10.1145/1016850.1016882}.

\bibitemdeclare{incollection}{hinze:03:gh-applications}
\bibitem{hinze:03:gh-applications}
\bibinfo{author}{Ralf \surnamestart Hinze\surnameend} \& \bibinfo{author}{Johan
  \surnamestart Jeuring\surnameend} (\bibinfo{year}{2003}):
  \emph{\bibinfo{title}{Generic {Haskell}: {Applications}}}.
\newblock In \bibinfo{editor}{Roland \surnamestart Backhouse\surnameend} \&
  \bibinfo{editor}{Jeremy \surnamestart Gibbons\surnameend}, editors: {\sl
  \bibinfo{booktitle}{Generic Programming: {Advanced} Lectures}}, {\sl
  \bibinfo{series}{Lecture Notes in Computer Science}} \bibinfo{volume}{2793},
  \bibinfo{publisher}{Springer Berlin / Heidelberg}, pp.
  \bibinfo{pages}{57--96}, \doi{10.1007/978-3-540-45191-4\_2}.

\bibitemdeclare{inproceedings}{hinze:03:gh-theory}
\bibitem{hinze:03:gh-theory}
\bibinfo{author}{Ralf \surnamestart Hinze\surnameend} \& \bibinfo{author}{Johan
  \surnamestart Jeuring\surnameend} (\bibinfo{year}{2003}):
  \emph{\bibinfo{title}{Generic Haskell: Practice and Theory}}.
\newblock In \bibinfo{editor}{Roland \surnamestart Backhouse\surnameend} \&
  \bibinfo{editor}{Jeremy \surnamestart Gibbons\surnameend}, editors: {\sl
  \bibinfo{booktitle}{Generic Programming, Advanced Lectures}},
  \bibinfo{volume}{2793}, \bibinfo{publisher}{Springer-Verlag}, pp.
  \bibinfo{pages}{1--56}, \doi{10.1007/978-3-540-45191-4\_1}.

\bibitemdeclare{inproceedings}{hinze:2006:comparing-gp-approaches}
\bibitem{hinze:2006:comparing-gp-approaches}
\bibinfo{author}{Ralf \surnamestart Hinze\surnameend}, \bibinfo{author}{Johan
  \surnamestart Jeuring\surnameend} \& \bibinfo{author}{Andres \surnamestart
  L\"{o}h\surnameend} (\bibinfo{year}{2007}): \emph{\bibinfo{title}{Comparing
  Approaches to Generic Programming in Haskell}}.
\newblock In: {\sl \bibinfo{booktitle}{Proceedings of the 2006 International
  Conference on Datatype-generic Programming}}, \bibinfo{series}{SSDGP'06},
  \bibinfo{publisher}{Springer-Verlag}, \bibinfo{address}{Berlin, Heidelberg},
  pp. \bibinfo{pages}{72--149}, \doi{10.1007/978-3-540-76786-2\_2}.
\newblock \urlprefix\url{http://dl.acm.org/citation.cfm?id=1782894.1782896}.

\bibitemdeclare{inproceedings}{hinze:06:syb-revolutions}
\bibitem{hinze:06:syb-revolutions}
\bibinfo{author}{Ralf \surnamestart Hinze\surnameend} \&
  \bibinfo{author}{Andres \surnamestart L\"{o}h\surnameend}
  (\bibinfo{year}{2006}): \emph{\bibinfo{title}{``Scrap Your Boilerplate''
  Revolutions}}.
\newblock In: {\sl \bibinfo{booktitle}{Proceedings of the 8th International
  Conference on Mathematics of Program Construction}},
  \bibinfo{series}{MPC'06}, \bibinfo{publisher}{Springer-Verlag},
  \bibinfo{address}{Berlin, Heidelberg}, pp. \bibinfo{pages}{180--208},
  \doi{10.1007/11783596\_13}.

\bibitemdeclare{article}{generic-programming-in-3d}
\bibitem{generic-programming-in-3d}
\bibinfo{author}{Ralf \surnamestart Hinze\surnameend} \&
  \bibinfo{author}{Andres \surnamestart L{\"o}h\surnameend}
  (\bibinfo{year}{2009}): \emph{\bibinfo{title}{Generic Programming in {3D}}}.
\newblock {\sl \bibinfo{journal}{Science of Computer Programming}}
  \bibinfo{volume}{74}(\bibinfo{number}{8}), pp. \bibinfo{pages}{590--628},
  \doi{10.1016/j.scico.2007.10.006}.

\bibitemdeclare{inproceedings}{hinze:06:syb-reloaded}
\bibitem{hinze:06:syb-reloaded}
\bibinfo{author}{Ralf \surnamestart Hinze\surnameend}, \bibinfo{author}{Andres
  \surnamestart L\"{o}h\surnameend} \& \bibinfo{author}{Bruno~C.d.S.
  \surnamestart Oliveira\surnameend} (\bibinfo{year}{2006}):
  \emph{\bibinfo{title}{``{Scrap Your Boilerplate}'' Reloaded}}.
\newblock In \bibinfo{editor}{Masami \surnamestart Hagiya\surnameend} \&
  \bibinfo{editor}{Philip \surnamestart Wadler\surnameend}, editors: {\sl
  \bibinfo{booktitle}{Proceedings of the Eighth International Symposium on
  Functional and Logic Programming (FLOPS 2006)}}, {\sl
  \bibinfo{series}{Lecture Notes in Computer Science}} \bibinfo{volume}{3945},
  \bibinfo{publisher}{Springer Berlin / Heidelberg}, pp.
  \bibinfo{pages}{13--29}, \doi{10.1007/11737414\_3}.

\bibitemdeclare{inproceedings}{jansson:97:polyp}
\bibitem{jansson:97:polyp}
\bibinfo{author}{Patrik \surnamestart Jansson\surnameend} \&
  \bibinfo{author}{Johan \surnamestart Jeuring\surnameend}
  (\bibinfo{year}{1997}): \emph{\bibinfo{title}{PolyP -- a Polytypic
  Programming Language Extension}}.
\newblock In: {\sl \bibinfo{booktitle}{Proceedings of the 24th ACM
  SIGPLAN-SIGACT Symposium on Principles of Programming Languages}},
  \bibinfo{series}{POPL '97}, \bibinfo{publisher}{ACM}, \bibinfo{address}{New
  York, NY, USA}, pp. \bibinfo{pages}{470--482}, \doi{10.1145/263699.263763}.

\bibitemdeclare{inproceedings}{jansson:99:gp-intro}
\bibitem{jansson:99:gp-intro}
\bibinfo{author}{Patrik \surnamestart Jansson\surnameend},
  \bibinfo{author}{Johan \surnamestart Jeuring\surnameend} \&
  \bibinfo{author}{Lambert \surnamestart Meertens\surnameend}
  (\bibinfo{year}{1999}): \emph{\bibinfo{title}{Generic programming: An
  introduction}}.
\newblock In: {\sl \bibinfo{booktitle}{3rd International Summer School on
  Advanced Functional Programming}}, \bibinfo{publisher}{Springer-Verlag}, pp.
  \bibinfo{pages}{28--115}, \doi{10.1007/10704973\_2}.

\bibitemdeclare{inproceedings}{karvonen:2007:generics-ml}
\bibitem{karvonen:2007:generics-ml}
\bibinfo{author}{Vesa~A.J. \surnamestart Karvonen\surnameend}
  (\bibinfo{year}{2007}): \emph{\bibinfo{title}{Generics for the Working
  ML'er}}.
\newblock In: {\sl \bibinfo{booktitle}{Proceedings of the 2007 Workshop on
  Workshop on ML}}, \bibinfo{series}{ML '07}, \bibinfo{publisher}{ACM},
  \bibinfo{address}{New York, NY, USA}, pp. \bibinfo{pages}{71--82},
  \doi{10.1145/1292535.1292547}.

\bibitemdeclare{article}{lammel:05:syb-with-class}
\bibitem{lammel:05:syb-with-class}
\bibinfo{author}{Ralf \surnamestart L\"{a}mmel\surnameend} \&
  \bibinfo{author}{Simon~Peyton \surnamestart Jones\surnameend}
  (\bibinfo{year}{2005}): \emph{\bibinfo{title}{Scrap Your Boilerplate with
  Class: Extensible Generic Functions}}.
\newblock {\sl \bibinfo{journal}{SIGPLAN Not.}}
  \bibinfo{volume}{40}(\bibinfo{number}{9}), pp. \bibinfo{pages}{204--215},
  \doi{10.1145/1090189.1086391}.

\bibitemdeclare{inproceedings}{lammel:03:SYB}
\bibitem{lammel:03:SYB}
\bibinfo{author}{Ralf \surnamestart L\"{a}mmel\surnameend} \&
  \bibinfo{author}{Simon \surnamestart Peyton~Jones\surnameend}
  (\bibinfo{year}{2003}): \emph{\bibinfo{title}{Scrap Your Boilerplate: A
  Practical Design Pattern for Generic Programming}}.
\newblock In: {\sl \bibinfo{booktitle}{Proceedings of the 2003 ACM SIGPLAN
  International Workshop on Types in Languages Design and Implementation}},
  \bibinfo{series}{TLDI '03}, \bibinfo{publisher}{ACM}, \bibinfo{address}{New
  York, NY, USA}, pp. \bibinfo{pages}{26--37}, \doi{10.1145/604174.604179}.

\bibitemdeclare{manual}{ocaml-manual}
\bibitem{ocaml-manual}
\bibinfo{author}{Xavier \surnamestart Leroy\surnameend},
  \bibinfo{author}{Damien \surnamestart Doligez\surnameend},
  \bibinfo{author}{Alain \surnamestart Frisch\surnameend},
  \bibinfo{author}{Jacques \surnamestart Garrigue\surnameend},
  \bibinfo{author}{Didier \surnamestart Rémy\surnameend} \&
  \bibinfo{author}{Jérôme \surnamestart Vouillon\surnameend}
  (\bibinfo{year}{2017}): \emph{\bibinfo{title}{The {OC}aml system, release
  4.06, Documentation and user’s manual}}.
\newblock \bibinfo{organization}{INRIA}.
\newblock
  \urlprefix\url{http://caml.inria.fr/pub/docs/manual-ocaml/extn.html#sec261}.

\bibitemdeclare{article}{dyn:leroy}
\bibitem{dyn:leroy}
\bibinfo{author}{Xavier \surnamestart Leroy\surnameend} \&
  \bibinfo{author}{Michel \surnamestart Mauny\surnameend}
  (\bibinfo{year}{1993}): \emph{\bibinfo{title}{Dynamics in {ML}}}.
\newblock {\sl \bibinfo{journal}{Journal of Functional Programming}}
  \bibinfo{volume}{3}(\bibinfo{number}{4}), pp. \bibinfo{pages}{431--463},
  \doi{10.1017/S0956796800000848}.

\bibitemdeclare{inproceedings}{Magalhaes:10:generic-deriving}
\bibitem{Magalhaes:10:generic-deriving}
\bibinfo{author}{Jos{\'e}~Pedro \surnamestart Magalh\~{a}es\surnameend},
  \bibinfo{author}{Atze \surnamestart Dijkstra\surnameend},
  \bibinfo{author}{Johan \surnamestart Jeuring\surnameend} \&
  \bibinfo{author}{Andres \surnamestart L\"{o}h\surnameend}
  (\bibinfo{year}{2010}): \emph{\bibinfo{title}{A Generic Deriving Mechanism
  for Haskell}}.
\newblock In: {\sl \bibinfo{booktitle}{Proceedings of the Third ACM Haskell
  Symposium on Haskell}}, \bibinfo{series}{Haskell '10},
  \bibinfo{publisher}{ACM}, \bibinfo{address}{New York, NY, USA}, pp.
  \bibinfo{pages}{37--48}, \doi{10.1145/1863523.1863529}.

\bibitemdeclare{inproceedings}{Magalhaes:11:GPID}
\bibitem{Magalhaes:11:GPID}
\bibinfo{author}{Jos{\'e}~Pedro \surnamestart Magalh\~{a}es\surnameend} \&
  \bibinfo{author}{Johan \surnamestart Jeuring\surnameend}
  (\bibinfo{year}{2011}): \emph{\bibinfo{title}{Generic Programming for Indexed
  Datatypes}}.
\newblock In: {\sl \bibinfo{booktitle}{Proceedings of the Seventh ACM SIGPLAN
  Workshop on Generic Programming}}, \bibinfo{series}{WGP '11},
  \bibinfo{publisher}{ACM}, \bibinfo{address}{New York, NY, USA}, pp.
  \bibinfo{pages}{37--46}, \doi{10.1145/2036918.2036924}.

\bibitemdeclare{article}{mcbride:applicative}
\bibitem{mcbride:applicative}
\bibinfo{author}{Conor \surnamestart Mcbride\surnameend} \&
  \bibinfo{author}{Ross \surnamestart Paterson\surnameend}
  (\bibinfo{year}{2008}): \emph{\bibinfo{title}{Applicative Programming with
  Effects}}.
\newblock {\sl \bibinfo{journal}{J. Funct. Program.}}
  \bibinfo{volume}{18}(\bibinfo{number}{1}), pp. \bibinfo{pages}{1--13},
  \doi{10.1017/S0956796807006326}.

\bibitemdeclare{article}{Meertens1992}
\bibitem{Meertens1992}
\bibinfo{author}{Lambert \surnamestart Meertens\surnameend}
  (\bibinfo{year}{1992}): \emph{\bibinfo{title}{Paramorphisms}}.
\newblock {\sl \bibinfo{journal}{Formal Aspects of Computing}}
  \bibinfo{volume}{4}(\bibinfo{number}{5}), pp. \bibinfo{pages}{413--424},
  \doi{10.1007/BF01211391}.

\bibitemdeclare{inproceedings}{mitchell:Uniplate}
\bibitem{mitchell:Uniplate}
\bibinfo{author}{Neil \surnamestart Mitchell\surnameend} \&
  \bibinfo{author}{Colin \surnamestart Runciman\surnameend}
  (\bibinfo{year}{2007}): \emph{\bibinfo{title}{Uniform Boilerplate and List
  Processing}}.
\newblock In: {\sl \bibinfo{booktitle}{Proceedings of the ACM SIGPLAN Workshop
  on Haskell Workshop}}, \bibinfo{series}{Haskell '07},
  \bibinfo{publisher}{ACM}, \bibinfo{address}{New York, NY, USA}, pp.
  \bibinfo{pages}{49--60}, \doi{10.1145/1291201.1291208}.

\bibitemdeclare{inproceedings}{norell:03:polytypic}
\bibitem{norell:03:polytypic}
\bibinfo{author}{Ulf \surnamestart Norell\surnameend} \&
  \bibinfo{author}{Patrik \surnamestart Jansson\surnameend}
  (\bibinfo{year}{2004}): \emph{\bibinfo{title}{Polytypic Programming in
  Haskell}}.
\newblock In: {\sl \bibinfo{booktitle}{Proceedings of the 15th International
  Conference on Implementation of Functional Languages}},
  \bibinfo{series}{IFL'03}, \bibinfo{publisher}{Springer-Verlag},
  \bibinfo{address}{Berlin, Heidelberg}, pp. \bibinfo{pages}{168--184},
  \doi{10.1007/978-3-540-27861-0\_11}.

\bibitemdeclare{article}{Multiplate}
\bibitem{Multiplate}
\bibinfo{author}{Russell \surnamestart O'Connor\surnameend}
  (\bibinfo{year}{2011}): \emph{\bibinfo{title}{Functor is to Lens as
  Applicative is to Biplate: Introducing Multiplate}}.
\newblock {\sl \bibinfo{journal}{CoRR}} \bibinfo{volume}{abs/1103.2841}.
\newblock \urlprefix\url{http://arxiv.org/abs/1103.2841}.

\bibitemdeclare{inproceedings}{Oliveira:2005:typecase}
\bibitem{Oliveira:2005:typecase}
\bibinfo{author}{Bruno C. d.~S. \surnamestart Oliveira\surnameend} \&
  \bibinfo{author}{Jeremy \surnamestart Gibbons\surnameend}
  (\bibinfo{year}{2005}): \emph{\bibinfo{title}{TypeCase: A Design Pattern for
  Type-indexed Functions}}.
\newblock In: {\sl \bibinfo{booktitle}{Proceedings of the 2005 ACM SIGPLAN
  Workshop on Haskell}}, \bibinfo{series}{Haskell '05},
  \bibinfo{publisher}{ACM}, \bibinfo{address}{New York, NY, USA}, pp.
  \bibinfo{pages}{98--109}, \doi{10.1145/1088348.1088358}.

\bibitemdeclare{article}{rodriguez:08:comparing-gp-libraries}
\bibitem{rodriguez:08:comparing-gp-libraries}
\bibinfo{author}{Alexey \surnamestart Rodriguez\surnameend},
  \bibinfo{author}{Johan \surnamestart Jeuring\surnameend},
  \bibinfo{author}{Patrik \surnamestart Jansson\surnameend},
  \bibinfo{author}{Alex \surnamestart Gerdes\surnameend}, \bibinfo{author}{Oleg
  \surnamestart Kiselyov\surnameend} \& \bibinfo{author}{Bruno C. d.~S.
  \surnamestart Oliveira\surnameend} (\bibinfo{year}{2008}):
  \emph{\bibinfo{title}{Comparing Libraries for Generic Programming in
  Haskell}}.
\newblock {\sl \bibinfo{journal}{SIGPLAN Not.}}
  \bibinfo{volume}{44}(\bibinfo{number}{2}), pp. \bibinfo{pages}{111--122},
  \doi{10.1145/1543134.1411301}.

\bibitemdeclare{article}{swierstra:08:alacarte}
\bibitem{swierstra:08:alacarte}
\bibinfo{author}{Wouter \surnamestart Swierstra\surnameend}
  (\bibinfo{year}{2008}): \emph{\bibinfo{title}{Data types {\`a} la carte}}.
\newblock {\sl \bibinfo{journal}{J. Funct. Program.}}
  \bibinfo{volume}{18}(\bibinfo{number}{4}), pp. \bibinfo{pages}{423--436},
  \doi{10.1017/S0956796808006758}.

\bibitemdeclare{inproceedings}{Wadler92theessence}
\bibitem{Wadler92theessence}
\bibinfo{author}{Philip \surnamestart Wadler\surnameend}
  (\bibinfo{year}{1992}): \emph{\bibinfo{title}{The Essence of Functional
  Programming}}.
\newblock In: {\sl \bibinfo{booktitle}{Proceedings of the 19th ACM
  SIGPLAN-SIGACT Symposium on Principles of Programming Languages}},
  \bibinfo{series}{POPL '92}, \bibinfo{publisher}{ACM}, \bibinfo{address}{New
  York, NY, USA}, pp. \bibinfo{pages}{1--14}, \doi{10.1145/143165.143169}.

\bibitemdeclare{misc}{wadler:98:expression-problem}
\bibitem{wadler:98:expression-problem}
\bibinfo{author}{Philip \surnamestart Wadler\surnameend}
  (\bibinfo{year}{1998}): \emph{\bibinfo{title}{The expression problem}}.
\newblock \bibinfo{howpublished}{Posted on the Java Genericity mailing list}.
\newblock
  \urlprefix\url{http://homepages.inf.ed.ac.uk/wadler/papers/expression/expression.txt}.

\bibitemdeclare{inproceedings}{weirich:06:replib}
\bibitem{weirich:06:replib}
\bibinfo{author}{Stephanie \surnamestart Weirich\surnameend}
  (\bibinfo{year}{2006}): \emph{\bibinfo{title}{RepLib: A Library for Derivable
  Type Classes}}.
\newblock In: {\sl \bibinfo{booktitle}{Proceedings of the 2006 ACM SIGPLAN
  Workshop on Haskell}}, \bibinfo{series}{Haskell '06},
  \bibinfo{publisher}{ACM}, \bibinfo{address}{New York, NY, USA}, pp.
  \bibinfo{pages}{1--12}, \doi{10.1145/1159842.1159844}.

\bibitemdeclare{inproceedings}{modular-implicits}
\bibitem{modular-implicits}
\bibinfo{author}{Leo \surnamestart White\surnameend},
  \bibinfo{author}{Fr{\'{e}}d{\'{e}}ric \surnamestart Bour\surnameend} \&
  \bibinfo{author}{Jeremy \surnamestart Yallop\surnameend}
  (\bibinfo{year}{2014}): \emph{\bibinfo{title}{Modular implicits}}.
\newblock In: {\sl \bibinfo{booktitle}{Proceedings {ML} Family/OCaml Users and
  Developers workshops, ML/OCaml 2014, Gothenburg, Sweden, September 4-5,
  2014.}}, pp. \bibinfo{pages}{22--63}, \doi{10.4204/EPTCS.198.2}.

\bibitemdeclare{inproceedings}{Yallop:2007:Deriving}
\bibitem{Yallop:2007:Deriving}
\bibinfo{author}{Jeremy \surnamestart Yallop\surnameend}
  (\bibinfo{year}{2007}): \emph{\bibinfo{title}{Practical Generic Programming
  in OCaml}}.
\newblock In: {\sl \bibinfo{booktitle}{Proceedings of the 2007 Workshop on
  Workshop on ML}}, \bibinfo{series}{ML '07}, \bibinfo{publisher}{ACM},
  \bibinfo{address}{New York, NY, USA}, pp. \bibinfo{pages}{83--94},
  \doi{10.1145/1292535.1292548}.

\bibitemdeclare{inproceedings}{Yallop:2016:staging-generic-programming}
\bibitem{Yallop:2016:staging-generic-programming}
\bibinfo{author}{Jeremy \surnamestart Yallop\surnameend}
  (\bibinfo{year}{2016}): \emph{\bibinfo{title}{Staging Generic Programming}}.
\newblock In: {\sl \bibinfo{booktitle}{Proceedings of the 2016 ACM SIGPLAN
  Workshop on Partial Evaluation and Program Manipulation}},
  \bibinfo{series}{PEPM '16}, \bibinfo{publisher}{ACM}, \bibinfo{address}{New
  York, NY, USA}, pp. \bibinfo{pages}{85--96}, \doi{10.1145/2847538.2847546}.

\bibitemdeclare{inbook}{higher-kinded-polymorphism}
\bibitem{higher-kinded-polymorphism}
\bibinfo{author}{Jeremy \surnamestart Yallop\surnameend} \&
  \bibinfo{author}{Leo \surnamestart White\surnameend} (\bibinfo{year}{2014}):
  \emph{\bibinfo{title}{Functional and Logic Programming: 12th International
  Symposium, FLOPS 2014, Kanazawa, Japan, June 4-6, 2014. Proceedings}},
  chapter \bibinfo{chapter}{Lightweight Higher-Kinded Polymorphism}, pp.
  \bibinfo{pages}{119--135}.
\newblock \bibinfo{publisher}{Springer International Publishing},
  \bibinfo{address}{Cham}, \doi{10.1007/978-3-319-07151-0\_8}.

\end{thebibliography}

\end{document}